\documentclass[12pt,reqno]{article}

\usepackage{mathtools}
\usepackage{dirtytalk}
\usepackage{amsmath}
\usepackage{amsfonts}
\usepackage{amssymb}
\usepackage{amsxtra}
\usepackage{amsthm}
\usepackage{graphicx}
\usepackage{caption}
\usepackage{ushort}
\usepackage{color}
\usepackage{hyperref}
\usepackage[parsep]{collref}
\hypersetup{linktocpage=true}
\usepackage{enumitem}
\usepackage{titlesec}
\usepackage{cite}
\usepackage{braket}
\usepackage{flexisym}
\usepackage{dsfont}
\usepackage{titlesec}
\usepackage{romannum}
\usepackage{caption}
\usepackage{subcaption}
\usepackage{mathrsfs}
\usepackage[bottom]{footmisc}

\newcommand\numberthis{\addtocounter{equation}{1}\tag{\theequation}}
\newcommand{\mf}{\mathfrak}
\newcommand{\mb}{\mathbb}
\newcommand{\mc}{\mathcal}
\newcommand{\bd}{\boldsymbol}

\newcommand{\ov}{\overline}
\newcommand{\ph}{\phantom}

\DeclareMathOperator{\im}{Im}
\DeclareMathOperator{\re}{Re}

\DeclareMathOperator{\SU}{SU}

\DeclareMathOperator{\vol}{vol}

\DeclareMathOperator{\tr}{tr}

\newcommand{\imperial}{\it The Blackett Laboratory, Imperial College London\\
Prince Consort Road, London SW7 2AZ}

\newcommand{\auth}{K. C. Matthew Cheung and Rahim Leung}

\let\oldabstract\abstract
\let\oldendabstract\endabstract
\makeatletter
\renewenvironment{abstract}
{%
               {\list{}{\addtolength{\leftmargin}{1em} 
                        \listparindent 1.5em%
                        \itemindent    \listparindent%
                        \rightmargin   \leftmargin%
                        \parsep        \z@ \@plus\p@}%
                \item\relax}%
               {\endlist}%
\oldabstract}
{\oldendabstract}
\makeatother

\numberwithin{equation}{subsection}

\let\oldsection\section
\renewcommand{\section}{\renewcommand{\theequation}{\thesection.\arabic{equation}}\oldsection}
\let\oldsubsection\subsection
\renewcommand{\subsection}{\renewcommand{\theequation}{\thesubsection.\arabic{equation}}\oldsubsection}

\titlespacing*{\section}
{0pt}{5.5ex plus 1ex minus .2ex}{4.3ex plus .2ex}
\titlespacing*{\subsection}
{0pt}{5.5ex plus 1ex minus .2ex}{4.3ex plus .2ex}

\begin{document}
\setcounter{page}{0}
\thispagestyle{empty}

\begin{center}  

{\Large {\bf Wrapped NS5-Branes, Consistent Truncations and In\"{o}n\"{u}-Wigner Contractions}}   

\vspace{15pt}

\auth

\vspace{7pt}
\imperial

\end{center} 

\vspace{10pt}

\begin{abstract}

\noindent We construct consistent Kaluza-Klein truncations of type \Romannum{2}A supergravity on (\romannum{1}) $\Sigma_2\times S^3$ and (\romannum{2}) $\Sigma_3\times S^3$, where $\Sigma_2 = S^2/\Gamma$, $\mb{R}^2/\Gamma$, or $\mb{H}^2/\Gamma$, and $\Sigma_3 = S^3/\Gamma$, $\mb{R}^3/\Gamma$, or $\mb{H}^3/\Gamma$, with $\Gamma$ a discrete group of symmetries, corresponding to NS5-branes wrapped on $\Sigma_2$ and $\Sigma_3$. The resulting theories are a $D=5$, $\mc{N}=4$ gauged supergravity coupled to three vector multiplets with scalar manifold $SO(1,1)\times SO(5,3)/(SO(5)\times SO(3))$ and gauge group $SO(2)\times\left(SO(2)\ltimes_{\Sigma_2}\mb{R}^4\right)$ which depends on the curvature of $\Sigma_2$, and a $D=4$, $\mc{N}=2$ gauged supergravity coupled to one vector multiplet and two hypermultiplets with scalar manifold $SU(1,1)/U(1)\times G_{2(2)}/SO(4)$ and gauge group $\mb{R}^+\times\mb{R}^+$ for truncations (\romannum{1}) and (\romannum{2}) respectively. Instead of carrying out the truncations at the 10-dimensional level, we show that they can be obtained directly by performing In\"{o}n\"{u}-Wigner contractions on the 5 and 4-dimensional gauged supergravity theories that come from consistent truncations of 11-dimensional supergravity associated with M5-branes wrapping $\Sigma_2$ and $\Sigma_3$. This suggests the existence of a broader class of lower-dimensional gauged supergravity theories related by group contractions that have a 10 or 11-dimensional origin.

\end{abstract}
\vfill\leftline{}\vfill
\pagebreak

\tableofcontents
\addtocontents{toc}{\protect\setcounter{tocdepth}{2}}
\pagenumbering{arabic}
\setcounter{page}{1}
\setcounter{footnote}{0}

\section{Introduction}\label{introduction}

Supergravity theories in 10/11 dimensions are low-energy approximations of string/M-theory, and the studies of these theories provide invaluable insight into the rich structure of their high-energy counterparts. The construction of solutions of higher-dimensional supergravity theories, however, is a difficult task. A particularly powerful framework that has been developed over the years to tackle this problem is consistent Kaluza-Klein (KK) reductions. These truncations reduce the higher-dimensional equations of motion to a set of lower-dimensional equations obtainable from a lower-dimensional supergravity theory, which are much easier to solve. 

In this paper, we present two new consistent KK truncations of $D=10$ type \Romannum{2}A supergravity on (\romannum{1}) $\Sigma_2\times S^3$, where $\Sigma_2=S^2$, $\mb{R}^2$, $\mb{H}^2$ or a quotient thereof, to a gauged $\mc{N}=4$ supergravity theory in $D=5$, and (\romannum{2}) $\Sigma_3\times S^3$, where $\Sigma_3=S^3$, $\mb{R}^3$, $\mb{H}^3$ or a quotient thereof, to a gauged $\mc{N}=2$ supergravity theory in $D=4$, at the level of the bosonic fields. The $S^3$ factor common to both truncations corresponds to the standard $S^3$ truncation of type \Romannum{2}A supergravity to the maximal $ISO(4)$ gauged supergravity in $D=7$. Within this $D=7$ theory is a \say{vacuum} solution that uplifts to the NS5-brane near horizon, linear dilaton solution. The further truncations on $\Sigma_2$ and $\Sigma_3$ then correspond to the worldvolume of the NS5-brane wrapping these geometries. In particular, $\Sigma_2$ and $\Sigma_3$ are interpreted as a Slag/K\"ahler 2-cycle and Slag 3-cycle of a Calabi--Yau 2- and 3-fold ($CY_2$, and $CY_3$) respectively. A motivation for truncation (\romannum{1}) stems from the supergravity solution in \cite{Gauntlett:2001ps} and \cite{Bigazzi:2001aj}, which describes the near horizon limit of NS5-branes wrapping an $S^2$, embedded inside a $CY_2$. The dual $D=4$ theory can be viewed as the IR limit of the little string theory compactified on $S^2$ with a topological twist. For truncation (\romannum{2}), we similarly note the supergravity solution presented in \cite{Gauntlett:2001ur}, which describes the near-horizon limit of NS5-branes wrapping an $S^3$, embedded inside a $CY_3$. The dual $D=3$ theory can be viewed as the IR limit of the little string theory compactified on $S^3$ with a topological twist. 

The topological twist is an important feature of wrapped brane solutions. Schematically, the Killing spinor equation on the worldvolume of a brane wrapped on a cycle $\Sigma$ is $\left(d + \omega_{(1)}- A_{(1)}\right)\epsilon = 0$, with $\omega_{(1)}$ the spin connection on $\Sigma$, and $A_{(1)}$ the gauge field that couples to the R-symmetry current. This equation, in general, does not admit covariantly constant spinors, in which case supersymmetry is broken. An elegant solution to this, as pioneered in \cite{Witten:1988ze,Bershadsky:1995qy} and applied to constructing supergravity solutions by \cite{Maldacena:2000mw,Maldacena:2000yy,Brinne:2000fh,Acharya:2000mu,Gauntlett:2000ng,Nieder:2000kc,Gauntlett:2001jj,Nunez:2001pt,Edelstein:2001pu,Gomis:2001vg,Hernandez:2001bh,Schvellinger:2001ib,Naka:2002jz,Gauntlett:2001qs,Benini:2013cda,Suh:2018tul,Kim:2019fsg}, is to set the gauge field to be equal to the spin connection on the cycle --- the \say{twist}, so that the Killing spinor equation admits covariantly constant spinors. This topological twist will be fully incorporated in both our truncations in order to preserve supersymmetry. 

\begin{figure}[h]
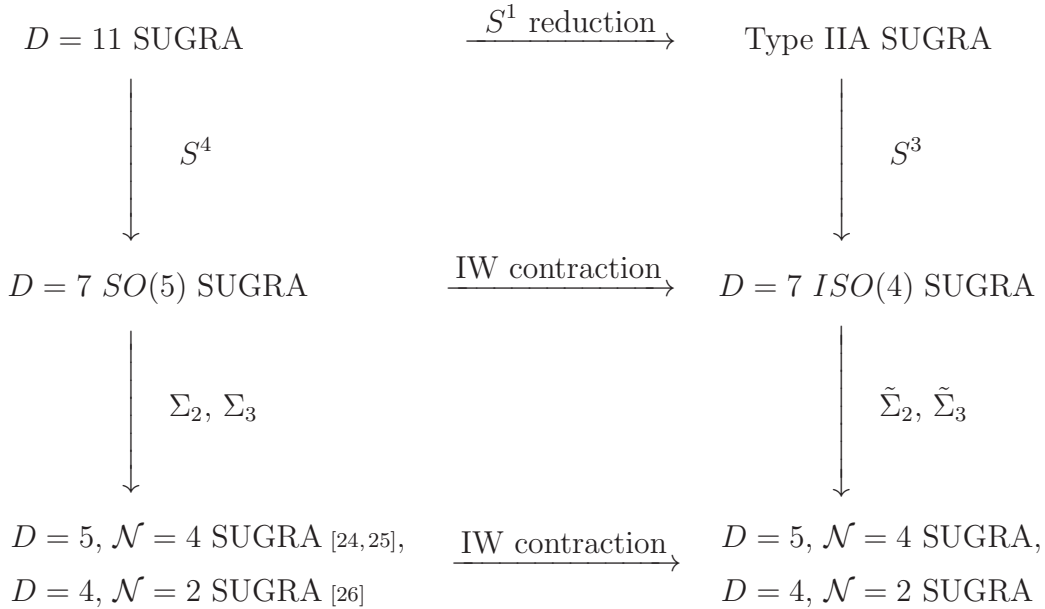

\begin{equation}
    \begin{aligned}
        & \text{\,\,\,\,\,$D=11$ SUGRA}
        & \xrightarrow{\text{$\,\,\,\,$$S^1$ reduction$\,\,\,$}}\hspace{4.2mm}
        & \text{\,\,\,\,\, Type \Romannum{2}A SUGRA}\\
        & \hspace{12.0mm}\rotatebox[origin=c]{270}{$\xrightarrow[\phantom{\text{expansion1}}]{\phantom{\text{integration1}}}$}\text{ $S^4$}
        & \hspace{2.15mm}\resizebox{10mm}{7.5mm}{$\phantom{xx}$}\hspace{4.25mm}
        & \hspace{12.0mm}\rotatebox[origin=c]{270}{$\xrightarrow[\phantom{\text{expansion1}}]{\phantom{\text{integration1}}}$}\text{ $S^3$}\\
        & \text{\,\,$D=7$ $SO(5)$ SUGRA}
        & \xrightarrow{\text{ IW contraction }}\hspace{4.25mm}
        &  \text{\,\,$D=7$ $ISO(4)$ SUGRA}\\
        & \hspace{12.0mm}\rotatebox[origin=c]{270}{$\xrightarrow[\phantom{\text{expansion1}}]{\phantom{\text{integration1}}}$}\text{$\Sigma_2$, $\Sigma_3$}
        & \hspace{2.15mm}\resizebox{10mm}{7.5mm}{$\phantom{xx}$}\hspace{4.25mm}
        & \hspace{12.0mm}\rotatebox[origin=c]{270}{$\xrightarrow[{\phantom{integration1}}]{\phantom{\text{expansion1}}}$}\text{$\tilde{\Sigma}_2$, $\tilde{\Sigma}_3$}\\
        &\begin{array}{lll}\text{$D=5$, $\mc{N}=4$ SUGRA\scriptsize\cite{Cheung:2019pge,Cassani:2019vcl}}, \\\text{$D=4$, $\mc{N}=2$ SUGRA\scriptsize\cite{Donos:2010ax}}\end{array}
        & \xrightarrow{\text{ IW contraction }}\hspace{3.5mm}
        & \begin{array}{lll}\text{$D=5$, $\mc{N}=4$ SUGRA}, \\\text{$D=4$, $\mc{N}=2$ SUGRA}\end{array}\nonumber
    \end{aligned}
\end{equation}
\caption{The possible routes of truncation. The IW contraction can be performed at any of the specified points, but it is computationally easiest at the 4/5-dimensional level.}\label{fig:truncation_diagram}
\end{figure}

To carry out truncations (\romannum{1}) and (\romannum{2}), the straightforward method would be to first reduce the type \Romannum{2}A theory on $S^3$ to obtain the maximal $ISO(4)$ gauged supergravity in $D = 7$, and then further reducing on a Riemann surface $\Sigma_2$ to obtain the $D=5$ theory, or on a Slag 3-cycle $\Sigma_3$ to obtain the $D=4$ theory. Instead, we will show that the truncations can be carried out consistently starting from the $D=5$ and $D=4$ theories obtained from an M5-brane wrapping $\Sigma_2$ and $\Sigma_3$, which we will discuss below, by performing In\"{o}n\"{u}-Wigner (IW) contractions. In terms of the 11-dimensional supergravity theory where the M5-brane lives, the IW contraction corresponds to the group contraction which takes $S^4 \to S^3\times\mb{R}$, where $S^4$ is the internal 4-sphere of the M5-brane. The opening of an isometry direction along $\mb{R}$ allows for the truncation of the 11-dimensional theory to the type \Romannum{2}A theory, as well as the interpretation of the M5-brane becoming the NS5-brane. This was shown in \cite{Cvetic:2003xr} as corresponding to a consistent transition from the $D=7$ maximal $SO(5)$ gauged supergravity theory to the $D=7$ maximal $ISO(4)$ gauged supergravity theory. Our full truncation procedure is summarised in figure \ref{fig:truncation_diagram}.

The key message from figure \ref{fig:truncation_diagram} is that by virtue of the consistency of the IW contraction, once the theory describing an M5-brane wrapped on a calibrated cycle is known, the theory describing an NS5-brane wrapping on the same cycle can be obtained accordingly. To be concrete, we will first describe our procedure for truncation (\romannum{1}). We start from the KK truncation in $D = 11$, first by reducing on $S^4$ to the maximal $SO(5)$ gauged supergravity in $D = 7$ and then further reducing on the Riemann surface $\Sigma_2$. The result is a gauged $D = 5$, $\mc{N} = 4$ (16 supercharges) supergravity theory coupled to three vector multiplets with gauge group $SO(2)\times ISO(3)$, corresponding to the aforementioned wrapped M5-brane truncation on a Riemann surface described in \cite{Cheung:2019pge} and \cite{Cassani:2019vcl}. Here, at the 5-dimensional level, we perform the IW contraction given in \cite{Cvetic:2003xr} to obtain a new $D=5$, $\mc{N}=4$ gauged supergravity theory coupled to three vector multiplets with scalar manifold $SO(1,1)\times SO(5,3)/(SO(5)\times SO(3))$. The scalar manifold of this new $D=5$ theory is the same as the $D=5$ theory in \cite{Cheung:2019pge,Cassani:2019vcl}. However, this should not come as a surprise since the IW contraction procedure retains the same number of degrees of freedom. Along with the stringent condition set by $\mc{N}=4$ supersymmetry, this ensures that the scalar manifold must remain the same. As we will show, the gauge group of the truncated $D=5$ theory is $SO(2)\times G_{A^{100}_{5,17}}$ when $\Sigma_2=\mb{R}^2/\Gamma$, and $SO(2)\times G_{A^{0}_{5,18}}$ when $\Sigma_2=S^2/\Gamma$ or $\mb{H}^2/\Gamma$, where $G_{A^{100}_{5,17}}$ and $G_{A^{0}_{5,18}}$ are two, five-dimensional matrix groups whose Lie algebras are listed in \cite{Patera:1976ud}. The groups $G_{A^{100}_{5,17}}$ and $G_{A^{0}_{5,18}}$ are isomorphic to $SO(2)\ltimes_{\Sigma_2}\mb{R}^4$, where the action of the semi-direct product depends on the curvature of $\Sigma_2$. As a consequence of the appearance of these unconventional gauge groups, the precise details of the gauging, such as the embedding tensors, as well as the vacuum structure of the theory, are entirely different from that of \cite{Cheung:2019pge,Cassani:2019vcl}. The method for truncation (\romannum{2}) proceeds analogously. We first truncate the $D = 7$ maximal theory on a Slag 3-cycle $\Sigma_3$ as described in \cite{Donos:2010ax} to obtain a $D = 4$, $\mc{N} = 2$ gauged supergravity theory coupled to a single vector multiplet and two hypermultiplets with gauge group $U(1)\times\mb{R}^+$. Then, at the 4-dimensional level, we perform the same IW contraction and obtain a new $D=4$, $\mc{N}=2$ gauged supergravity theory coupled to one vector multiplet and two hypermultiplets with scalar manifold $SU(1,1)/U(1)\times G_{2(2)}/SO(4)$ and gauge group $\mb{R}^+\times\mb{R}^+$ \footnote{For $\Sigma_3=S^3$, truncation (\romannum{2}) corresponds to a KK truncation of type \Romannum{2}A on $S^3\times S^3$. We note that the resulting $D=4$, $\mc{N}=2$ theory is not related to the $D=4$, $\mc{N}=4$ Freedman-Schwarz model \cite{Freedman:1978ra} which can also be obtained from reducing type \Romannum{2}A on $S^3\times S^3$ \cite{Chamseddine:1997mc} (for more details see \cite{Chamseddine:1997mc,Cowdall:1998bu}), as the precise details of the two truncation procedures are different.}. Similar to the $D=5$ case, the scalar manifold of our new $D=4$ theory is the same as the $D=4$ theory in \cite{Donos:2010ax}, but the precise details of the gauging and the vacuum structure of the theories are entirely different.


The plan of the rest of the paper is as follows. In Section \ref{Review of $S^4$ and $S^3$ reductions} we review the $S^4$ reduction of 11-dimensional supergravity to the maximal $SO(5)$ gauged supergravity in $D=7$, and how it relates to the $S^3$ reduction of type \Romannum{2}A to the maximal $ISO(4)$ theory through the IW contraction. Following this, in Sections \ref{KK_truncations_Riemann} and \ref{KK_truncations_Slag}, we review the theories describing wrapped M5-branes on Riemann surfaces and Slag 3-cycles, and construct the analogous theories describing wrapped NS5-branes using the IW contractions introduced in Section \ref{Review of $S^4$ and $S^3$ reductions}. There, we will also consider consistent sub-truncations of the wrapped NS5-brane theories, and present new and reproduce old solutions. We conclude with a few final remarks in Section \ref{conclusion}, and collect some useful results in the appendices.

\section{Review of $S^4$ and $S^3$ reductions and the $D=7$ maximal $ISO(4)$ gauged theory}\label{Review of $S^4$ and $S^3$ reductions}

In this section, we will review the $D=7$ maximal $SO(5)$ gauged supergravity and the $D=7$ maximal $ISO(4)$ gauged supergravity, which arise from consistent KK truncations of $D=11$ supergravity and type \Romannum{2}A supergravity respectively, and remind readers how the two theories are related through the IW contraction.

The $D=7$ maximal $SO(5)$ gauged supergravity can be obtained by performing a Pauli reduction of $D=11$ supergravity on $S^4$. The details of this truncation, including the explicit demonstration of its consistency, are given in \cite{Nastase:1999cb,Nastase:1999kf}. The bosonic field content of the theory is comprised of a metric, $SO(5)$ Yang-Mills one-forms $A^{ij}_{(1)}$ transforming in the $\bd{10}$ of $SO(5)$, three-forms $S^i_{(3)}$ transforming in the $\bd{5}$ of $SO(5)$, and fourteen scalar fields given by a symmetric unimodular matrix $T^{ij}$ that parametrises the coset $SL(5,\mb{R})/SO(5)$. The Lagrangian of the bosonic sector of the theory is given by \footnote{Our convention for the Hodge dual is $\ast(dx^{m_1}\wedge\cdots\wedge dx^{m_p}) = \frac{1}{q!}\sqrt{|g|}\epsilon_{n_1\cdots n_q}^{\ph{n_1\cdots n_q}m_1\cdots m_p} dx^{n_1}\wedge\cdots\wedge dx^{n_q}$, where $\epsilon_{n_1\cdots n_D}$ with lowered indices is numerical, and $q = D-p$.}
\begin{align}
\begin{split}
\mc{L}_{(7)}&=R\vol_7-\frac{1}{4}T^{-1}_{ij}{\ast D}T_{jk}\wedge T^{-1}_{kl} DT_{li}-\frac{1}{4}T^{-1}_{ik}T^{-1}_{jl} {\ast F^{ij}_{(2)}}\wedge F^{kl}_{(2)}-\frac{1}{2}T_{ij}\, {\ast S^i_{(3)}}\wedge S^j_{(3)}\\
&\quad+\frac{1}{2g}S^i_{(3)}\wedge DS^i_{(3)}-\frac{1}{8g}\epsilon_{ij_1j_2j_3j_4}\,S^i_{(3)}\wedge F^{j_1j_2}_{(2)}\wedge F^{j_3j_4}_{(2)}+\frac{1}{g}\Omega_{(7)}- V\vol_7 \,,
\end{split}
\end{align}
with 
\begin{align}
\begin{split}
DT_{ij}&\equiv dT_{ij}+gA^{ik}_{(1)}T_{kj}+gA^{jk}_{(1)}T_{ik} \,,\\
DS^i_{(3)}&\equiv dS^i_{(3)}+gA^{ij}_{(1)}\wedge S^j_{(3)}\,,\\
F^{ij}_{(2)}&\equiv dA^{ij}_{(1)}+gA^{ik}_{(1)}\wedge A^{kj}_{(1)} \,,
\end{split}
\end{align}
where $g$ is the coupling constant. The scalar potential is given by
\begin{equation}
V=\frac{1}{2}{g}^2\bigg(2\mathrm{Tr}(T^2)-(\mathrm{Tr}T)^2\bigg)
\end{equation} 
and $\Omega_{(7)}$ denotes the Chern-Simons terms for the Yang-Mills fields, which will not be important for our discussions here.

Any solution to the $D=7$ maximal theory lifts to a solution of $D=11$ supergravity, and the uplift formulae are provided in \cite{Nastase:1999cb,Nastase:1999kf}. Most notably, the $AdS_7$ vacuum solution with $A^{ij}_{(1)}=S^i_{(3)}=0$ and $T_{ij}=\delta_{ij}$ uplifts to the maximally supersymmetric $AdS_7\times S^4$ solution, which describes the near horizon limit of a stack of M5-branes. In \cite{Maldacena:2000mw}, a half-maximal supersymmetric $AdS_5\times \mb{H}^2/\Gamma$ solution was found, with $\Gamma$ a Fuchsian subgroup of $\mb{H}^2$. The uplift of this solution has the form $AdS_5\times \mb{H}^2/\Gamma\times S^4$, with the $S^4$ non-trivially fibred over $\mb{H}^2/\Gamma$. The solution is dual to an $\mc{N} = 2$ superconformal field theory in four dimensions that arises from the non-compact part of M5-branes wrapping a Riemann surface that is embedded in a $CY_2$. In \cite{Gauntlett:2000ng}, a supersymmetric $AdS_4\times \mb{H}^3/\Gamma$ solution preserving 8 supercharges was found. This uplifts to $AdS_4\times \mb{H}^3/\Gamma\times S^4$, where the $S^4$ is non-trivially fibred over $\mb{H}^3/\Gamma$. The solution is dual to $\mc{N} = 2$ superconformal field theory in three dimensions that arises from the non-compact part of M5-branes wrapping a Slag 3-cycle that is embedded in a $CY_3$. These solutions in \cite{Maldacena:2000mw} and \cite{Gauntlett:2000ng} motivate the constructions of consistent truncations of the $D=7$ maximal theory in \cite{Cheung:2019pge,Cassani:2019vcl,Donos:2010ax}, which we will briefly review in Sections \ref{M5CY2truncations_review} and \ref{M5CY3truncations_review}.

We are interested in the analogous story involving NS5-branes. The natural setting for this is the $D=7$ maximal $ISO(4)$ gauged supergravity theory, which can either be obtained by performing a Pauli reduction of type \Romannum{2}A supergravity on $S^3$ \footnote{The existence of a consistent truncation of IIA supergravity on $S^3$ leading to $D = 7$, $ISO(4)$ supergravity was first suggested in \cite{Boonstra:1998mp}.}, interpreted as the internal 3-sphere of a stack of NS5-branes \cite{Cvetic:2000dm,Cvetic:2000ah}, or by taking an IW contraction of the $D=7$ maximal theory which brings the $SO(5)$ gauge group to $ISO(4)$ \cite{Cvetic:2003xr}. The IW contraction procedure, as outlined in \cite{Cvetic:2003xr}, involves decomposing the $SO(5)$ vector indices in a $4+1$ split, then rescaling all the fields by a contraction parameter $k$ which is taken to zero at the end so that the gauge group becomes $ISO(4)$. To be precise, the decomposition and rescaling of the fields in the maximal theory is given by:
\begin{equation}
\begin{split}\label{7d_IW_ansatz}
&g=k^2\tilde{g}\,,\quad A_{(1)}^{5A}=k^3\tilde A^{5A}\,,\quad A_{(1)}^{AB}=k^{-2}\tilde{A}_{(1)}^{AB}\,,\quad S^5_{(3)}=k^{-4}\tilde{S}_{(3)}\,,\quad S^A_{(3)}=k \tilde S^A_{(3)} \,,\\
&T_{ij} = \begin{pmatrix} k^{-2}\tilde\Phi^{1/4} \tilde T_{AB} & -k^3\tilde\Phi^{1/4}(\tilde T\tilde\tau)^A \\ -k^3\tilde\Phi^{1/4}(\tilde T\tilde\tau)^A & k^8\tilde\Phi^{-1} + k^8 \tilde\Phi^{1/4}\tilde T_{CD}\tilde\tau^C\tilde\tau^D \end{pmatrix} \,,\quad g_{mn}=\tilde g_{mn} \,,
\end{split}
\end{equation}
where $A,B\in\{1,\ldots,4\}$. Compared to \cite{Cvetic:2003xr}, our $5^{\text{th}}$ index is their $0^{\text{th}}$ index. We note that there is an error in the $(0,0)$ component (our $(5,5)$ component) of the decomposition of $T_{ij}$ given in \cite{Cvetic:2003xr}, which rendered $\det T\neq1$. We have fixed this issue in \eqref{7d_IW_ansatz}. 

After substituting \eqref{7d_IW_ansatz} into the equations of motion of the $SO(5)$ gauged theory and taking the limit $k\rightarrow 0$, one obtains the $D=7$ maximal $ISO(4)$ gauged supergravity, whose equations of motion are recorded in appendix \ref{7dequations}. In terms of the $D=11$ and type \Romannum{2}A theories, the IW contraction corresponds to taking the $S^4$ on which the $D=11$ theory is reduced on, and turning it into $S^3\times\mb{R}$, with $\mb{R}$ an isometry direction. To see this, let $\mu^i$, $i\in\{1,\dots,5\}$, be the embedding coordinates of $S^4$ in $\mb{R}^5$ given by 
\begin{equation}\label{s4constraint}
\mu^i\mu^i = 1 \,.
\end{equation} 
The IW contraction in \eqref{7d_IW_ansatz}, now interpreted as a set of singular rescalings of the metric and 4-form flux in the $D=11$ theory, is accompanied by an additional rescaling of the $\mu^i$ coordinates \cite{Cvetic:2003xr}. We split $\mu^i$ into $\mu^A$ and $\mu^5$ with $A\in\{1,\dots,4\}$, and rescale 
\begin{equation}
\mu^A = \tilde\mu^A\,,\quad \mu^5 = k^5\tilde\mu^5 \,.
\end{equation} 
The original constraint \eqref{s4constraint} in the limit $k\to0$ now reads 
\begin{equation}
\tilde\mu^A\tilde\mu^A = 1\,,
\end{equation} 
with $\tilde\mu^5$ unconstrained. The resulting topology is $S^3\times\mb{R}$, with $\tilde\mu^A$ parameterising the $S^3$, and $\tilde\mu^5$ parameterising $\mb{R}$.

The bosonic field content of the $ISO(4)$ gauged theory consists of a metric, $SO(4)$ Yang-Mills 1-forms $\tilde{A}^{AB}_{(1)}$ transforming in the $\bd 6$ of $SO(4)$, four 1-forms $\tilde A^{5A}_{(1)}$ in the $\bd 4$ of $SO(4)$, 3-forms $\tilde{S}^A_{(3)}$ transforming in the $\bd{4}$ of $SO(4)$, a 3-form $\tilde S_{(3)}$, four scalars $\tilde\tau^A$ in the $\bd{4}$ of $SO(4)$, and ten scalar fields given by $\tilde\Phi$ and a symmetric unimodular matrix $\tilde{T}^{AB}$ parametrising the coset $SL(4,\mb{R})/SO(4)$. By defining the Yang-Mills field strength
\begin{equation}\label{7dym}
\tilde{F}^{AB}_{(2)}\equiv d\tilde{A}^{AB}_{(1)}+\tilde{g}\tilde{A}^{AC}_{(1)}\wedge \tilde{A}^{CB}_{(1)}\,, 
\end{equation}
the covariant derivatives 
\begin{equation}\label{7dcovderiv}
\begin{split}
\tilde D\tilde{S}^A_{(3)}&\equiv d\tilde{S}^A_{(3)}+\tilde{g}\tilde{A}^{AB}_{(1)}\wedge \tilde{S}^B_{(3)}\,,\\
\tilde D\tilde{A}^{5A}_{(1)}&\equiv d\tilde{A}^{5A}_{(1)}+\tilde{g}\tilde{A}^{AB}_{(1)}\wedge \tilde{A}^{5B}_{(1)}\,,\\
\tilde D\tilde{T}_{AB}&\equiv d\tilde{T}_{AB}+\tilde{g}\tilde{A}^{AC}_{(1)}\tilde{T}_{CB}+\tilde{g}\tilde{A}^{BC}_{(1)}\tilde{T}_{AC}\,,\\
\tilde D\tilde\tau^A &\equiv d\tilde\tau^A + \tilde g\tilde\tau^B\tilde A^{AB}_{(1)} \,,
\end{split}
\end{equation} 
the following combinations of fundamental fields
\begin{equation}\label{7dcombinations}
\begin{split}
&\tilde G^A_{(3)} = \tilde S^A_{(3)} - \tilde\tau^A \tilde S_{(3)} \,,\\
&\tilde G^A_{(2)} = \tilde D\tilde A^{5A}_{(1)} + \tilde\tau^B \tilde F^{BA}_{(2)} \,,\\
&\tilde G^A_{(1)} = \tilde D\tilde\tau^A - \tilde g \tilde A^{5A}_{(1)} \,,
\end{split}
\end{equation} 
and making use of \eqref{7dbianchi} to integrate $\tilde S_{(3)}$ as 
\begin{equation}
\tilde S_{(3)} = d\tilde B_{(2)} + \frac{1}{8}\epsilon_{ABCD}\left(\tilde F^{AB}_{(2)}\wedge\tilde A^{CD}_{(1)} - \frac{1}{3}\tilde g\tilde A^{AB}_{(1)}\wedge\tilde A^{CE}_{(1)}\wedge\tilde A^{ED}_{(1)}\right) \,,
\end{equation}
the Lagrangian of the bosonic sector is given by
\begin{equation}\label{7dmaximal}
\begin{split}
\mc{L}_{(7)}&=\tilde{R}\,\tilde\vol_7-\frac{5}{16}\Phi^{-2}{\tilde{\ast}d\tilde\Phi}\wedge d\tilde\Phi-\frac{1}{4}\tilde{T}^{-1}_{AB}\tilde{T}^{-1}_{CD} {\tilde{\ast} \tilde D}\tilde{T}_{BC}\wedge \tilde D\tilde{T}_{DA}\\
&\quad- \frac{1}{2}\tilde\Phi^{5/4}\tilde T_{AB}\tilde\ast\tilde G^A_{(1)}\wedge\tilde G^B_{(1)}-\frac{1}{4}\tilde\Phi^{-1/2}\tilde{T}^{-1}_{AC}\tilde{T}^{-1}_{BD} {\tilde{\ast}\tilde{F}^{AB}_{(2)}}\wedge \tilde{F}^{CD}_{(2)} \\
&\quad-\frac{1}{2}\tilde\Phi^{3/4}\tilde T^{-1}_{AB}\tilde\ast\tilde G^A_{(2)}\wedge\tilde G^B_{(2)} -\frac{1}{2}\tilde\Phi^{-1} {\tilde{\ast} \tilde{S}_{(3)}}\wedge \tilde{S}_{(3)} -\frac{1}{2}\tilde\Phi^{1/4}\tilde T_{AB}\tilde\ast G^A_{(3)}\wedge\tilde G^B_{(3)} \\
&\quad- \tilde{V}\tilde\vol_7 + \frac{1}{2\tilde g}\tilde D\tilde S^A_{(3)}\wedge\tilde S^A_{(3)} +\tilde S^A_{(3)}\wedge\tilde S^{}_{(3)}\wedge\tilde A^{5A}_{(1)}+\frac{1}{\tilde{g}}\tilde{\Omega}_{(7)}\\
&\quad +\frac{1}{2\tilde g}\epsilon_{ABCD}\tilde S^A_{(3)}\wedge\tilde D\tilde A^{5B}_{(1)}\wedge\tilde F^{CD}_{(2)} + \frac{1}{4}\epsilon_{ABCD}\tilde S^{}_{(3)}\wedge\tilde F^{AB}_{(2)}\wedge\tilde A^{5C}_{(1)}\wedge\tilde A^{5D}_{(1)} \,.
\end{split}
\end{equation}
The scalar potential is 
\begin{equation}
\tilde{V}=\frac{1}{2}{\tilde{g}}^2\tilde\Phi^{1/2}\left(2\mathrm{Tr}(\tilde{T}^2)-(\mathrm{Tr}\tilde{T})^2\right)\,,
\end{equation}
and $\tilde{\Omega}_{(7)}$ denotes the Chern-Simons terms depending only on $\tilde A^{AB}_{(1)}$ and $\tilde A^{5A}_{(1)}$, which again, will not be important for our following discussions. There is a consistent truncation of this maximal theory to a half-maximal $SO(4)$ gauged theory obtained by setting 
\begin{equation}\label{halfmax}
\tilde\tau^A = 0 \,,\quad \tilde A^{5A}_{(1)} = 0 \,,\quad \tilde S^A_{(3)} = 0 \,,
\end{equation}
where the removal of the $\tilde A^{5A}_{(1)}$ fields breaks the $ISO(4)$ gauge group to $SO(4)$. We will call this the half-maximal truncation, and it will be used in our next sections. In the context of the type IIA theory in 10 dimensions, the half-maximal truncation corresponds to the removal of the RR sector. 

Any solution to the $D=7$ maximal theory lifts to a solution of $D=10$ type \Romannum{2}A supergravity, and the uplift formulae are provided in \cite{Cvetic:2000dm,Cvetic:2000ah}. Most notably, the linear dilaton solution with $\tilde{A}^{AB}_{(1)}=\tilde{S}_{(3)}=0$ and $\tilde{T}_{AB}=\delta_{AB}$ preserves 16 supercharges and uplifts to the supersymmetric $D=10$ solution, which describes the near horizon limit of a stack of NS5-branes. Similar to the M5 case, supersymmetric solutions corresponding to NS5-branes wrapping calibrated cycles, like an $S^2$ in $CY_2$ and an $S^3$ in $CY_3$, were constructed in \cite{Gauntlett:2001ps,Bigazzi:2001aj} and \cite{Gauntlett:2001ur} respectively. The uplift of these solutions to the type \Romannum{2}A theory are dual to compactifying little string theory on the calibrated cycles with a topological twist, and the geometry of the solutions has the internal $S^3$ non-trivially fibred over the cycles. These solutions motivate our construction of the consistent truncations of the maximal $D=7$ theory on the calibrated cycles. By virtue of the consistency of the IW contraction, we can obtain such truncations by directly applying \eqref{7d_IW_ansatz} to the corresponding truncations describing M5-branes wrapping on the appropriate cycles. This will be done in the next two sections. We will also make use of the half-maximal truncation in \eqref{halfmax} to obtain the corresponding truncations of the NSNS sector of the type \Romannum{2}A/B theory, or equivalently, the type \Romannum{1} theory.

\section{Consistent KK truncations on Riemann surfaces in $CY_2$}\label{KK_truncations_Riemann}

\subsection{Review of wrapped M5-branes on Riemann surfaces in $CY_2$}\label{M5CY2truncations_review}

We first summarise the consistent KK ansatz used in \cite{Cheung:2019pge} for M5-branes wrapped on Riemann surfaces. The ansatz for the $D = 7$ metric is given by
\begin{equation}
ds^2_7=e^{-4\phi}ds^2_{5}+e^{6\phi}ds^2(\Sigma_2) \,,
\end{equation} 
where $\phi$ is a real scalar field defined on the five-dimensional spacetime. We denote \{$\bar{e}^m;m\in\{0,\ldots,4\}$\} and \{$\bar{e}^a;a\in\{1,2\}$\} as the orthonormal frames for both $ds^2_{5}$ and $ds^2(\Sigma_2)$ respectively, and let $\bar{\omega}^{m}_{\phantom{m}n}$ and $\bar{\omega}^{a}_{\phantom{a}b}$ be the corresponding spin connections. The metric of the Riemann surface is normalised to $R_{ab}=lg^2\delta_{ab}$, with $l=0,\pm1$. The $SO(5)$ fields are decomposed via $SO(5)\to SO(2)\times SO(3)$, where the vector index of $SO(5)$ accordingly decomposes as $i = (a,\alpha)$, with $a\in\{1,2\}$ and $\alpha\in\{3,4,5\}$. The decomposition of the fields are given by
\begin{equation}
\begin{split}
&A^{ab}_{(1)}=\frac{1}{g}\bar{\omega}^{ab}+\epsilon^{ab}A_{(1)}\,,\\
&A^{a\alpha}_{(1)}=-A^{\alpha a}_{(1)}=\psi^{1\alpha} \bar{e}^a-\epsilon^{ab}\psi^{2\alpha} \bar{e}^b\,,\\
&A^{\alpha\beta}_{(1)}=A^{\alpha \beta}_{(1)} \,.
\end{split}
\end{equation}
This incorporates the spin connection $\bar{\omega}^{ab}$ in the expression for $A^{ab}_{(1)}$, which corresponds to the topological twist condition that ensures the preservation of supersymmetry on the wrapped M5-brane with worldvolume $\mb{R}^{1,3}\times \Sigma_2$. The ansatz is also comprised of three scalar fields $\psi^{1\alpha}$, another three scalar fields $\psi^{2\alpha}$, a 1-form $A_{(1)}$ and three 1-forms $A^{\alpha \beta}_{(1)}$, all defined on the $D=5$ spacetime. The 3-form fields $S^i_{(3)}$ are split into
\begin{equation}
\begin{split}
&S^a_{(3)}=K^1_{(2)}\wedge\bar{e}^a-\epsilon^{ab}K^2_{(2)}\wedge \bar{e}^b\,,\\
&S^\alpha_{(3)}=h^{\alpha}_{(3)}+\chi^{\alpha}_{(1)}\wedge \mathrm{vol}(\Sigma_2) \,,
\end{split}
\end{equation}
where $K^1_{(2)}$, $K^2_{(2)}$, $h^\alpha_{(3)}$, $\chi^\alpha_{(1)}$ are 2-, 2-, 3- and 1-forms in $D=5$ respectively. The $SL(5,\mb{R})/SO(5)$ scalars $T_{ij}$ are taken to be
\begin{equation}
T^{ab}=e^{-6\lambda}\delta^{ab}\,,\quad T^{a\alpha}=0\,,\quad T^{\alpha\beta}=e^{4\lambda}\mc{T}^{\alpha \beta} \,,
\end{equation} 
where $\lambda$ is a scalar, and $\mc{T}^{\alpha\beta}$ is a symmetric, unimodular matrix parametrising the coset $SL(3,\mb{R})/SO(3)$, all in $D=5$. For future convenience, we will call these $D=5$ fields the M5 fields. The resulting 5-dimensional theory is an $\mc{N}=4$ gauged supergravity theory with three vector multiplets with gauge group $SO(2)\times ISO(3)$ and scalar manifold $SO(1,1)\times SO(5,3)/(SO(5)\times SO(3))$.

\subsection{NS5-branes wrapped on Riemann surfaces in $CY_2$}

For NS5-branes, we turn to the maximal $ISO(4)$ theory. The analogous ansatz for NS5-branes wrapped on Riemann surfaces is the following. The $D=7$ metric is given by  
\begin{equation}\label{ns57dmetric}
ds^2_7=e^{-4\tilde{\phi}}d\tilde{s}^2_{5}+e^{6\tilde{\phi}}ds^2(\tilde{\Sigma}_2) \,.
\end{equation}
We introduce orthonormal frames \{$\bar{\tilde e}^m;m\in\{0,\ldots,4\}$\} and \{$\bar{\tilde e}^a;a\in\{1,2\}$\} for both $d\tilde{s}^2_{5}$ and $ds^2(\tilde{\Sigma}_2)$ respectively, and let $\bar{\tilde{\omega}}^{m}_{\phantom{m}n}$ and $\bar{\tilde{\omega}}^{a}_{\phantom{a}b}$ be the corresponding spin connections. The metric of the Riemann surface satisfies $\tilde{R}_{ab}=l\tilde{g}^2\delta_{ab}$ with $l=0,\pm1$. The fields are decomposed via $SO(4)\to SO(2)_1\times SO(2)_2$, where the $SO(4)$ vector index decomposes accordingly as $i = (a,\alpha)$, with $a\in\{1,2\}$ and $\alpha\in\{3,4\}$. The fields are given by
\begin{equation}
\begin{split}
&\tilde{A}^{ab}_{(1)}=\frac{1}{\tilde{g}}\bar{\tilde{\omega}}^{ab}+\epsilon^{ab}\tilde{A}_{(1)}\,,\\
&\tilde{A}^{a\alpha}_{(1)}=-\tilde{A}^{\alpha a}_{(1)}=\tilde{\psi}^{1\alpha} \bar{\tilde{e}}^a-\epsilon^{ab}\tilde{\psi}^{2\alpha} \bar{\tilde{e}}^b\,,\\
&\tilde{A}^{\alpha\beta}_{(1)}=\epsilon^{\alpha\beta}\tilde{\mc{A}}_{(1)} \,,\\
&\tilde A^{a5}_{(1)} = \tilde\Psi^1\bar{\tilde{e}}^a-\epsilon^{ab}\tilde{\Psi}^{2} \bar{\tilde{e}}^b \,,\\
&\tilde A^{\alpha 5}_{(1)} = \tilde V^\alpha_{(1)} \,.
\end{split}
\end{equation}
This again incorporates the spin connection $\bar{\tilde{\omega}}^{ab}$ in the expression for $\tilde{A}^{ab}_{(1)}$, which corresponds to the topological twist condition that ensures the preservation of supersymmetry on the wrapped NS5-brane. The 3-forms are taken to be
\begin{equation}
\begin{split}
&\tilde{S}^a_{(3)} = \tilde{K}^1_{(2)}\wedge\bar{\tilde{e}}^a - \epsilon^{ab}\tilde K^2_{(2)}\wedge\bar{\tilde{e}}^b \\
&\tilde S^\alpha_{(3)} = \tilde h^{\alpha}_{(3)} + \tilde{\chi}^\alpha_{(1)}\wedge \vol(\tilde{\Sigma}_2) \\
&\tilde{S}^5_{(3)}=\tilde{H}_{(3)}+\tilde{X}_{(1)}\wedge \vol(\tilde{\Sigma}_2) 
\end{split}
\end{equation} 
For the $SL(4,\mb{R})/SO(4)$ scalars $\tilde{T}_{ij}$ and the $\bd{4}$ scalars $\tilde \tau^A$, we take 
\begin{equation}\label{ns57dscalars}
\begin{split}
&\tilde\tau^a = 0\,,\quad \tilde\tau^\alpha=\tilde\tau^\alpha \,,\\
&\tilde{T}^{ab}=e^{-6\tilde{\lambda}}\delta^{ab}\,,\quad \tilde{T}^{a\alpha}=0\,,\quad \tilde{T}^{\alpha\beta}=e^{6\tilde{\lambda}}\tilde{\mc{T}}^{\alpha\beta} \,,
\end{split}
\end{equation} 
where the symmetric, unimodular matrix $\tilde{\mc{T}}^{\alpha \beta}$ parametrises $SL(2,\mb{R})/SO(2)$, all in $D=5$. We will call these $D=5$ fields the NS5 fields, which are distinguished notationally from the M5 fields by a tilde.

We can now substitute the ansatz directly into the $D=7$ equations of motion to obtain a $D=5$ theory. However, as explained in Section \ref{introduction}, it is quicker, and perhaps more instructive to utilise the IW contraction that connects the $SO(5)$ and $ISO(4)$ theories. To do so, we must identify our NS5 fields in terms of the M5 fields presented in the previous section using the IW contraction procedure. 

Using \eqref{7d_IW_ansatz}, we have 
\begin{equation} \label{5d_identification}
\begin{split}
&g=k^2\tilde{g}\,,\quad\bar{e}^a={k^{-2}}\bar{\tilde{e}}^a\,,\quad \bar{e}^m=k^{4/3}\bar{\tilde{e}}^m\,,\quad \phi=\tilde\phi + \frac{2}{3}\log k\,,\quad\psi^{a\alpha}=\tilde{\psi}^{a\alpha} \,,\\
&\psi^{a5} = k^5\tilde\Psi^a \,,\quad \lambda=\tilde\lambda - \frac{1}{24}\log\tilde\Phi + \frac{1}{3}\log k\,,\quad \mc{T}^{\alpha\beta}=k^{-10/3}\tilde\Phi^{5/12}e^{2\tilde{\lambda}}\tilde{\mc{T}}^{\alpha\beta} \\
&\mc{T}^{\alpha5} = -k^{5/3}\tilde\Phi^{5/12}e^{2\tilde\lambda}(\tilde{\mc{T}}\tilde\tau)^\alpha \,,\quad \mc{T}^{55}=k^{20/3}\left(\tilde\Phi^{-5/6}e^{-4\tilde{\lambda}} + \tilde\Phi^{5/12}e^{2\tilde\lambda}\tilde\tau\tilde{\mc{T}}\tilde\tau\right)\\
&A_{(1)}=k^{-2}\tilde{A}_{(1)}\,,\quad A_{(1)}^{\alpha\beta}=k^{-2}\epsilon^{\alpha\beta}\tilde{\mc{A}}_{(1)} \,,\quad A^{\alpha5}_{(1)} = k^3\tilde V^{\alpha}_{(1)}\,,\quad \chi^\alpha_{(1)} = k^5\tilde\chi^\alpha_{(1)} \,,\\
&\chi^5_{(1)}=\tilde{X}_{(1)}\,,\quad K^a_{(2)} = k^3\tilde K^a_{(2)} \,,\quad h^\alpha_{(3)}=k\tilde h^\alpha_{(3)} \,,\quad h^5_{(3)}=k^{-4}\tilde{H}_{(3)}\,.
\end{split}
\end{equation} 
We now substitute \eqref{5d_identification} into the $D=5$ equations of motion obtained from the M5-brane wrapping a Riemann surface to obtain a new set of $D=5$ equations after taking $k\to0$. These equations are collected in appendix \ref{5deomssection}. To present the five-form Lagrangian that encodes these equations, we define the $SO(2)\times SO(2)$ covariant derivatives
\begin{equation}
\begin{split}
\tilde D\tilde h^\alpha_{(3)} &\equiv d\tilde h^\alpha_{(3)} + \tilde g\epsilon_{\alpha\beta}\tilde{\mc{A}}_{(1)}\wedge\tilde h^\beta_{(3)} \,,\\
\tilde D\tilde V^\alpha_{(1)} &\equiv d\tilde V^\alpha_{(1)} + \tilde g\epsilon_{\alpha\beta}\tilde{\mc{A}}_{(1)}\wedge\tilde V^\beta_{(1)} \,,\\
\tilde D\tilde \chi^\alpha_{(1)} &\equiv d\tilde \chi^\alpha_{(1)} + \tilde g\epsilon_{\alpha\beta}\tilde{\mc{A}}_{(1)}\wedge\tilde \chi^\beta_{(1)} \,,\\
\tilde D\tilde\psi^{a\alpha} &\equiv d\tilde\psi^{a\alpha} + \tilde g\epsilon_{ab}\tilde\psi^{b\alpha}\tilde A_{(1)} + \tilde g\epsilon_{\alpha\beta}\tilde\psi^{a\beta}\tilde{\mc{A}}_{(1)} \,,\\
\tilde D\tilde{\mc{T}}_{\alpha\beta} &\equiv d\tilde{\mc{T}}_{\alpha\beta} + \tilde g\epsilon_{\alpha\gamma}\tilde{\mc{T}}_{\gamma\beta}\tilde{\mc{A}}_{(1)} + \tilde g\epsilon_{\beta\gamma}\tilde{\mc{T}}_{\alpha\gamma}\tilde{\mc{A}}_{(1)} \,, \\
\tilde D\tilde\tau^\alpha &\equiv d\tilde\tau^\alpha + \tilde g\epsilon_{\alpha\beta}\tilde\tau^\beta\tilde{\mc{A}}_{(1)} \,,\\
\tilde D\tilde\Psi^a &\equiv d\tilde\Psi^a + \tilde g\epsilon_{ab}\tilde\Psi^b\tilde{{A}}_{(1)} \,,
\end{split}
\end{equation}
the field strengths
\begin{equation}
\tilde F_{(2)} \equiv d\tilde A_{(1)} \,,\quad \tilde{\mc{F}}_{(2)} \equiv d\tilde{\mc{A}}_{(1)} \,,
\end{equation}
the following combinations of our fundamental fields
\begin{equation}
\begin{split}\label{5ddefs}
\tilde G^\alpha_{(3)} &\equiv (\tilde{\mc{T}}\tilde h_{(3)})^\alpha - (\tilde{\mc{T}}\tilde\tau)^\alpha \tilde H_{(3)} \,, \\
\tilde J^\alpha_{(2)} &\equiv \tilde D\tilde V^\alpha_{(1)} + \epsilon^{\alpha\beta}\tilde\tau^\beta\tilde{\mc{F}}_{(2)} \,, \\
\tilde\sigma^\alpha_{(1)} &\equiv (\tilde{\mc{T}}\tilde\chi_{(1)})^\alpha - (\tilde{\mc{T}}\tilde\tau)^\alpha \tilde X_{(1)} \,,\\
\tilde P^a_{(1)} &\equiv \tilde D\tilde\Psi^a - \tilde g\tilde V^{\alpha}_{(1)}\tilde\psi^{a\alpha} + \tilde\tau^\alpha\tilde D\tilde\psi^{a\alpha} \,,\\
\tilde Q^\alpha_{(1)} &\equiv \tilde D\tilde\tau^\alpha + \tilde g\tilde V^\alpha_{(1)} \,,\\
\tilde{R}^a &\equiv \tilde\Psi^a +\tilde\tau^{\alpha}\tilde\psi^{a\alpha} \,,
\end{split}
\end{equation}
and integrate \eqref{5dH3bianchi} and \eqref{5dXbianchi} to write
\begin{equation}
\begin{split}\label{eq:H_X_integration}
&\tilde H_{(3)} = d\tilde\Gamma_{(2)} + \frac{1}{2}\tilde{\mc{A}}_{(1)}\wedge \tilde F_{(2)} + \frac{1}{2}\tilde A_{(1)}\wedge\tilde{\mc{F}}_{(2)} \,,\\
&\tilde X_{(1)} = d\tilde\Xi+\epsilon_{\alpha\beta}\tilde{\psi}^{a\alpha}\tilde D\tilde{\psi}^{a\beta}+\tilde{g}l\tilde{\mc{A}}_{(1)} \,.
\end{split}
\end{equation}
The five-form Lagrangian is then \footnote{We note that when we vary the action to obtain the equations of motion, we must also include a Lagrange multiplier to enforce the constraint, $\det \tilde{\mc{T}} = 1$. This is done at the level of the action by introducing an auxiliary 5-form $q_{(5)}$ with $\mc{L}^{aux}_{(5)} = q_{(5)}(\det \tilde{\mc{T}} - 1)$. The on-shell value of $q_{(5)}$ is found by tracing the $\tilde{\mc{T}}_{\alpha\beta}$ equation of motion.}
\begin{equation}\label{5dlagrangianoriginal}
\mc{L}_{(5)}=\tilde{R}\,\tilde{\text{vol}}_5+\mc{L}^{kin}_{(5)}+\mc{L}^{pot}_{(5)}+\mc{L}^{top}_{(5)} \,,
\end{equation} 
where $\tilde{R}$ is the Ricci scalar of the $D = 5$ metric, the kinetic terms are 
\begin{equation}\label{5dkineticterms}
\begin{split}
\mc{L}^{kin}_{(5)}&=-30\tilde{\ast}d\tilde{\phi}\wedge d\tilde{\phi}-36\tilde{\ast}d\tilde{\lambda}\wedge d\tilde{\lambda}-\frac{5}{16}\tilde\Phi^{-2}\tilde{\ast}d{\tilde\Phi}\wedge d{\tilde\Phi}\\
&\quad-\frac{1}{4}\tilde{\mc{T}}^{-1}_{\alpha\beta}\tilde{\mc{T}}^{-1}_{\gamma\rho}\tilde{\ast}\tilde{D}\tilde{\mc{T}}_{\beta\gamma}\wedge \tilde{D}\tilde{\mc{T}}_{\rho\alpha}-\tilde\Phi^{-1/2}e^{-6\tilde{\phi}}\tilde{\mc{T}}^{-1}_{\alpha\beta}\tilde{\ast}\tilde{D}\tilde{\psi}^{a\alpha}\wedge \tilde{D}\tilde{\psi}^{a\beta}\\
&\quad-\tilde\Phi^{3/4}e^{6\tilde{\lambda}-6\tilde{\phi}}{\tilde{\ast}\tilde{P}^a_{(1)}\wedge\tilde{P}^a_{(1)}}-\frac{1}{2}\tilde\Phi^{5/4}e^{6\tilde\lambda}\tilde{\mc{T}}_{\alpha\beta}\tilde\ast\tilde Q_{(1)}^\alpha\wedge\tilde Q^\beta_{(1)} \\
&\quad-\frac{1}{2}\tilde\Phi^{-1}e^{-12\tilde{\phi}}\tilde{\ast}\tilde{X}_{(1)}\wedge\tilde{X}_{(1)}-\frac{1}{2}\tilde\Phi^{1/4}e^{6\tilde{\lambda}-12\tilde{\phi}}\tilde{\mc{T}}^{-1}_{\alpha\beta}\tilde{\ast}\tilde{\sigma}^\alpha_{(1)}\wedge\tilde{\sigma}^\beta_{(1)}\\
&\quad-\frac{1}{2}\tilde\Phi^{-1/2}e^{4\tilde{\phi}+12\tilde{\lambda}}\tilde{\ast}\tilde{F}_{(2)}\wedge \tilde{F}_{(2)}-\frac{1}{2}\tilde\Phi^{-1/2}e^{4\tilde{\phi}-12\tilde{\lambda}}\tilde{\ast}\tilde{\mc{F}}_{(2)}\wedge \tilde{\mc{F}}_{(2)}\\
&\quad-\frac{1}{2}\tilde\Phi^{3/4}e^{4\tilde{\phi}-6\tilde{\lambda}}\tilde{\mc{T}}^{-1}_{\alpha\beta}\tilde{\ast}\tilde{J}^\alpha_{(2)}\wedge \tilde{J}^\beta_{(2)}-\tilde{\Phi}^{1/4}e^{-6\tilde{\lambda}-2\tilde{\phi}}{\tilde{\ast}\tilde{K}^a_{(2)}}\wedge\tilde{K}^a_{(2)} \\
&\quad-\frac{1}{2}\tilde\Phi^{-1}e^{8\tilde{\phi}}\tilde{\ast}\tilde{H}_{(3)}\wedge\tilde{H}_{(3)}-\frac{1}{2}\tilde\Phi^{1/4}e^{6\tilde{\lambda}+8\tilde{\phi}}\tilde{\mc{T}}^{-1}_{\alpha\beta}\tilde{\ast}\tilde{G}^\alpha_{(3)}\wedge\tilde{G}^\beta_{(3)}\,,
\end{split}
\end{equation} 
the potential terms are
\begin{align*}\label{5doriginalpotential}
\mc{L}^{pot}_{(5)}=&-\tilde{g}^2\left\{e^{-10\tilde{\phi}}\left(e^{12\tilde{\lambda}}(\tilde{\psi}\tilde{\mc{T}}\tilde{\psi})-2(l+\tilde{\psi}^2)+e^{-12\tilde{\lambda}}(\tilde{\psi}\tilde{\mc{T}}^{-1}\tilde{\psi})+\tilde{\Phi}^{5/4}e^{-6\tilde{\lambda}}\tilde{R}^2\right)\right.\\
&\left.\phantom{+\tilde{g}^2\{}+\frac{1}{2}\tilde\Phi^{1/2}e^{-4\tilde{\phi}}\left(2e^{12\tilde{\lambda}}\text{Tr}(\tilde{\mc{T}}^2)-e^{12\tilde{\lambda}}(\text{Tr}\tilde{\mc{T}})^2-4\text{Tr}\tilde{\mc{T}}\right)\right.\\
&\left.\phantom{+\tilde{g}^2\{}+\tilde\Phi^{-1/2}e^{-12\tilde{\lambda}-16\tilde{\phi}}\epsilon^{ab}\epsilon^{cd}(\tilde{\psi}^{a}\tilde{\mc{T}}^{-1}\tilde{\psi}^c)(\tilde{\psi}^{b}\tilde{\mc{T}}^{-1}\tilde{\psi}^d)+\frac{1}{2}\tilde\Phi^{-1/2}e^{12\tilde{\lambda}-16\tilde{\phi}}(l-\tilde{\psi}^2)^2\right.\\
&\left.\phantom{+\tilde{g}^2\{}+2\tilde\Phi^{3/4}e^{-6\tilde{\lambda}-16\tilde{\phi}}\epsilon^{ab}\epsilon^{cd}(\tilde{\psi}^{a}\tilde{\mc{T}}^{-1}\tilde{\psi}^c)\tilde{R}^b\tilde{R}^d\right\}\tilde{\text{vol}}_5 \,,\numberthis
\end{align*}
where $\tilde\psi^2\equiv\tilde\psi^{a\alpha}\tilde\psi^{a\alpha}$ and $\tilde{R}^2\equiv \tilde{R}^a\tilde{R}^a$, and the topological terms are given by 
\begin{align*}\label{5dtopological}
\mc{L}^{top}_{(5)}&=\frac{1}{\tilde{g}}\epsilon_{ab}\tilde{K}^a_{(2)}\wedge \tilde{D}\tilde{K}^b_{(2)}+2\epsilon_{ab}\tilde{R}^a\tilde{K}^b_{(2)}\wedge \tilde{H}_{(3)}+2\epsilon_{ab}\tilde{\psi}^{a\alpha}\tilde{K}^b_{(2)}\wedge (\tilde{\mc{T}}^{-1}\tilde{G}_{(3)})^\alpha\\
&\quad+\frac{2}{\tilde{g}}\epsilon_{\alpha\beta}\tilde{D}\tilde{\psi}^{a\alpha}\wedge \tilde{J}^\beta_{(2)}\wedge \tilde{K}_{(2)}^a+\frac{2}{\tilde{g}}\tilde{P}^a_{(1)}\wedge \tilde{K}_{(2)}^a\wedge \tilde{\mc{F}}_{(2)} - \frac{1}{\tilde g}\tilde Q^\alpha_{(1)}\wedge(\tilde{\mc{T}}^{-1}\tilde\sigma_{(1)})^\alpha \wedge \tilde H_{(3)} \\
&\quad -\frac{2}{\tilde g}\epsilon_{\alpha\beta}\tilde R^a\tilde D\tilde\psi^{a\alpha}\wedge\tilde Q^\beta_{(1)}\wedge\tilde H_{(3)}+ \tilde R^2\tilde{\mc{F}}_{(2)}\wedge\tilde H_{(3)} + \frac{1}{2\tilde g}(l-\tilde\psi^2)\epsilon_{\alpha\beta}\tilde Q^\alpha_{(1)}\wedge\tilde Q^\beta_{(1)}\wedge\tilde H_{(3)} \\
&\quad + \frac{1}{\tilde g}\tilde D((\tilde{\mc{T}}^{-1}\tilde\sigma_{(1)})^\alpha)\wedge (\tilde{\mc{T}}^{-1}\tilde{G}_{(3)})^\alpha - \frac{1}{\tilde g}\epsilon_{\alpha\beta}(\tilde{\mc{T}}^{-1}\tilde\sigma_{(1)})^\alpha\wedge\tilde J^\beta_{(2)}\wedge\tilde F_{(2)} \\
&\quad -\frac{2}{\tilde g}\epsilon_{\alpha\beta}(\tilde{\mc{T}}^{-1}\tilde{G}_{(3)})^\alpha\wedge\left(\tilde D\tilde\psi^{a\beta}\wedge\tilde P^a_{(1)} + \frac{1}{2}\tilde g (l-\tilde\psi^2)\tilde J^\beta_{(2)} +\tilde{g} \epsilon_{ab}\tilde\psi^{a\beta}\tilde{R}^b\tilde F_{(2)}\right) \\
&\quad+ \frac{1}{\tilde g}(\tilde{\mc{T}}^{-1}\tilde{G}_{(3)})^\alpha\wedge \tilde Q^\alpha_{(1)}\wedge \tilde X_{(1)} + \frac{1}{\tilde g}\epsilon_{ab}\tilde R^a\tilde D\tilde R^b\wedge\tilde{\mc{F}}_{(2)} \wedge\tilde{\mc{F}}_{(2)}\\
&\quad+ \frac{1}{2\tilde g^2}\epsilon_{\alpha\beta}\tilde Q^\alpha_{(1)}\wedge\tilde Q^\beta_{(1)}\wedge\left(d\tilde \Xi + \tilde g l\tilde{\mc{A}}_{(1)}\right) \wedge\tilde F_{(2)} +\frac{2}{\tilde g}\epsilon_{ab}\epsilon_{\alpha\beta}\tilde\psi^{a\alpha}\tilde P^b_{(1)}\wedge\tilde J^\beta_{(2)}\wedge\tilde{\mc{F}}_{(2)} \\
&\quad -\frac{2}{\tilde g}\tilde R^a\tilde D\tilde\psi^{a\alpha}\wedge\tilde J^\alpha_{(2)}\wedge\tilde F_{(2)} +  \frac{l}{\tilde g}\tilde Q^\alpha_{(1)}\wedge\tilde J^\alpha_{(2)}\wedge\tilde F_{(2)} - \frac{1}{\tilde g}\tilde\psi^{a\alpha}\tilde\psi^{a\beta}\tilde Q^\alpha_{(1)}\wedge\tilde J^\beta_{(2)}\wedge\tilde F_{(2)} \\
&\quad  -\frac{1}{\tilde g}\epsilon_{ab}\epsilon_{\beta\gamma}\tilde\psi^{a\alpha}\tilde\psi^{b\beta}\tilde Q^\alpha_{(1)}\wedge\tilde J^\gamma_{(2)}\wedge\tilde{\mc{F}}_{(2)}  - \frac{1}{\tilde g}\epsilon_{ab}\epsilon_{\alpha\beta}\epsilon_{\gamma\eta}\tilde\psi^{a\beta}\tilde D\tilde\psi^{b\gamma}\wedge\tilde J^\alpha_{(2)}\wedge\tilde J^\eta_{(2)} \,.\numberthis
\end{align*}
Any solution of the equations of motion in \ref{5deomssection} can be uplifted to type IIA supergravity. This can be done by first using \eqref{ns57dmetric}-\eqref{ns57dscalars} to uplift to the $ISO(4)$ gauged theory in $D=7$, then using the uplift formulae in \cite{Cvetic:2003xr} which connect the $ISO(4)$ gauged theory and the type IIA theory. 

As we shall show, this is a $SO(2)\times\left(SO(2)\ltimes_{\Sigma_2}\mb{R}^4\right)$ gauged $D=5$, $\mc{N}=4$ supergravity coupled to three vector multiplets with scalar manifold $SO(1,1)\times SO(5,3)/(SO(5)\times SO(3))$.

\subsection{Field redefinitions}

In order to make contact with the canonical language of $D = 5$, $\mc{N}=4$ supergravity in the next section, we find it convenient to make the following field redefinitions. We first replace $(\tilde{\mc{T}}^{-1}\tilde\sigma_{(1)})^\alpha$ by introducing two one-forms $\tilde{\mathscr{A}}^\alpha_{(1)}$ and two Stueckelberg scalar fields $\tilde{\xi}^\alpha$,
\begin{equation}
(\tilde{\mc{T}}^{-1}\tilde\sigma_{(1)})^\alpha = \tilde D\tilde\xi^\alpha + \tilde g\tilde{\mathscr{A}}^\alpha_{(1)}-\tilde{\tau}^\alpha\tilde{D}\tilde{\Xi}+\tilde{g}\tilde{\Xi}\tilde{V}^\alpha_{(1)} - \epsilon_{\alpha\beta}\left(\tilde\psi^{a\beta}\tilde\psi^{a\gamma}\tilde Q^\gamma_{(1)}+2\tilde R^a\tilde D\tilde\psi^{a\beta} \right) \,,
\end{equation}
where 
\begin{equation}
\tilde D\tilde\xi^\alpha\equiv d\tilde{\xi}^\alpha+\tilde{g}\epsilon_{\alpha\beta}\tilde{\xi}^\beta\tilde{\mc{A}}_{(1)} \,,\quad\tilde D\tilde\Xi \equiv d\tilde\Xi + \tilde gl\tilde{\mc{A}}_{(1)} \,,
\end{equation} 
and we note that the $SO(2)$ gauge symmetry is non-linearly realised by $\tilde{\Xi}$. Substituting this into \eqref{tsigma1eqn}, we deduce that
\begin{align}\label{dualG3}
\begin{split}
\tilde\Phi^{1/4}e^{6\tilde\lambda+8\tilde\phi}\tilde\ast\tilde G^\alpha_{(3)} &=\tilde{D}\tilde{\mathscr{A}}^\alpha_{(1)}-l\epsilon_{\alpha\gamma}\tilde{D}\tilde{V}^\gamma_{(2)}-l\tilde{D}\tilde{\tau}^\alpha\wedge\tilde{\mc{A}}_{(1)}+\epsilon_{\alpha\gamma}\tilde{\xi}^\gamma\tilde{\mc{F}}_{(2)}+2\epsilon^{ab}\tilde{\psi}^{a\alpha}\tilde{K}^b_{(2)}\\
&\quad+2\tilde\psi^{a\alpha}\tilde R^a\tilde{\mc{F}}_{(2)} + \epsilon_{\beta\gamma}\tilde\psi^{a\alpha}\tilde\psi^{a\beta}\tilde J^\gamma_{(2)}+\tilde\Xi\tilde J^\alpha_{(2)}\,,
\end{split}
\end{align}
where
\begin{equation}
\tilde D\tilde{\mathscr{A}}^{\alpha}_{(1)}\equiv d\tilde{\mathscr{A}}^{\alpha}_{(1)} + \tilde g\epsilon_{\alpha\beta}\tilde{\mc{A}}_{(1)}\wedge\tilde{\mathscr{A}}^{\beta}_{(1)} - l\tilde{\mc{A}}_{(1)}\wedge\tilde Q^\alpha_{(1)} \,,
\end{equation}
and we note again that the $SO(2)$ gauge symmetry is non-linearly realised by $\tilde{\mathscr{A}}^\alpha_{(1)}$. We also need to dualise $\tilde H_{(3)}$. There are two ways of doing this, the first being to integrate \eqref{starH32} directly, and the second, easier way is to add to the original Lagrangian the term
\begin{equation}
\mc{L}^{\text{dual}}_{(5)} = \mathscr{\tilde{{B}}}_{(1)}\wedge\left(d\tilde H_{(3)} - \tilde{\mc{F}}_{(2)}\wedge\tilde{{F}}_{(2)}\right) \,,
\end{equation}
with $\mathscr{\tilde{{B}}}_{(1)}$ a Lagrange multiplier that enforces the Bianchi identity $d\tilde H_{(3)}=\tilde{\mc{F}}_{(2)}\wedge\tilde{{F}}_{(2)}$. Treating $\tilde H_{(3)}$ as a fundamental field, the variation of the total Lagrangian $\mc{L}^{}_{(5)} + \mc{L}^{\text{dual}}_{(5)}$ with respect to $\tilde H_{(3)}$ yields the algebraic relation
\begin{equation}\label{dualH3}
\begin{split}
\tilde\Phi^{-1}e^{8\tilde\phi}\tilde\ast\tilde H_{(3)} &= d\mathscr{\tilde{{B}}}_{(1)}- \tilde Q_{(1)}^\alpha\wedge\left(  \tilde{\mathscr{A}}^\alpha_{(1)}+\frac{1}{\tilde g}\tilde D\tilde\xi^\alpha-\frac{1}{\tilde{g}}\tilde\tau^\alpha\tilde{D}\tilde\Xi-\frac{1}{\tilde{g}}\tilde\Xi\tilde{D}\tilde\tau^\alpha \right)\\
&\quad+ 2\epsilon_{ab}\tilde R^a\tilde K^b_{(2)} + \tilde R^2\tilde{\mc{F}}_{(2)}+ \frac{l}{2\tilde g}\epsilon_{\alpha\beta}\tilde Q^\alpha_{(1)}\wedge\tilde Q^\beta_{(1)}\,,
\end{split}
\end{equation}
which we will substitute back into the total Lagrangian. Finally, we also find it convenient to redefine
\begin{equation}
\tilde K^a_{(2)} = -\frac{1}{\sqrt{2}}\epsilon_{ab}\tilde L^b_{(2)}+\epsilon_{ab}\tilde R^b\tilde{\mc{F}}_{(2)} + \epsilon_{ab}\epsilon_{\alpha\beta}\tilde\psi^{b\alpha}\tilde J^\beta_{(2)}\,.
\end{equation}
Making use of the above redefinitions, the kinetic terms for the vectors can be rewritten as
\begin{equation}\label{eq_ke_vectors_after_redef}
\begin{split}
\mc{L}^{V}=& -\frac{1}{2}\tilde\Phi^{-1/2}e^{4\tilde{\phi}+12\tilde{\lambda}}\tilde{\ast}\tilde{F}_{(2)}\wedge \tilde{F}_{(2)}-\tilde{\Phi}^{1/4}e^{-6\tilde{\lambda}-2\tilde{\phi}}{\tilde{\ast}\tilde{K}^a_{(2)}}\wedge\tilde{K}^a_{(2)}\\
&-\frac{1}{2}\tilde\Phi^{-1/2}e^{4\tilde{\phi}-12\tilde{\lambda}}\tilde{\ast}\tilde{\mc{F}}_{(2)}\wedge \tilde{\mc{F}}_{(2)}-\frac{1}{2}\tilde\Phi^{3/4}e^{4\tilde{\phi}-6\tilde{\lambda}}\tilde{\mc{T}}^{-1}_{\alpha\beta}\tilde{\ast}\tilde{J}^\alpha_{(2)}\wedge \tilde{J}^\beta_{(2)}\\
&+\frac{1}{2}\tilde\Phi^{-1}e^{8\tilde{\phi}}\tilde{\ast}\tilde{H}_{(3)}\wedge\tilde{H}_{(3)} +\frac{1}{2}\tilde\Phi^{1/4}e^{6\tilde{\lambda}+8\tilde{\phi}}\tilde{\mc{T}}^{-1}_{\alpha\beta}\tilde{\ast}\tilde{G}^\alpha_{(3)}\wedge\tilde{G}^\beta_{(3)} \,.
\end{split}
\end{equation}
We note that the positive signs in the $\tilde H_{(3)}$ and $\tilde G^\alpha_{(3)}$ terms do not indicate the presence of ghosts, as when we consider the dualised fields \eqref{dualG3} and \eqref{dualH3} which encodes the fundamental degrees of freedom, we obtain a sign flip from applying the Hodge star twice. The topological terms greatly simplify to
\begin{equation}
\begin{split}
\mc{L}^{T}&= \frac{1}{2\tilde g}\epsilon_{ab}\tilde L^a_{(2)}\wedge\tilde D\tilde L^b_{(2)} -\frac{\tilde g l}{2}\epsilon_{\alpha\beta}\tilde V^\alpha_{(1)}\wedge\tilde V^\beta_{(1)}\wedge\tilde{\mc{A}}_{(1)}\wedge\tilde F_{(2)} \\
&\quad- \epsilon_{\alpha\beta}\left(\tilde D\tilde{\mathscr{A}}^{\alpha}_{(1)}-l\epsilon_{\alpha\gamma}\tilde{J}^\gamma_{(2)}\right)\wedge\tilde V^\beta_{(1)}\wedge\tilde F_{(2)}\\
&\quad -\tilde{F}_{(2)}\wedge\tilde{\mc{F}}_{(2)}\wedge\left(\tilde{\mathscr{B}}_{(1)}- \tilde\tau^\alpha\left[\tilde{\mathscr{A}}^\alpha_{(1)}-\frac{l}{\tilde{g}}\epsilon_{\alpha\beta}\tilde{Q}^\beta_{(1)}+ \frac{l}{2\tilde g}\epsilon_{\alpha\beta}d\tilde\tau^\beta\right] + \frac{1}{\tilde g}\tilde\xi^\alpha\tilde Q^\alpha_{(1)} \right)\\
&\quad +\frac{1}{\tilde g}\epsilon_{\alpha\beta}\left(l\tilde\tau^\alpha + \epsilon_{\alpha\gamma}\tilde\Xi\tilde\tau^\gamma\right)\tilde F_{(2)}\wedge\tilde{\mc{F}}_{(2)}\wedge\tilde Q^\beta_{(1)} - \frac{l}{\tilde g}\epsilon_{\alpha\beta}d\tilde\tau^\alpha\wedge\tilde Q^\beta_{(1)}\wedge\tilde{\mc{A}}_{(1)}\wedge\tilde F_{(2)} \,.
\end{split}
\end{equation} 
Up to total derivatives, these can be rewritten as
\begin{align*}\label{eq:5dtopological_useful}
\mc{L}^{T}&=  \frac{1}{2\tilde g}\epsilon_{ab}\tilde L^a_{(2)}\wedge\tilde D\tilde L^b_{(2)} -\frac{\tilde g l}{2}\epsilon_{\alpha\beta}\tilde V^\alpha_{(1)}\wedge\tilde V^\beta_{(1)}\wedge\tilde{\mc{A}}_{(1)}\wedge\tilde F_{(2)} \numberthis \\
&\quad- \epsilon_{\alpha\beta}\left( d\left[\tilde{\mathscr{A}}^\alpha_{(1)}-l\epsilon_{\alpha\beta}\tilde{V}^\beta_{(1)}\right]+\tilde{g}\epsilon_{\alpha\gamma}\tilde{\mc{A}}_{(1)}\wedge \left[\tilde{\mathscr{A}}^\gamma_{(1)}-l\epsilon_{\gamma\rho}\tilde{V}^\rho_{(1)}\right]+\tilde{g}l\tilde{V}^\alpha_{(1)}\wedge \tilde{\mc{A}}_{(1)}\right)\wedge\tilde V^\beta_{(1)}\wedge\tilde F_{(2)}\\
&\quad -\tilde{F}_{(2)}\wedge\tilde{\mc{F}}_{(2)}\wedge\left(\mathscr{\tilde{{B}}}_{(1)}- \tilde\tau^\alpha\left[\tilde{\mathscr{A}}^\alpha_{(1)}-l\epsilon_{\alpha\beta}\tilde{V}^\beta_{(1)}- \frac{l}{2\tilde g}\epsilon_{\alpha\beta}d\tilde\tau^\beta\right] + \frac{1}{2\tilde g}\tilde\tau^2 d\tilde\Xi -\tilde\Xi\tilde\tau^\alpha\tilde V^\alpha_{(1)}+ \frac{1}{\tilde g}\tilde\xi^\alpha\tilde Q^\alpha_{(1)} \right)\,,
\end{align*} 
which, as we will see, is the form in which the $\mc{N}=4$ supersymmetry is manifest.

\subsection{$D=5$, $\mc{N}=4$ supersymmetry}

In this section we first summarise the general structure of $\mc{N}=4$ gauged supergravity in $D=5$, coupled to $n=3$ vector multiplets, mostly following the conventions of \cite{Schon:2006kz} (which generalised the results in \cite{DallAgata:2001wgl}). For a cleaner presentation, we will remove all tildes from the NS5 fields for the rest of this section. 

The ungauged theory, as discussed in \cite{Awada:1985ep}, has a global symmetry group given by $SO(1,1)\times SO(5,n=3)$. The bosonic field content consists of a metric, $6+n=9$ Abelian vector fields and $1+5n=16$ scalar fields. The nine vector fields can be written as $\mathcal{A}^0_{(1)}$ and $\mathcal{A}^M_{(1)}$, with $M=1,\dots,8$, which transform as a scalar and vector with respect to $SO(5,3)$, respectively. The scalar manifold is given by $SO(1,1)\times SO(5,3)/(SO(5)\times SO(3))$, with the $SO(1,1)$ part described by a real scalar field $\Sigma$, and the coset $SO(5,3)/(SO(5)\times SO(3))$ parametrised by the $8\times 8$ matrix $\mathcal{V}^A{}_M$. The matrix $\mathcal{V}^A{}_M$ is an element of $SO(5,3)$ satisfying
\begin{equation}
\mathcal{V}^T\eta \mathcal{V} =\eta\,,
\end{equation}
where $\eta$ is the invariant metric tensor of $SO(5,3)$. Global $SO(5,3)$ transformations are chosen to act on the right, while local $SO(5)\times SO(3)$ transformations act on the left via
\begin{equation}
\mathcal{V}\longrightarrow h(x)\mathcal{V}g \,,\quad\quad g\in SO(5,3)\,,\quad h\in SO(5)\times SO(3)\,.
\end{equation}
The coset can also be parametrised by a symmetric positive definite matrix $\mathcal{M}_{MN}$ defined by
\begin{equation}\label{emmmat}
\mathcal{M}_{MN}\equiv\left(\mathcal{V}^T \mathcal{V}\right)_{MN}\,,
\end{equation}
with $\mathcal{M}_{MN}$ an element of $SO(5,3)$. We can raise indices using $\eta$, and the inverse of $\mc{M}_{MN}$, which we denote by $\mathcal{M}^{MN}$, is given by 
\begin{equation}
\mathcal{M}^{MN}\equiv \eta^{MP}\eta^{NQ}\mathcal{M}_{PQ}=\left(\mathcal{M}^{-1}\right)^{MN}\,.
\end{equation}
We will work in a basis in which $\eta$ is not diagonal, but instead given by
\begin{equation}\label{etadef}
\eta = 
\begin{pmatrix}
0 & 0 & \mathds{1}_3\\
0 & -\mathds{1}_2 & 0\\
\mathds{1}_3 & 0 & 0
\end{pmatrix}\,.
\end{equation}
In order to work in a basis in which $\eta$ is diagonal with the first five entries $-1$ and the last three entries $+1$, as in \cite{Schon:2006kz}, we can apply a similarity transformation using the matrix
\begin{equation}\label{usimmat}
\mathcal{U} =
\begin{pmatrix}
-U & 0 & U\\
0 & \mathds{1}_2 & 0\\
U & 0 & U
\end{pmatrix}\,,
\quad \text{with}\quad
U=\frac{1}{\sqrt{2}}\begin{pmatrix}
0 & 0 & 1\\
0 & 1 & 0\\
1& 0 & 0
\end{pmatrix}\,,
\end{equation}
which satisfies $\mathcal{U} = \mathcal{U}^T =\mathcal{U}^{-1}$ and $\det\,\mathcal{U}=1$. To evaluate the expression for the $\mc{N}=4$ scalar potential \eqref{eq:N=4_potential_SW}, we will also need the following antisymmetric tensor
\begin{equation}\label{mfiveddefu}
\mathcal{M}_{M_1\dots M_5}\equiv \epsilon_{m_1\dots m_5}(\mathcal{U}\cdot\mathcal{V})^{m_1}{}_{M_1}\dots
(\mathcal{U}\cdot\mathcal{V})^{m_5}{}_{M_5}\,,
\end{equation}
with the indices $m_1,\dots, m_5$ running from 1 to 5.

The general $D=5$, $\mc{N}=4$ gauged theory \cite{Schon:2006kz} is specified by a set of embedding tensors $f_{MNP}=f_{[MNP]}$, $\xi_{MN}=\xi_{[MN]}$ and $\xi_M$, which specify both the gauge group in $SO(1,1)\times SO(5,3)$ as well assigning specific vector fields to the generators of the gauge group. The covariant derivative is given by \footnote{\label{foot2}Here the terms involving the generators differ by a factor of two with the analogous expression in \cite{Schon:2006kz}. However, the generators that we use in \eqref{explicgen} below, also differ by a factor of two, so our covariant derivative stays the same as \cite{Schon:2006kz}.}
\begin{equation}\label{eq:5dN=4_cov_derv}
D_\mu=\nabla_\mu-\frac{1}{2}g\left(\mathcal{A}^M_{(1)\mu}f_{M}^{\phantom{M}NP}t_{NP}+\mathcal{A}^0_{(1)\mu}\xi^{NP}t_{NP}
+\mathcal{A}^M_{(1)\mu}\xi^Nt_{MN}+\mathcal{A}^M_{(1)\mu}\xi_{M}t_{0}\right)\,,
\end{equation}
where $t_{MN}=t_{[MN]}$ are the generators for $SO(5,3)$, $t_0$ is the generator for $SO(1,1)$, we have raised indices using $\eta$, and $\nabla_\mu$ is the Levi-Civita connection. To ensure the closure of the gauge algebra, the embedding tensors must satisfy the following algebraic constraints
\begin{align}\label{algconstraints}
3f_{R[MN}f_{PQ]}{}^R&=2f_{[MNP}\xi_{Q]}\,,\quad \xi_M{}^Qf_{QNP}=\xi_M\xi_{NP}-\xi_{[N}\xi_{P]M}\,,\nonumber\\
\xi_M\xi^M&=0\,,\quad\xi_{MN}\xi^N=0\,,\quad f_{MNP}\xi^P=0\,.
\end{align}
Associated with the vector fields $\mathcal{A}^0_{(1)}$ and $\mathcal{A}^M_{(1)}$, we also need to introduce
two-form gauge fields $\mathcal{B}_{(2)0}$ and $\mathcal{B}_{(2)M}$. In the ungauged theory these appear
on-shell as the Hodge duals of the fields strengths of the vectors. In the gauged theory the two-forms are introduced
as off-shell degrees of freedom, but the equations of motion ensure that the suitably defined covariant field strengths are still Hodge dual. In particular, the two-forms appear in the covariant field strengths for the vector fields, $\mathcal{H}^0_{(2)}$ and
$\mathcal{H}^M_{(2)}$, via
\begin{align}\label{tfmdefbee}
\mathcal{H}^M_{(2)}=&\,d\mathcal{A}_{(1)}^M-\frac{1}{2}g f_{NP}{}^M\mathcal{A}_{(1)}^N\wedge \mathcal{A}_{(1)}^P
-\frac{1}{2}g \xi_P{}^M\mathcal{A}_{(1)}^0\wedge \mathcal{A}_{(1)}^P
+\frac{1}{2}g \xi_P\mathcal{A}_{(1)}^M\wedge \mathcal{A}_{(1)}^P\nonumber\\
&+\frac{1}{2}g \xi^{MN}\mathcal{B}_{(2)N}
-\frac{1}{2}g \xi^{M}\mathcal{B}_{(2)0}\,,\nonumber\\
\mathcal{H}^0_{(2)}=&\,
d\mathcal{A}_{(1)}^0+\frac{1}{2}g \xi_M\mathcal{A}_{(1)}^M\wedge \mathcal{A}_{(1)}^0
+\frac{1}{2}g \xi^{M}\mathcal{B}_{(2)M}\,.
\end{align}
Using the $\mc{N}=4$ language of \cite{Schon:2006kz} \footnote{Note that we have multiplied the
Lagrangian in \cite{Schon:2006kz} by a factor of two.}, the Lagrangian of the bosonic sector of the theory can be written as 
\begin{equation}\label{n4lag1}
\mathcal{L}_{\mc{N}=4}={R}\,\text{vol}_{5}+\mathcal{L}^S_{\mc{N}=4}+\mathcal{L}^{pot}_{\mc{N}=4}+\mathcal{L}^V_{\mc{N}=4}+\mathcal{L}^T_{\mc{N}=4}\,.
\end{equation}
The scalar kinetic energy terms are given by
\begin{equation}
\mathcal{L}^S_{\mc{N}=4}=-3\Sigma^{-2}{\ast d\Sigma}\wedge d\Sigma
+\frac{1}{8}{\ast D \mathcal{M}_{MN}}\wedge D  \mathcal{M}^{MN}\,,
\end{equation}
and the scalar potential is given by
\begin{align}\label{eq:N=4_potential_SW}
\mathcal{L}^{pot}_{\mc{N}=4}=&-\frac{1}{2}{g}^2\left\{f_{MNP}f_{QRS}\Sigma^{-2}\left(\frac{1}{12}\mathcal{M}^{MQ}\mathcal{M}^{NR}\mathcal{M}^{PS}-\frac{1}{4}\mathcal{M}^{MQ}\eta^{NR}\eta^{PS}+\frac{1}{6}\eta^{MQ}\eta^{NR}\eta^{PS}\right)\right.\nonumber\\
&\phantom{-\frac{1}{2}{g}^2\Big\{}\left.+\frac{1}{4}\xi_{MN}\xi_{PQ}\Sigma^4\Big(\mathcal{M}^{MP}\mathcal{M^{NQ}}-\eta^{MP}\eta^{NQ}\Big)+\xi_{M}\xi_{N}\Sigma^{-2}\mathcal{M}^{MN}\right.\nonumber\\
&\phantom{-\frac{1}{2}{g}^2\Big\{}\left.+\frac{1}{3}\sqrt{2}f_{MNP}\xi_{QR}\Sigma\mathcal{M}^{MNPQR}\right\}\text{vol}_{5}\,.
\end{align}
The kinetic terms for the vectors, which also involve two-form contributions via \eqref{tfmdefbee}, are given by
\begin{equation}\label{eq:ke_vectors_SW}
\mathcal{L}^{V}_{\mc{N}=4}=-\Sigma^{-4}{\ast \mathcal{H}^0_{(2)}}\wedge \mathcal{H}^0_{(2)}
-\Sigma^{2}\mathcal{M}_{MN} {\ast \mathcal{H}^M_{(2)}}\wedge \mathcal{H}^N_{(2)}\,.
\end{equation}
In order to present the topological part of the Lagrangian in \eqref{n4lag1}, we introduce the calligraphic index $\mathcal{M}=(0,M)$ which allows us to package the 9 vector fields and 9 two-forms into the $\mc{A}_{(1)}^{\mc{M}}$ and $\mc{B}_{(2)\mc{M}}$, each transforming in the fundamental representation of $SO(1,1)\times SO(5,3)$. In the conventions of this paper \footnote{In an orthonormal frame, we take $\epsilon_{01234}=+1$ so that $\epsilon=\text{vol}_{5}$. We assume that \cite{Schon:2006kz} have taken $\epsilon_{01234}=-1$ and then the expression for the topological term given here agrees with that in \cite{Schon:2006kz} up to an overall factor of 2.}, we have
\begin{align}\label{eq:n4lag5_topological}
\mathcal{L}^{T}_{\mc{N}=4}=&
-\frac{1}{\sqrt{2}}gZ^{\mathcal{M}\mathcal{N}}\mathcal{B}_{\mathcal{M}}\wedge D\mathcal{B}_{\mathcal{N}}-\sqrt{2}gZ^{\mathcal{M}\mathcal{N}}\mathcal{B}_{\mathcal{M}}\wedge d_{\mathcal{N}\mathcal{P}\mathcal{Q}}\mathcal{A}^{\mathcal{P}}\wedge d\mathcal{A}^{\mathcal{Q}}\nonumber\\
&-\frac{\sqrt{2}}{3}{g}^2 Z^{\mathcal{M}\mathcal{N}}\mathcal{B}_{\mathcal{M}}\wedge d_{\mathcal{N}\mathcal{P}\mathcal{Q}}\mathcal{A}^{\mathcal{P}}\wedge X_{\mathcal{R}\mathcal{S}}^{\phantom{\mathcal{R}\mathcal{S}}\mathcal{Q}}\mathcal{A}^{\mathcal{R}}\wedge \mathcal{A}^{\mathcal{S}}+\frac{\sqrt{2}}{3}d_{\mathcal{M}\mathcal{N}\mathcal{P}}\mathcal{A}^{\mathcal{M}}\wedge d\mathcal{A}^{\mathcal{N}}\wedge d\mathcal{A}^{\mathcal{P}}\nonumber\\
&+\frac{1}{2\sqrt{2}}{g}d_{\mathcal{M}\mathcal{N}\mathcal{P}}X_{\mathcal{Q}\mathcal{R}}^{\phantom{\mathcal{Q}\mathcal{R}}\mathcal{M}}\mathcal{A}^{\mathcal{N}}\wedge \mathcal{A}^{\mathcal{Q}}\wedge \mathcal{A}^{\mathcal{R}}\wedge d\mathcal{A}^{\mathcal{P}}\nonumber\\
&+\frac{1}{10\sqrt{2}}{g}^2d_{\mathcal{M}\mathcal{N}\mathcal{P}}X_{\mathcal{Q}\mathcal{R}}^{\phantom{\mathcal{Q}\mathcal{R}}\mathcal{M}}X_{\mathcal{S}\mathcal{T}}^{\phantom{\mathcal{Q}\mathcal{R}}\mathcal{P}}\mathcal{A}^{\mathcal{N}}\wedge \mathcal{A}^{\mathcal{Q}}\wedge \mathcal{A}^{\mathcal{R}}\wedge \mathcal{A}^{\mathcal{S}}\wedge \mathcal{A}^{\mathcal{T}}\,.
\end{align}
Here the symmetric tensor $d_{\mathcal{M}\mathcal{N}\mathcal{P}}=d_{(\mathcal{M}\mathcal{N}\mathcal{P})}$ has non-zero components
\begin{equation}
d_{0MN}=d_{M0N}=d_{MN0}=\eta_{MN}\,,
\end{equation}
the antisymmetric tensor $Z^{\mathcal{M}\mathcal{N}}=Z^{[\mathcal{M}\mathcal{N}]}$ has components
\begin{equation}
Z^{MN}=\frac{1}{2}\xi^{MN}\,,\quad
Z^{0M}=-Z^{M0}=\frac{1}{2}\xi^{M}\,,\quad
\end{equation}
and the only non-zero components of $X_{\mathcal{M}\mathcal{N}}{}^{\mathcal{P}}$ are given by
\begin{equation}
X_{MN}{}^P=-f_{MN}{}^P-\tfrac{1}{2}\eta_{MN}\xi^P+\delta^P_{[M}\xi_{N]}\,,\quad
X_{M0}{}^0=\xi_M\,,\quad
X_{0M}{}^N=-\xi_M{}^N\,.
\end{equation}


\subsubsection{Scalar manifold}
We take the generators of $SO(5,3)$ to be given by the  $8\times 8$ matrices
\begin{equation}\label{explicgen}
(t_{MN})^A\,_B = \delta^A_M\eta_{BN}-\delta^A_N\eta_{MB}\,,
\end{equation}
with a non-diagonal $\eta$ given in \eqref{etadef}. In order to parametrise the coset $SO(5,3)/(SO(5)\times SO(3))$, we exponentiate a suitable solvable subalgebra of the Lie algebra. Following, for example \cite{Lu:1998xt}, the three non-compact Cartan generators $H^i$ and the twelve positive root generators, with positive weights under $H^i$, are given by \footnote{To compare with (3.31) of \cite{Lu:1998xt} we should make the identifications $(T^1,T^2,T^3)=(E_1{}^2,E_1{}^3,E_2{}^3)$, $(T^4,T^5,T^6)=(V^{12},V^{13},V^{23})$, $(T^7,T^8,T^9)=(U_1^1,U_1^2,U_1^3)$ and $(T^{10},T^{11},T^{12})=(U_2^1,U_2^2,U_2^3)$.}
\begin{align}\label{eq:generators}
&H^{1}=\sqrt{2}t_{16}\,, \quad H^{2}=\sqrt{2}t_{27}\,,\quad H^{3}=\sqrt{2}t_{38}\,,\\\nonumber
&T^{1}=-t_{26}\,,\quad T^{2}=-t_{36}\,,\quad T^{3}=-t_{37}\,, \quad T^{4}=t_{12}\,,\quad T^{5}=t_{13}\,,\quad T^{6}=t_{23}\,,\\\nonumber
&T^{7}=-t_{14}\,,\,\,T^{8}=-t_{24}\,, \,\,T^{9\phantom{0}}=-t_{34}\,,\,\,
T^{10}=-t_{15}\,,\,\,T^{11}=-t_{25}\,,\, \,T^{12}=-t_{35}\,.
\end{align}
We note that $\text{Tr}\left(T^i(T^j)^T\right)=2\delta^{ij}$ and $\text{Tr}\left(H^mH^n\right)=4\delta^{mn}$ with $H^m=\left(H^m\right)^T$. To make contact with the scalar fields in the reduced equations of motion, we first need an explicit embedding of the coset $SL(2,\mb{R})/SO(2)$ inside $SO(5,3)/(SO(5)\times SO(3))$. This is achieved by first defining 
\begin{equation}
\mc{H}=H^2-H^1,\quad \mc{E}=T^1\,.
\end{equation} 
In addition, we define $\hat{\mc{H}}=-H^1-H^2$ which commutes with the above two generators. We introduce three scalar fields $\{\varphi_1,\varphi_2,\rho\}$ to form the following coset representative
\begin{equation} 
\mc{V}_{(s)}=e^{\frac{1}{\sqrt{2}}\varphi_1\mc{H}+\frac{1}{\sqrt{2}}\varphi_2\hat{\mc{H}}}e^{\rho\mc{E}}=\left(
\begin{array}{ccccc}
 e^{-\varphi _2} V^{-T}&0 &0 &0&0\\
0&1&0&0&0\\
 0 &0& \mathds{1}_{2} & 0 &0\\
 0 & 0 & 0& e^{\varphi _2} V &0\\
 0 & 0 & 0& 0&1\\
\end{array}
\right)\,,
\end{equation} 
where the $2\times 2$ matrix $V$ parametrises the coset $SL(2,\mb{R})/SO(2)$ in the standard upper triangular gauge
\begin{equation} 
 V=\left(
\begin{array}{cc}
 e^{\varphi_1} & e^{\varphi_1} \rho \\
 0 & e^{-\varphi_1}
\end{array} 
\right) \,.
\end{equation} 
We can identify the scalar fields in the $2\times 2$ matrix ${\mc{T}}_{\alpha\beta}$ in the truncated theory as
\begin{equation}
{\mc{T}}_{\alpha\beta}=( V^T  V)_{\alpha\beta}
=\left(
\begin{array}{cc}
 e^{2\varphi_1} & e^{2\varphi_1} \rho \\
  e^{2\varphi_1}\rho  & e^{-2\varphi_1}+e^{2\varphi_1}\rho^2
\end{array}
\right) \,.
\end{equation} 
Collecting our results, the appropriate parametrisation of the coset $SO(5,3)/(SO(5)\times SO(3))$ is given by
\begin{equation}
\begin{split}\label{eq:mcV_parametrisation}
\mc{V}
&=\mc{V}_{(s)}e^{\frac{1}{\sqrt{2}}\varphi_3 H^3} e^{{\tau}^3 T^2}e^{{\tau}^4 T^3}e^{({\Xi}-{\psi}^{13}{\psi}^{14}-{\psi}^{23}{\psi}^{24}) T^4}\\
&\quad\cdot e^{\left({\xi}^4+[{\Psi}^1+{R}^1]{\psi}^{13}+[{\Psi}^2+{R}^2]{\psi}^{23}\right) T^5} e^{\left(-{\xi}^3+[{\Psi}^1+{R}^1]{\psi}^{14}+[{\Psi}^2+{R}^2]{\psi}^{24}\right) T^6}\\
&\quad\cdot e^{\sqrt{2}{\psi}^{13}T^7}e^{\sqrt{2}{\psi}^{14}T^8}e^{-{\sqrt{2}}{\Psi}^{1}T^9}e^{\sqrt{2}{\psi}^{23}T^{10}}e^{\sqrt{2}{\psi}^{24}T^{11}}e^{-{\sqrt{2}}{\Psi}^{2}T^{12}}\,,
\end{split}
\end{equation} 
where we identify ${\varphi}_2$ and ${\varphi}_3$ as
\begin{equation}
\begin{split} \label{eq:definitions_of_varphi2_and_Delta}
\varphi_2=3{\phi}+\frac{1}{4}\log\Phi \,,\\
\varphi_3=3{\lambda}-3{\phi}+\frac{3}{8}\log{\Phi}\,.
\end{split}
\end{equation} 
The remaining $SO(1,1)$ part of the scalar manifold is described by a real scalar field $\Sigma$,
\begin{equation}\label{eq:definitions_of_Sigma}
\Sigma={\Phi}^{1/8}e^{-{\phi}-3{\lambda}}\,.
\end{equation} 

\subsubsection{Gauge group}\label{sect:gauge_group}
In this section, we will demonstrate that the gauge group of the reduced $D=5$ theory is $SO(2)\times\left(SO(2)\ltimes_{\Sigma_2}\mb{R}^4\right)$, where the action of the semi-direct product depends on the curvature of $\Sigma_2$. Specifically, it is $SO(2)\times G_{A^{100}_{5,17}}$ when $l=0$, and $SO(2)\times G_{A^{0}_{5,18}}$ when $l=\pm 1$, where $G_{A^{100}_{5,17}}$ and $G_{A^{0}_{5,18}}$ are two five-dimensional matrix Lie groups with Lie algebras $A^{100}_{5,17}$ and $A^{0}_{5,18}$ respectively. We will elaborate more on this in the discussion below. The compact $SO(2)$ subgroup of the gauge group is generated by
\begin{align}
\mathfrak{g}_0=t_{45}\,,
\end{align}
which is associated with the gauge field ${A}_{(1)}$, and the non-compact part of the gauge group, $SO(2)\ltimes_{\Sigma_2}\mb{R}^4$, is generated by
\begin{equation}\label{eq:weird_generators}
\begin{split}
\mathfrak{g}_1=-t_{23}\,,\quad \mathfrak{g}_2=t_{13}\,, \quad \mathfrak{g}_3=-t_{36}\,,\quad\mathfrak{g}_4=-t_{37}\,, \quad \mathfrak{g}_5=t_{26}-t_{17}+lt_{12}\,,
\end{split}
\end{equation}
which are associated with the gauge fields ${\mathscr{A}}^\alpha_{(1)}$, ${V}^\alpha_{(1)}$ and ${\mc{A}}_{(1)}$ respectively (see \eqref{eq:5d_precise_gauging}). We note that the one-form ${\mathscr{B}}_{(1)}$ does not participate in the gauging. The generators in \eqref{eq:weird_generators} satisfy the following commutation relations
\begin{equation}
\begin{split}
\left[\mathfrak{g}_1,\mathfrak{g}_5\right]=-\mathfrak{g}_2\,,\quad \left[\mathfrak{g}_2,\mathfrak{g}_5\right]=\mathfrak{g}_1\,,\quad \left[\mathfrak{g}_3,\mathfrak{g}_5\right]=-l\mathfrak{g}_1-\mathfrak{g}_4\,,\quad \left[\mathfrak{g}_4,\mathfrak{g}_5\right]=-l\mathfrak{g}_2+\mathfrak{g}_3\,.
\end{split}
\end{equation}
Rather remarkably, the algebra associated to $l=0$ is not isomorphic to that associated to $l=\pm 1$. These two distinct algebras belong to two different families of five-dimensional real Lie algebras, namely $A^{spq}_{5,17}$ and $A^{p}_{5,18}$, which are listed in \cite{Patera:1976ud}. The subscripts $m$ and $n$ in $A^{p}_{m,n}$ denote the dimension of the Lie algebra and the $n$-th algebra on the list of \cite{Patera:1976ud} respectively, and the superscript $p$ in $A^{p}_{m,n}$ denotes the continuous parameter(s) on which the algebra can depend on. Specifically, when $l=0$, the algebra is described by $A^{100}_{5,17}$, and its minimal matrix group representation is given by \cite{Ghanam:2014} \footnote{The minimal matrix group representation of $A^{spq}_{5,17}$ is in general a $5\times 5$ matrix, however the minimal representation is reduced to a $4\times 4$ matrix when $p=q$ \cite{Ghanam:2014}.}
\begin{equation}\label{eq:matrix_rep_A517}
M_{A^{100}_{5,17}} = \begin{pmatrix} \cos\theta & \sin\theta &x_1 &x_3 \\ -\sin\theta & \cos\theta &x_2 &x_4  \\ 0 & 0 &1 &0 \\0 & 0 &0 &1  \\\end{pmatrix} \,,
\end{equation}
while the algebras for the $l=\pm 1$ cases are described by $A^{0}_{5,18}$ and their minimal matrix group representations are given by \cite{Ghanam:2014}
\begin{equation}\label{eq:matrix_rep_A518}
M_{A^{0}_{5,18}} = \begin{pmatrix} \cos\theta & \sin\theta &x_1 &x_3 \\ -\sin\theta & \cos\theta &x_2 &x_4  \\ 0 & 0 &1 &-l\theta \\0 & 0 &0 &1  \\\end{pmatrix} \,,
\end{equation}
where $\theta,x_1,x_2,x_3,x_4$ are the real parameters. We also provide an explicit representation of the generators \eqref{eq:weird_generators} in appendix \ref{appendix:min_rep}, which after exponentiation recovers \eqref{eq:matrix_rep_A517} and \eqref{eq:matrix_rep_A518}.

To understand the structure of the gauge group, let's focus on $l=0$, and consider two elements
\begin{equation}
M_1 = \begin{pmatrix} \cos\theta & \sin\theta &x_1 &x_3 \\ -\sin\theta & \cos\theta &x_2 &x_4  \\ 0 & 0 &1 &0 \\0 & 0 &0 &1  \\\end{pmatrix} \,,\quad M_2 = \begin{pmatrix} \cos\phi & \sin\phi &y_1 &y_3 \\ -\sin\phi & \cos\phi &y_2 &y_4  \\ 0 & 0 &1 &0 \\0 & 0 &0 &1  \\\end{pmatrix} \,.
\end{equation}
The composition $M_3 = M_1M_2$ sends
\begin{equation}
\theta \mapsto \theta + \phi \,,\quad x_i \mapsto x_i + R(\theta)y_i \,,\quad R(\theta) = \begin{pmatrix} \cos\theta & \sin\theta \\ -\sin\theta & \cos\theta \end{pmatrix}\oplus\begin{pmatrix} \cos\theta & \sin\theta \\ -\sin\theta & \cos\theta \end{pmatrix} \,,
\end{equation}
where $i\in\{1,2,3,4\}$. From this, we observe that this group is isomorphic to $SO(2)\ltimes\mb{R}^4$. When $l=\pm1$, the above map becomes a bit more complicated, but the overall $SO(2)\ltimes\mb{R}^4$ structure remains the same. Putting it all together, we conclude that the gauge group of the reduced $D=5$ theory is $SO(2)\times\left(SO(2)\ltimes_{\Sigma_2}\mb{R}^4\right)$.

\subsubsection{Embedding tensors}
The components of the embedding tensor are specified by
\begin{equation}\label{eq:embedding_tensor_comp}
\begin{split}
&\xi^M=0\,,\quad \xi^{45}=-\sqrt{2}\,,\\
&f_{178}=\sqrt{2}\,,\quad f_{268}=-\sqrt{2}\,,\quad f_{678}=-\sqrt{2}l\,,
\end{split}
\end{equation}
and the remaining components are zero. These satisfy the algebraic constraints given in \eqref{algconstraints}. With \eqref{eq:embedding_tensor_comp}, we can identify the gauge fields and two-forms of the $\mc N =4$ theory with those of our reduced theory via 
\begin{align*}
\mc{A}^0_{(1)} &= \frac{1}{\sqrt{2}}A_{(1)} \,,\quad \mc{A}^1_{(1)} =\frac{1}{\sqrt{2}}\left({\mathscr{A}}^3_{(1)}-l{V}^4_{(1)}\right) \,,\quad \mc{A}^2_{(1)} =\frac{1}{\sqrt{2}}\left({\mathscr{A}}^4_{(1)}+l{V}^3_{(1)}\right) \,,\\
\mc{A}^3_{(1)} &=-\frac{1}{\sqrt{2}}\left(\mathscr{{B}}_{(1)}- \tau^\alpha\left[{\mathscr{A}}^\alpha_{(1)}-l\epsilon_{\alpha\beta}{V}^\beta_{(1)}- \frac{l}{2g}\epsilon_{\alpha\beta}d\tau^\beta\right]+\frac{1}{2g}{\tau}^2d{\Xi}-{\Xi}{\tau}^\alpha{V}^\alpha_{(1)}+\frac{1}{g}\xi^\alpha Q^\alpha_{(1)}\right) \,,\\
\mc{B}^4_{(2)} &= \frac{1}{g}L^2_{(2)} \,,\quad \mc{B}^5_{(2)} = -\frac{1}{g}L^1_{(2)} \,,\\
\mc{A}^6_{(1)} &=-\frac{1}{\sqrt{2}}V^4_{(1)} \,,\quad \mc{A}^7_{(1)} =\frac{1}{\sqrt{2}}V^3_{(1)} \,, \quad \mc{A}^8_{(1)} =\frac{1}{\sqrt{2}}{\mc{A}}_{(1)} \,,\numberthis
\end{align*}
and the remaining components of $\mc{A}^M_{(1)}$ and $\mc{B}^M_{(2)}$ are all zero. For completeness, the corresponding covariant 2-form field strengths of the $\mc N=4$ theory are given by
\begin{align*}\label{eq:canonical_field_strengths}
\mc{H}^0_{(2)}&=\frac{1}{\sqrt{2}}{{F}}_{(2)}\,,\\
\mc{H}^1_{(2)}&=\frac{1}{\sqrt{2}}\left(d\left[{\mathscr{A}}^3_{(1)}-l{V}^4_{(1)}\right]+{g}{\mc{A}}_{(1)}\wedge \left[{\mathscr{A}}^4_{(1)}+l{V}^3_{(1)}\right]+{g}l{V}^3_{(1)}\wedge {\mc{A}}_{(1)}\right)\,,\\
\mc{H}^2_{(2)}&=\frac{1}{\sqrt{2}}\left(d\left[{\mathscr{A}}^4_{(1)}+l{V}^3_{(1)}\right]-{g}{\mc{A}}_{(1)}\wedge\left[{\mathscr{A}}^3_{(1)}-l{V}^4_{(1)}\right]+{g}l{V}^4_{(1)}\wedge {\mc{A}}_{(1)}\right)\,,\\
\mc{H}^3_{(2)}&=-\frac{1}{\sqrt{2}}d\left(\mathscr{{B}}_{(1)}- \tau^\alpha\left[{\mathscr{A}}^\alpha_{(1)}-l\epsilon_{\alpha\beta}{V}^\beta_{(1)}- \frac{l}{2g}\epsilon_{\alpha\beta}d\tau^\beta\right] + \frac{1}{2g}\tau^2 d\Xi -\Xi\tau^\alpha V^\alpha_{(1)}+ \frac{1}{g}\xi^\alpha Q^\alpha_{(1)}\right)\\
&\quad-\frac{1}{\sqrt{2}}{g}\left[{\mathscr{A}}^\alpha_{(1)}-l\epsilon_{\alpha\beta}{V}^\beta_{(1)}\right]\wedge \tilde{V}^\alpha_{(1)} +\frac{1}{2\sqrt{2}}{g}l\epsilon_{\alpha\beta}{V}^\alpha_{(1)}\wedge {V}^\beta_{(1)} \,,\\
\mc{H}^4_{(2)}&=\frac{1}{\sqrt{2}}{L}^1_{(2)}\,,\quad \mc{H}^5_{(2)}=\frac{1}{\sqrt{2}}{L}^2_{(2)}\,,\\
\mc{H}^6_{(2)}&=-\frac{1}{\sqrt{2}}{D}{V}^4_{(1)}\,,\quad \mc{H}^7_{(2)}=\frac{1}{\sqrt{2}}{D}{V}^3_{(1)}\,,\quad \mc{H}^8_{(2)}=\frac{1}{\sqrt{2}}{\mc{F}}_{(2)}\,.\numberthis
\end{align*}
With the above identifications, we find that the Lagrangian of our $D=5$ theory is equivalent to the canonical $\mc N=4$ Lagrangian. We have presented a few details of this calculation in appendix \ref{app:matching_N=4}.

\subsection{Half-maximal truncation and $D=5$, $\mc N=2$ supersymmetry}\label{5dN=2SUSY}

From \eqref{halfmax}, the half-maximal truncation of our $D=5$, $\mc N=4$ theory is obtained by setting 
\begin{equation}
\tau^\alpha = \xi^\alpha =  \Psi^a = {\mathscr{A}}^\alpha_{(1)} =  V^\alpha_{(1)} = K^a_{(2)} = 0 \,.
\end{equation}
The truncated theory is a $D=5$, $\mc{N}=2$ supergravity theory coupled to two vector multiplets, with very special real manifold $SO(1,1)\times SO(1,1)$, and two hypermultiplets parameterising the quaternionic K\"{a}hler manifold $SO(4,2)/(SO(4)\times SO(2))$ with gauging in the hypermultiplet sector. The gauge group is $SO(2)\times\mb{R}^+$ when $l=\pm1$ and $SO(2)\times SO(2)$ when $l=0$. To make the $\mc{N}=2$ supersymmetry explicit, we will rewrite the 5-dimensional theory in the canonical form. A review of the canonical Lagrangian of generic $D=5$, $\mc{N}=2$ supergravity coupled to matter is given in \cite{Gunaydin:1999zx,Ceresole:2000jd,Bergshoeff:2004kh}, and we will follow mostly the conventions of \cite{Bergshoeff:2004kh}.

Using the $\mc{N}=2$ language of \cite{Bergshoeff:2004kh}, the Lagrangian of the bosonic sector of the theory can be written as 
\begin{align}\label{eq:5d_canonical_lagrangian}
\begin{split}
\mc{L}_{(5)} &= {R}\,{{\text{vol}}}_5-\frac{1}{2}a_{IJ}{{\ast} H^{I}}\wedge H^{J}-\frac{1}{2}g_{xy}{{\ast} D\phi^{x}}\wedge D\phi^y-\frac{1}{2}g_{XY}{{\ast} Dq^X}\wedge Dq^Y\\
&\quad-\frac{1}{3\sqrt{3}}\mc{C}_{IJK}A^I\wedge F^J\wedge F^K+\mc{L}_{\mc{N}=2}^{pot}\,{{\text{vol}}}_5 \,,
\end{split}
\end{align} 
where
\begin{equation}
\begin{split}
&D_{\mu}\phi^x\equiv\partial_{\mu}\phi^x+{g}A^{I}_{\mu}K^x_{I}\,,\\
&D_{\mu}q^X\equiv\partial_{\mu}q^X+{g}A^{I}_{\mu}k^X_{I}\,,\\
&H_{(2)}^I\equiv dA_{(1)}^{I}+\frac{1}{2}{g}\bar f_{JK}^{\phantom{JK}I}A_{(1)}^J\wedge A_{(1)}^K \,.
\end{split}
\end{equation} 
Here, $A_{(1)}^I$, with $I\in\{0,1,2\}$, label the graviphoton as well as the two vector fields of the two vector multiplets, and $\phi^x$, with $x\in\{1,2\}$, denotes the two real scalar fields of the two vector multiplets that parametrise a very special real manifold $SO(1,1)\times SO(1,1)$. The $q^X$, with $X\in\{1,\dots,8\}$, are the eight real scalar fields of the two hypermultiplets that parametrise the quaternionic K\"{a}hler manifold $SO(4,2)/(SO(4)\times SO(2))$. Within the covariant derivatives, $K^x_{I}$ and $k^X_{I}$ are sets of three Killing vectors on the very special real manifold and on the quaternionic K\"{a}hler manifold respectively. The structure constants of the gauge group are given by $\bar f_{JK}^{\phantom{JK}I}$. We explain below how our truncated Lagrangian can be recast in the form of \eqref{eq:5d_canonical_lagrangian} with gauging only in the hypermultiplet sector.

We start with the vector multiplets. The very special real geometry is determined by a real, symmetric, constant tensor $\mathcal{C}_{IJK}$ which specifies the embedding of $SO(1,1)\times SO(1,1)$ in a three-dimensional space with coordinates $h^I$ through
\begin{equation}
\mathcal{C}_{IJK}h^I h^J h^K=1\,.
\end{equation}
Defining $h_I=\mathcal{C}_{IJK}h^J h^K$, the kinetic terms for the vectors are defined by
\begin{equation}\label{aijmet}
a_{IJ}=-2\mathcal{C}_{IJK}h^K+3h_I h_J\,.
\end{equation}
Indices can be lowered and raised using $a_{IJ}$ and its inverse $a^{IJ}$. Moreover, the pull-back of $a_{IJ}$ gives the metric for the scalar fields $\phi^x$
\begin{equation}\label{gxymet}
g_{xy}=3\partial_x h^I\partial_y h^Ja_{IJ}\,.
\end{equation}
With these definitions, we can immediately identify the field strengths and vector fields as
\begin{equation}
A_{(1)}^{I}=\left({A}_{(1)},{\mc{A}}_{(1)},\mathscr B_{(1)}\right) \quad \Longleftrightarrow \quad H_{(2)}^{I}=\left({F}_{(2)},{\mc{F}}_{(2)},{\mathscr{G}}_{(2)}\right) \,, 
\end{equation} 
where $\mathscr G_{(2)} = d\mathscr B_{(1)}$, and the non-zero components of both symmetric tensors $a_{IJ}$ and $\mc{C}_{IJK}$ are
\begin{equation}
a_{00}=\Phi^{-1/2}e^{4{\phi}+12{\lambda}},\quad a_{11}=\Phi^{-1/2}e^{4{\phi}-12{\lambda}},\quad a_{22}=\Phi e^{-8{\phi}},\quad \mc{C}_{012}=\frac{\sqrt{3}}{2} \,.
\end{equation} 
The scalar coset of the vector multiplet sector is identified to be $SO(1,1)\times SO(1,1)$, and $g_{xy}=\delta_{xy}$. In \eqref{eq:ungauged_ke_scalar}, we provide the canonical expression of the ungauged kinetic term for the scalar fields, which allows us to identify the two scalar fields within the very special real manifold to be,
\begin{equation}
\phi^1=2\sqrt{6}\left({\phi}-\frac{1}{8}\log\Phi\right),\quad \phi^2=6\sqrt{2}{\lambda} \,,
\end{equation} 
with the embedding coordinates given by
\begin{align}
\begin{split}\label{embeddingcoordinate}
h^I&=\frac{1}{\sqrt{3}}\left(\Phi^{1/4}e^{-2{\phi}-6{\lambda}},\Phi^{1/4}e^{-2{\phi}+6{\lambda}},\Phi^{-1/2} e^{4{\phi}}\right)\\
&=\frac{1}{\sqrt{3}}\left(e^{-\phi^1/{\sqrt{6}}-\phi^2/{\sqrt{2}}},e^{-{\phi^1}/{\sqrt{6}}+{\phi^2}/{\sqrt{2}}},e^{{2\phi^1}/{\sqrt{6}}}\right)\,.
\end{split}
\end{align} 
We now turn to the hypermultiplets. The explicit parametrisation of the quaternionic K\"{a}hler manifold $SO(4,2)/(SO(4)\times SO(2))$ can be found in appendix \ref{coset_parametrisation}. We identify the coordinates on this quaternionic K\"{a}hler manifold to be
\begin{equation}
q^X=\left(\varphi_1,3{\phi}+\frac{1}{4}\log\Phi,\rho,\xi,{\psi}^{11},{\psi}^{21},{\psi}^{12},{\psi}^{22}\right)\,.
\end{equation}
The two scalar fields $\varphi_1$ and $\rho$ are introduced to parametrise the coset ${\mc{T}}_{\alpha\beta}$ as we explain in appendix \ref{coset_parametrisation}. The associated metric on this quaternionic K\"{a}hler manifold can be found in \eqref{eq:scalarcoset_metric}. The Killing vectors $k^X_{I}$, which are responsible for the gauging, are given by 
\begin{equation}
\begin{split}
k_0&=\epsilon^{ab}{\psi}^{b\alpha}\partial_{{\psi}^{a\alpha}} \,,\\
k_1&=l\partial_{\xi}+\epsilon^{\alpha\beta}{\psi}^{a\beta}\partial_{{\psi}^{a\alpha}}+\rho\partial_{\varphi_1}+\left(e^{-4\varphi_1}-\rho^2-1\right)\partial_{\rho} \,,\\
k_2&=0 \,.
\end{split}
\end{equation}
The gauging only involves two vectors, $A_{(1)}$ and ${\mc{A}}_{(1)}$, and the gauge group is $SO(2)\times SO(2)$ when $l=0$, and $SO(2)\times\mb{R}^+$ when $l=\pm1$. To completely demonstrate the $\mc{N}=2$ supersymmetry, it remains to check that the potential matches with $\mc{L}^{pot}_{\mc{N}=2}$, which is given in \eqref{eq:5d_pot}. We will show that this is indeed the case in appendix \ref{coset_parametrisation}.

We have shown that our truncation is an $\mc{N}=2$ supergravity theory in $D=5$ comprised of one gravity multiplet, two vector multiplets and two hypermultiplets with $SO(2)\times\mb{R}^+$ gauging when $l=\pm1$ and $SO(2)\times SO(2)$ gauging when $l=0$. The scalar manifold of this truncated theory is $SO(1,1)\times SO(1,1)\times SO(4,2)/(SO(4)\times SO(2))$, with $SO(1,1)\times SO(1,1)$ describing the vector multiplet sector and $SO(4,2)/(SO(4)\times SO(2))$ describing the hypermultiplet sector. In view of the scalar manifold of the $D=5$, $\mc N=4$ theory we obtained in the previous section, we find that it is decomposed as
\begin{equation}
SO(1,1) \times \frac{SO(5,3)}{SO(5)\times SO(3)} \longrightarrow SO(1,1)\times SO(1,1) \times\frac{SO(4,2)}{SO(4)\times SO(2)} 
\end{equation}
via the half-maximal truncation. We also observe that one of the $SO(1,1)$ factors in the scalar manifold is directly inherited from the $\mc N=4$ theory. The second $SO(1,1)$ factor in our truncated theory can be interpreted as a scalar that decoupled from the original $SO(5,3)/(SO(5)\times SO(3))$ coset. In terms of the non-linearly realised global symmetry of the scalar manifolds, we find that under the half-maximal truncation, 
\begin{equation}
SO(5,3) \longrightarrow SO(1,1)\times SO(4,2) \,,
\end{equation}
where we ignored the common factor of $SO(1,1)$. Under the half-maximal truncation, the gauge group becomes $SO(2)\times \mb{R}^+$ when $l=\pm1$, with $\mb{R}^+\subset G_{A^0_{5,18}}$, and $SO(2)\times SO(2)$ when $l=0$, with the second factor $SO(2)\subset G_{A^{100}_{5,17}}$.

\subsection{Further consistent sub-truncations}

For the $D=5$, $\mc{N}=2$ theory, there are two further sub-truncations of interest that arise from keeping sectors invariant under the subgroups of $SO(2)_1\times SO(2)_2$. As we will see, these sub-truncations do not form supergravity theories.

\subsubsection{$SO(2)_1$ invariant sector}

We consider keeping only the fields that are singlets under $SO(2)_1$. This amounts to setting $\mathscr B_{(1)}=0$, $\psi^{a\alpha}=0$ and $A_{(1)}=0$ in the equations of motion. For $l\neq0$, we find that $\Xi$ acts as a Stueckelberg field for ${\mc{A}}_{(1)}$, as it appears in the gauge invariant combination $X_{(1)} = d\Xi +  g l{\mc{A}}_{(1)}$. Its equation of motion is implied by the equation of motion for ${\mc{A}}_{(1)}$. The remaining fields, taking into account of $\Xi$ being pure gauge, consists of the metric, the scalars ${\mc{T}}_{\alpha\beta}, \phi,\lambda,\Phi$ and the 1-form ${\mc{A}}_{(1)}$. 

\subsubsection{$SO(2)_2$ invariant sector}

Here, we consider only singlets under $SO(2)_2$. It is consistent to set $\psi^{a\alpha} = 0$, ${\mc T}_{\alpha\beta} = \delta_{\alpha\beta}$, $\mathscr B_{(1)} =0$, and ${\mc A}_{(1)} = 0$. It is also consistent to set $A_{(1)} = 0$ as well as $\Xi = 0$ in the equations of motion. The remaining fields consist of the metric, $\phi$, $\lambda$ and $\Phi$. For $l=0$, we can set $\lambda=0$ and $\log\Phi = -12\phi$. The remaining field is $\phi$, with a negative-definite potential. For $l=\pm1$, it is impossible to truncate any linear combination of $\phi$, $\lambda$ and $\Phi$. The $l=\pm1$ cases cover the solutions in \cite{Gauntlett:2001ps,Bigazzi:2001aj}, whereas the $l=0$ case cover the NS5-brane near-horizon, linear dilaton solution. We will present these solutions below. 

\subsection{Some solutions}

To demonstrate the consistency of our truncation, we will reproduce the one-parameter family of $1/4$-BPS solutions obtained in \cite{Gauntlett:2001ps,Bigazzi:2001aj} corresponding to an NS5-brane wrapping on an $S^2$ or $\mb{H}^2$. The solutions with an $S^2$ are dual to pure $\mc{N}=2$ super Yang-Mills theory in $D=4$ in the IR, while the dual description of the solutions with an $\mb{H}^2$ is unclear. From the 5-dimensional perspective, the solutions lie in the $SO(2)_2$ invariant sector of the half-maximal theory with $l=\pm1$, where the only fields are the metric, $\phi$, $\lambda$, and $\Phi$. The metric is given by 
\begin{equation}
d s^2_5 = {g}^{\frac{4}{3}}z^{\frac{2}{3}}e^{-\frac{2}{3}(x-2l{g}^2z)}\left[ds^2(\mb{R}^{1,3}) + {g}^2e^{2x}dz^2\right] \,,
\end{equation} 
where $z$ is a radial variable, and the function $x(z)$ is defined as
\begin{equation}
e^{-2x} = 1-\frac{l(1+ce^{-2l{g}^2z})}{2{g}^2z} \,.
\end{equation} 
Here, $c$ is a real integration constant that, for $l=1$, parameterises the different flows from the UV to the IR. The values of the scalar fields are 
\begin{equation}
e^{6\phi} = {g}^2ze^{-\frac{2}{5}(x-2l{g}^2z)} \,,\quad \lambda = -\frac{x}{6} \,,\quad \Phi^{\frac{5}{4}} = e^{x-2l{g}^2z} \,.
\end{equation} 
The above solutions can easily be uplifted back to $D=10$ using our truncation procedure. The explicit uplift is given in  \cite{Gauntlett:2001ps,Bigazzi:2001aj}.

For $l=0$ in the $SO(2)_2$ invariant sector, we report the following domain wall solution 
\begin{equation}
e^{5\phi} = \frac{2 g}{3}r \,,\quad ds^2_5 = r^2 ds^2(\mb{R}^{1,3}) + dr^2\,,
\end{equation}
with $\lambda = 0$ and $\log\Phi = -12\phi$. This corresponds to the NS5-brane near-horizon, linear dilaton solution when uplifted to $D=10$. 

\section{Consistent KK truncations on Slag 3-cycles in $CY_3$}\label{KK_truncations_Slag}

\subsection{Review of wrapped M5-branes on Slag 3-cycles in $CY_3$}\label{M5CY3truncations_review}

As in Section \ref{KK_truncations_Riemann}, we will first summarise the consistent KK ansatz used in \cite{Donos:2010ax} to describe M5-branes wrapped on Slag 3-cycles. The $D = 7$ metric is given by 
\begin{equation}
ds^2_7=e^{-6\phi}ds^2_{4}+e^{4\phi}ds^2(\Sigma_3) \,,
\end{equation} 
where $\phi$ is a real scalar field defined in $D=4$. We denote \{$\bar{e}^m;m\in\{0,\ldots,3\}$\} and \{$\bar{e}^a;a\in\{1,2,3\}$\} as the orthonormal frames for both $ds^2_{4}$ and $ds^2(\Sigma_3)$ respectively, and let $\bar{\omega}^{m}_{\phantom{m}n}$ and $\bar{\omega}^{a}_{\phantom{a}b}$ be the corresponding spin connections. The metric of the Slag 3-cycle is normalised to satisfy $R_{ab}=lg^2\delta_{ab}$ with $l=0,\pm1$. The $SO(5)$ fields are decomposed via $SO(5)\to SO(3)\times SO(2)$, where the $SO(5)$ vector index is $i = (a,\alpha)$, with $a\in\{1,2,3\}$ and $\alpha\in\{4,5\}$, and the fields under this decomposition are given by
\begin{equation}
\begin{split}
&A^{ab}_{(1)}=\frac{1}{g}\bar{\omega}^{ab}+\beta\epsilon_{abc}\,\bar{e}^c\,,\\
&A^{a\alpha}_{(1)}=-A^{\alpha a}_{(1)}=\theta^\alpha \bar{e}^a\,,\\
&A^{\alpha\beta}_{(1)}=\epsilon^{\alpha\beta}A_{(1)}\,.
\end{split}
\end{equation} 
This ansatz incorporates the spin connection $\bar{\omega}^{ab}$ in the expression for $A^{ab}_{(1)}$, which corresponds to the topological twist condition that ensures the preservation of supersymmetry on the wrapped M5-brane with worldvolume $\mb{R}^{1,2}\times \Sigma_3$. The ansatz is also comprised of a scalar field $\beta$, two scalar fields $\theta^\alpha$ and a vector field $A_{(1)}$, all defined in $D = 4$. The 3-form fields $S^i_{(3)}$ are split into 
\begin{equation}
\begin{split}
&S^a_{(3)}=B_{(2)}\wedge\bar{e}^a+\epsilon^{abc}C_{(1)}\wedge \bar{e}^b\wedge \bar{e}^c\,,\\
&S^\alpha_{(3)}=h^{\alpha}_{(3)}+g\chi^{\alpha} \vol(\Sigma_3) \,,
\end{split}
\end{equation} 
where $B_{(2)}$, $C_{(1)}$, $h^\alpha_{(3)}$, $\chi^\alpha$ are 2-, 1-, 3- and 0-forms in $D=4$ respectively. The $SL(5,\mb{R})/SO(5)$ scalars $T_{ij}$ are taken to be
\begin{equation}
T^{ab}=e^{-4\lambda}\delta^{ab}\,,\quad T^{a\alpha}=0 \,,\quad T^{\alpha\beta}=e^{6\lambda}\mc{T}^{\alpha \beta} \,,
\end{equation} 
where $\lambda$ is a scalar, and the symmetric, unimodular matrix $\mc{T}^{\alpha \beta}$ parametrises $SL(2,\mb{R})/SO(2)$, all in $D=4$. The resulting 4-dimensional theory is a $U(1)\times \mb{R}^+$ gauged $\mc{N}=2$ supergravity theory coupled to a vector multiplet and two hypermultiplets with scalar manifold $SU(1,1)/U(1) \times G_{2(2)}/SO(4)$ and gauging in the hypermultiplet sector.

\subsection{NS5-branes wrapped on Slag 3-cycles in $CY_3$}

For NS5-branes wrapped on Slag 3-cycles, we will, as we did for Riemann surfaces, write down the ansatz in terms of the maximal $ISO(4)$ theory, then use the IW contraction in \eqref{7d_IW_ansatz} to relate the M5 and NS5 fields. In terms of the maximal theory, the ansatz for the $D = 7$ metric is given by 
\begin{equation}\label{ns57dmetric4d}
ds^2_7=e^{-6\tilde{\phi}}d\tilde{s}^2_{4}+e^{4\tilde{\phi}}ds^2(\tilde{\Sigma}_3) \,.
\end{equation} 
We introduce orthonormal frames \{$\bar{\tilde{e}}^m;m\in\{0,\ldots,3\}$\} and \{$\bar{\tilde{e}}^a;a\in\{1,2,3\}$\} for both $d\tilde{s}^2_{4}$ and $ds^2(\tilde{\Sigma}_3)$ respectively, and let $\bar{\tilde{\omega}}^{m}_{\phantom{m}n}$ and $\bar{\tilde{\omega}}^{a}_{\phantom{a}b}$ be the corresponding spin connections. We normalise the metric of the Slag 3-cycle to satisfy $\tilde{R}_{ab}=l\tilde{g}^2\delta_{ab}$ with $l=0,\pm1$. The fields are decomposed via $SO(4)\to SO(3)_1\times SO(3)_2\to SO(3)_1$, and are given by
\begin{equation}
\begin{split}
&\tilde{A}^{ab}_{(1)}=\frac{1}{\tilde{g}}\bar{\tilde{\omega}}^{ab}+\tilde\beta\epsilon_{abc} \bar{\tilde e}^c\,,\\
&\tilde{A}^{a4}_{(1)}=\tilde\Theta \bar{\tilde{e}}^a\,,\\
&\tilde{A}^{a5}_{(1)}=\tilde\theta \bar{\tilde{e}}^a \,,\\
&\tilde A^{45}_{(1)} = \tilde{{A}}_{(1)} \,.
\end{split}
\end{equation}
This again incorporates the spin connection $\bar{\tilde{\omega}}^{ab}$ in the expression for $\tilde{A}^{ab}_{(1)}$, which corresponds to the topological twist condition that ensures the preservation of supersymmetry on the wrapped NS5-brane. The 3-form fields are taken to be
\begin{equation}
\begin{split}
&\tilde{S}^a_{(3)} = \tilde{B}_{(2)}\wedge\bar{\tilde{e}}^a + \epsilon^{abc}\tilde C_{(1)}\wedge\bar{\tilde{e}}^b\wedge \bar{\tilde{e}}^c \,, \\
&\tilde S^4_{(3)} = \tilde h_{(3)} +\tilde g \tilde{\chi} \vol(\tilde{\Sigma}_3) \,, \\
&\tilde{S}^5_{(3)}=\tilde{H}_{(3)}+\tilde g\tilde{X} \vol(\tilde{\Sigma}_3) \,.
\end{split}
\end{equation} 
For the $SL(4,\mb{R})/SO(4)$ scalars $\tilde{T}_{ij}$ and the $\bd{4}$ scalars $\tilde\tau^A$, we take 
\begin{equation}\label{ns57dscalars4d}
\begin{split}
&\tilde\tau^a = 0\,,\quad \tilde\tau^4=\tilde\tau \,,\\
&\tilde{T}^{ab}=e^{-4\tilde{\lambda}}\delta^{ab}\,,\quad \tilde{T}^{a4}=0\,,\quad  \tilde{T}^{44}=e^{12\tilde{\lambda}} \,.
\end{split}
\end{equation} 
Using \eqref{7d_IW_ansatz}, we make the following identification of the M5 and NS5 fields 
\begin{equation} \label{4d_identification}
\begin{split}
&g=k^2\tilde{g}\,,\quad \bar{e}^a={k^{-2}}\bar{\tilde{e}}^a\,,\quad \bar{e}^m=k^{3}\bar{\tilde{e}}^m\,,\quad \phi=\tilde\phi + \log k\,,\quad \beta=\tilde\beta \,,\quad \theta^4 = \tilde\theta \,,\\
&\theta^5=k^5\tilde\Theta \,,\quad \chi^4 = k^5\tilde \chi \,,\quad \chi^5 =\tilde X \,,\quad \lambda=\tilde\lambda - \frac{1}{16}\log\tilde\Phi + \frac{1}{2}\log k\,,\\
&\mc{T}^{44}=k^{-5}e^{6\tilde\lambda}\tilde\Phi^{5/8}\,,\quad \mc{T}^{45} = -e^{6\tilde\lambda}\tilde\Phi^{5/8}\tilde\tau \,,\quad \mc{T}^{55}= k^5(e^{-6\tilde\lambda}\tilde\Phi^{-5/8}+e^{6\tilde\lambda}\tilde\Phi^{5/8}\tilde\tau^2) \,, \\
&A_{(1)}=k^{3}\tilde{A}_{(1)}\,,\quad C_{(1)} = k^5\tilde C_{(1)} \,,\quad B_{(2)} = k^3\tilde B_{(2)} \,,\quad h^4_{(3)} = k\tilde h_{(3)} \,,\quad h^5_{(3)} = k^{-4}\tilde H_{(3)} \,.
\end{split}
\end{equation} 
As in Section \ref{KK_truncations_Riemann}, we obtain the equations of motion governing the NS5 fields by substituting the above identification into the $D=4$ equations for the M5 fields and taking the $k\to0$ limit. The resulting equations of motion are collected in appendix \ref{4deomssection}. To present the Lagrangian, we define the following combinations of our fundamental fields,
\begin{equation}\label{4dcombinations}
\begin{split}
&\tilde\rho \equiv \tilde\Theta + \tilde\tau\tilde\theta \,,\\
&\tilde\sigma \equiv \tilde\chi - \tilde\tau\tilde X \,,\\
&\tilde P_{(1)} \equiv d\tilde\Theta + \tilde\tau d\tilde\theta - \tilde g\tilde\theta \tilde A_{(1)} \,,\\
&\tilde Q_{(1)} \equiv d\tilde\tau + \tilde g \tilde A_{(1)} \,,\\
&\tilde G_{(3)} \equiv \tilde h_{(3)} -\tilde\tau \tilde H_{(3)} \,,
\end{split}
\end{equation}
and integrate \eqref{H34d} and \eqref{X4d} to write
\begin{equation}
\begin{split}
&\tilde H_{(3)} = d\tilde\Sigma_{(2)} \,,\\
&\tilde X = \frac{3}{2}l\tilde\theta - \tilde\theta^3 - 3\tilde\beta^2\tilde\theta \,.
\end{split}
\end{equation} 
With this, the 4-form Lagrangian given by 
\begin{equation}\label{4doriginal}
\mc{L}_{(4)} = \tilde R\,\tilde\vol_4 + \mc{L}^{kin}_{(4)} + \mc{L}^{pot}_{(4)} + \mc{L}^{top}_{(4)} \,,
\end{equation}
where $\tilde R$ is the Ricci scalar of the $D=4$ metric, the kinetic terms are
\begin{align*}
\mc{L}^{kin}_{(4)} &= -30\tilde\ast d\tilde\phi\wedge d\tilde\phi - 48\tilde\ast d\tilde\lambda\wedge d\tilde\lambda - \frac{5}{16}\tilde\Phi^{-2}\tilde\ast d\tilde\Phi \wedge d\tilde\Phi - \frac{3}{2}\tilde\Phi^{-1/2}e^{8\tilde\lambda-4\tilde\phi}\tilde\ast d\tilde\beta\wedge d\tilde\beta \\
&\quad - \frac{3}{2}\tilde\Phi^{-1/2}e^{-8\tilde\lambda-4\tilde\phi}\tilde\ast d\tilde\theta\wedge d\tilde\theta  -6\tilde\Phi^{1/4}e^{-4\tilde\lambda-8\tilde\phi}\tilde\ast\tilde C_{(1)}\wedge\tilde C_{(1)}- \frac{1}{2}\tilde\Phi^{5/4}e^{12\tilde\lambda}\tilde\ast\tilde Q_{(1)}\wedge\tilde Q_{(1)}\\
&\quad -\frac{3}{2}\tilde\Phi^{3/4}e^{4\tilde\lambda-4\tilde\phi}\tilde\ast\tilde P_{(1)}\wedge\tilde P_{(1)}-\frac{1}{2}\tilde\Phi^{3/4}e^{-12\tilde\lambda+6\tilde\phi}\tilde\ast\tilde F_{(2)}\wedge\tilde F_{(2)} - \frac{1}{2}\tilde\Phi^{-1}e^{12\tilde\phi}\tilde\ast\tilde H_{(3)}\wedge\tilde H_{(3)} \\
&\quad-\frac{3}{2}\tilde\Phi^{1/4}e^{-4\tilde\lambda+2\tilde\phi}\tilde\ast\tilde B_{(2)}\wedge\tilde B_{(2)}- \frac{1}{2}\tilde\Phi^{1/4}e^{12\tilde\lambda+12\tilde\phi}\tilde\ast\tilde G_{(3)}\wedge\tilde G_{(3)} \,,\numberthis
\end{align*}
the potential terms are
\begin{align*}\label{4dpotentialterm}
\mc{L}^{pot}_{(4)} = \tilde g^2&\left\{3le^{-10\tilde\phi}-\frac{3}{8}\tilde\Phi^{-1/2}e^{8\tilde\lambda-14\tilde\phi}(l-2\tilde\beta^2-2\tilde\theta^2)^2 + \frac{1}{2}\tilde\Phi^{1/2}e^{-6\tilde\phi}(3e^{-8\tilde\lambda}-e^{24\tilde\lambda}+6e^{8\tilde\lambda})\right.\\
&\left. -\frac{3}{2}e^{-10\tilde\phi}(e^{16\tilde\lambda}-2+e^{-16\tilde\lambda})\tilde\theta^2 - \frac{3}{2}\tilde\Phi^{5/4}e^{-10\tilde\phi-4\tilde\lambda}\tilde\rho^2 - 6\tilde\Phi^{3/4}e^{4\tilde\lambda-14\tilde\phi}\tilde\beta^2\tilde\rho^2 \right.\\
&\left.-6\tilde\Phi^{-1/2}e^{-8\tilde\lambda-14\tilde\phi}\tilde\beta^2\tilde\theta^2- \frac{1}{2}\tilde\Phi^{-1}e^{-18\tilde\phi}\tilde X^2 - \frac{1}{2}\tilde\Phi^{1/4}e^{12\tilde\lambda-18\tilde\phi}\tilde\sigma^2 \right\}\tilde\vol_4 \,,\numberthis
\end{align*}
and the topological terms are
\begin{align*}
\mc{L}^{top}_{(4)} &= 6(\tilde\theta G_{(3)}+\tilde\rho\tilde H_{(3)})\wedge\tilde C_{(1)} - d\tilde\sigma\wedge\tilde G_{(3)} + \tilde\sigma\tilde Q_{(1)}\wedge\tilde H_{(3)}-\tilde X\tilde Q_{(1)}\wedge\tilde G_{(3)} \\
&\quad +2\tilde\beta^3\tilde F_{(2)}\wedge\tilde F_{(2)} - 3\tilde\rho^2 d\tilde\theta\wedge\tilde H_{(3)} - 3\tilde\beta\tilde B_{(2)}\wedge\tilde B_{(2)} + 6\tilde\beta\tilde\rho d\tilde\beta\wedge\tilde G_{(3)} \numberthis\\
&\quad +\frac{6}{\tilde g}\tilde B_{(2)}\wedge d\tilde C_{(1)} - \frac{6}{\tilde g}\tilde B_{(2)}\wedge d\tilde\theta\wedge\tilde P_{(1)} + \frac{6}{\tilde g}d\tilde\beta\wedge\tilde F_{(2)}\wedge\tilde C_{(1)} + \frac{6}{\tilde g}\tilde\beta\tilde F_{(2)}\wedge d\tilde\theta\wedge\tilde P_{(1)} \\
&\quad +\frac{3}{2}(l-2\tilde\beta^2-2\tilde\theta^2)\left(\tilde\beta\tilde F_{(2)}\wedge\tilde F_{(2)} + \tilde\rho\tilde Q_{(1)}\wedge\tilde H_{(3)} - \tilde P_{(1)}\wedge\tilde G_{(3)} - \tilde B_{(2)}\wedge\tilde F_{(2)}\right) \,.
\end{align*}
Any solution of the equations of motion in \ref{4deomssection} can be uplifted to type IIA supergravity. This can be done by first using \eqref{ns57dmetric4d}-\eqref{ns57dscalars4d} to uplift to the $ISO(4)$ gauged theory in $D=7$, then using the uplift formulae in \cite{Cvetic:2003xr} which connect the $ISO(4)$ gauged theory and the type IIA theory. 

As we shall show, this theory is a $D=4$, $\mc{N}=2$ supergravity theory coupled to a vector multiplet and two hypermultiplets with $\mb{R}^+\times\mb{R}^+$ gauging in the hypermultiplet sector, and a scalar manifold given by the symmetric space, $SU(1,1)/U(1) \times G_{2(2)}/SO(4)$.

\subsection{Field redefinitions}

In terms of the variables in \eqref{4doriginal}, it is not obvious what the fundamental degrees of freedom of our theory are. In order to properly identify these degrees of freedom, we need to change to new variables by integrating several Bianchi identities. Firstly, consider the 2-form
\begin{equation}\label{cF4d}
\tilde{\mc{F}}_{(2)} = \tilde\Phi^{1/4}e^{-4\tilde\lambda+2\tilde\phi}\tilde\ast \tilde B_{(2)} + 2\tilde\beta\tilde B_{(2)} -\tilde\beta^2\tilde F_{(2)} \,.
\end{equation}
From \eqref{B24d} and \eqref{starB24d}, we find that $\tilde{\mc{F}}_{(2)}$ is closed, so we can write 
\begin{equation}
\tilde{\mc{F}}_{(2)} = d\tilde B_{(1)} \,.
\end{equation}
With this, we find that \eqref{C14d} becomes 
\begin{equation}
d\tilde C_{(1)} -\frac{\tilde g}{2}\tilde{\mc{F}}_{(2)}  - \frac{\tilde g}{4}\big(l-2\tilde\theta^2\big)\tilde F_{(2)} - d\tilde\theta\wedge\tilde P_{(1)} = 0 \,,
\end{equation} 
allowing us to integrate for $\tilde C_{(1)}$, 
\begin{equation}
\tilde C_{(1)} = \tilde D\tilde C + \frac{1}{2}\tilde\theta\tilde P_{(1)} - \frac{1}{2}\tilde\rho\, d\tilde\theta \,,
\end{equation} 
where $\tilde C$ is a scalar, and we defined
\begin{equation}
\tilde D\tilde C \equiv d\tilde C + \frac{\tilde g}{2}\tilde B_{(1)} + \frac{\tilde gl}{4}\tilde A_{(1)} \,.
\end{equation}
Going further, by defining
\begin{equation}
\tilde G_{(1)} = \tilde\Phi^{1/4}e^{12\tilde\lambda+12\tilde\phi}\tilde\ast\tilde G_{(3)} \,,
\end{equation}
\eqref{starG34d} can be integrated to give
\begin{equation}
\tilde G_{(1)} = d\tilde G - 6\tilde\theta\tilde C_{(1)} + 2\tilde\theta^2\tilde P_{(1)} -2 \tilde\rho\tilde\theta d\tilde\theta\,,
\end{equation}
where $\tilde G$ is a scalar. This also allows us to integrate \eqref{sigma4d} for $\tilde\sigma$, 
\begin{equation}\label{sigmaredef}
\tilde\sigma = -\tilde G - \left(\frac{3}{2}l-\tilde\theta^2-3\tilde\beta^2\right)\tilde\rho \,.
\end{equation}
Finally, by defining
\begin{equation}
\tilde H_{(1)} = \tilde\Phi^{-1}e^{12\tilde\phi}\tilde\ast\tilde H_{(3)} \,,
\end{equation}
\eqref{starH34d} can be integrated to give
\begin{equation}
\tilde H_{(1)}= d\tilde\Gamma - (\tilde G+2\tilde\theta^2\tilde\rho)\tilde Q_{(1)} - 6\tilde\rho\,\tilde C_{(1)} - 3\tilde\rho^2 d\tilde\theta\,,
\end{equation}
where $\tilde\Gamma$ is a scalar. It is also more convenient to work in the following basis of the dilatons,
\begin{equation}
\tilde\varphi_0 = -4\tilde\lambda+2\tilde\phi+\frac{1}{4}\log\tilde\Phi \,,\quad \tilde\varphi_1 = 2\sqrt{3}\tilde\lambda + 4\sqrt{3}\tilde\phi - \frac{\sqrt{3}}{8}\log\tilde\Phi \,,\quad \tilde\varphi_2 = 6\tilde\lambda+\frac{5}{8}\log\tilde\Phi \,.
\end{equation}
Collecting our results, we observe that our theory contains a graviton, ten scalar degrees of freedom, given by $\tilde\varphi_0,\tilde\varphi_1,\tilde\varphi_2,\tilde\beta,\tilde\theta,\tilde\Gamma,\tilde\Theta,\tilde\tau,\tilde C,\tilde G$,
and two vector degrees of freedom given by $\tilde A_{(1)}$ and $\tilde B_{(1)}$. For clarity, we can re-express $\tilde B_{(2)}$ in terms of the field strengths $\tilde{\mc{F}}_{(2)}$ and $\tilde F_{(2)}$ by solving \eqref{cF4d}, 
\begin{equation}
\tilde B_{(2)} = \frac{1}{4\tilde\beta^2+e^{2\tilde\varphi_0}}\left[2\tilde\beta(\tilde{\mc{F}}_{(2)}+\tilde\beta^2\tilde F_{(2)}) - e^{\tilde\varphi_0}\tilde\ast(\tilde{\mc{F}}_{(2)}+\tilde\beta^2\tilde F_{(2)})\right] \,.
\end{equation}
By introducing the axion-dilaton
\begin{equation}\label{axiondilaton4d}
\tilde z = \tilde\beta + ie^{\tilde\varphi_0} \,,
\end{equation}
we observe that 
\begin{equation}
\tilde B_{(2)} = \re\left[(\tilde z+\tilde\beta)^{-1}\right](\tilde{\mc{F}}_{(2)}+\tilde\beta^2\tilde F_{(2)})+\im\left[(\tilde z+\tilde\beta)^{-1}\right]\tilde\ast(\tilde{\mc{F}}_{(2)}+\tilde\beta^2\tilde F_{(2)}) \,.
\end{equation}
The Lagrangian describing the dynamics of our new variables is given by
\begin{equation}\label{4dus}
\mc{L}_{(4)} = \tilde R\,\tilde\vol_4 + \mc{L}^{VM}_{(4)} + \mc{L}^{HM}_{(4)} + \mc{L}^{pot}_{(4)} \,,
\end{equation}
where
\begin{equation}\label{vectorus}
\begin{split}
\mc{L}^{VM}_{(4)} &= -\frac{3}{2}\tilde\ast d\tilde\varphi_0\wedge d\tilde\varphi_0 - \frac{3}{2}e^{-2\tilde\varphi_0}\tilde\ast d\tilde\beta\wedge d\tilde\beta \\
&\quad +\frac{3}{2}\im\left[(\tilde z+\tilde\beta)^{-1}\right]\tilde\ast (\tilde{\mc{F}}_{(2)}+\tilde\beta^2\tilde F_{(2)})\wedge(\tilde{\mc{F}}_{(2)}+\tilde\beta^2\tilde F_{(2)}) \\
&\quad +\frac{3}{2}\re\left[(\tilde z+\tilde\beta)^{-1}\right](\tilde{\mc{F}}_{(2)}+\tilde\beta^2\tilde F_{(2)})\wedge(\tilde{\mc{F}}_{(2)}+\tilde\beta^2\tilde F_{(2)}) \\
&\quad -\frac{1}{2}e^{3\tilde\varphi_0}\tilde\ast\tilde F_{(2)}\wedge\tilde F_{(2)}-3\tilde\beta\tilde{\mc{F}}_{(2)}\wedge\tilde F_{(2)} - \tilde\beta^3\tilde F_{(2)}\wedge\tilde F_{(2)} \,,
\end{split}
\end{equation}
and
\begin{equation}\label{hyperus}
\begin{split}
\mc{L}^{HM}_{(4)} &= -\frac{1}{2}\tilde\ast d\tilde\varphi_1\wedge d\tilde\varphi_1 - \frac{1}{2}\tilde\ast d\tilde\varphi_2 \wedge d\tilde\varphi_2 - \frac{3}{2}e^{-\frac{1}{\sqrt{3}}\tilde\varphi_1-\tilde\varphi_2}\tilde\ast d\tilde\theta\wedge d\tilde\theta \\
&\quad- \frac{1}{2}e^{-\sqrt{3}\tilde\varphi_1+\tilde\varphi_2}\tilde \ast\tilde H_{(1)}\wedge \tilde H_{(1)}- \frac{3}{2}e^{-\frac{1}{\sqrt{3}}\tilde\varphi_1+\tilde\varphi_2}\tilde\ast\tilde P_{(1)}\wedge\tilde P_{(1)} \\
&\quad -6e^{-\frac{2}{\sqrt{3}}\tilde\varphi_1}\tilde\ast C_{(1)}\wedge C_{(1)} - \frac{1}{2}e^{-\sqrt{3}\tilde\varphi_1-\tilde\varphi_2}\tilde\ast\tilde G_{(1)}\wedge \tilde G_{(1)}  \\
&\quad - \frac{1}{2}e^{2\tilde\varphi_2}\tilde\ast\tilde Q_{(1)}\wedge\tilde Q_{(1)} \,.
\end{split}
\end{equation}
and $\mc{L}^{pot}_{(4)}$ is given by \eqref{4dpotentialterm} with $\tilde\sigma$ taking its integrated form in \eqref{sigmaredef}. 

\subsection{$D=4$, $\mc{N}=2$ supersymmetry}

In this section, we will show that our theory corresponds to the bosonic sector of $D=4$, $\mc{N}=2$ supergravity coupled to a vector multiplet and two hypermultiplets with Abelian gauging in the hypermultiplet sector and scalar manifold $SU(1,1)/U(1)\times G_{2(2)}/SO(4)$. This will be done by recasting our Lagrangian \eqref{4dus} into the canonical form (see \cite{Andrianopoli:1996vr,Andrianopoli:1996cm})
\begin{equation}
\mc{L}_{(4)} =  R\vol_4 + \mc{L}^{VM}_{(4)} + \mc{L}^{HM}_{(4)} + \mc{L}^{pot}_{(4)} \,,
\end{equation}
where
\begin{equation}\label{vectorcan}
\mc{L}^{VM}_{(4)} = -G_{z\ov z}{\ast dz}\wedge d\ov z - \im\mc{N}_{IJ}{\ast F^I_{(2)}}\wedge F^J_{(2)} - \re\mc{N}_{IJ}F^I_{(2)}\wedge F^J_{(2)} 
\end{equation}
corresponds to the ungauged vector multiplet, 
\begin{equation}\label{hypercan}
\mc{L}^{HM}_{(4)} = -2h_{uv}{\ast Dq^u}\wedge Dq^v 
\end{equation}
corresponds to the gauged hypermultiplet, and $\mc L^{pot}_{(4)} = -g^2V$, with $V$ the scalar potential. For clarity, we have removed tildes from our NS5 fields for the rest of this section.

In \eqref{vectorcan}, $z$ is a complex coordinate, and $G_{z\ov z}$ a K\"ahler metric with K\"ahler potential $K_V$ on $SU(1,1)/U(1)$, $F^I_{(2)} = dA^I_{(1)}$, $I\in\{0,1\}$ are the Abelian field strengths of the vectors in the graviton and vector multiplet, and $\mc{N}_{IJ}$ is a $z$-dependent matrix. Taking $X^I$ to be the homogeneous coordinates on $SU(1,1)/U(1)$, and $\mc{F}$ to be a prepotential, the K\"ahler potential $K_V$ and matrix $\mc{N}_{IJ}$ are given by 
\begin{equation}
K_V = -\log\left(i\ov X^I\mc{F}_I - iX^I\ov{\mc{F}}_I\right) \,,
\end{equation}
and
\begin{equation}
\mc{N}_{IJ} = \ov{\mc{F}}_{IJ} + 2i\frac{(\im\mc{F}_{IK})(\im\mc{F}_{JL})X^KX^L}{(\im\mc{F}_{AB})X^AX^B} \,,
\end{equation}
where $\mc{F}_I = \partial_I\mc{F}$ and $\mc{F}_{IJ} = \partial_I\partial_J\mc{F}$. In \eqref{hypercan}, $h_{uv}$ is the Einstein metric on $G_{2(2)}/SO(4)$ normalised such that its Ricci tensor is $-2(2+n_H)=-8$ times itself, where $n_H= 2$ is the number of hypermultiplets. The covariant derivatives of the coordinates $q^u$ are defined in terms of $n_V+1=2$ specific Killing vectors $k_I^u$ of the metric $h_{uv}$ as 
\begin{equation}
Dq^u \equiv dq^u - gk^u_IA_{(1)}^I \,,
\end{equation}
where $n_V=1$ is the number of vector multiplets. Finally, for our case, where the gauging is restricted to the hypermultiplet sector, the scalar potential is given by
\begin{equation}\label{potentialcan}
V = 8e^{K_V}X^I\ov{X}^Jh_{uv}k^u_Ik^v_J - \left([(\im\mc{N})^{-1}]^{IJ} + 8e^{K_V}X^I\ov{X}^J\right)P^x_IP^x_J \,.
\end{equation}
Here, the six scalars $P^x_I$, $x\in\{1,2,3\}$, are moment maps defined by the relation
\begin{equation}\label{momentum}
dP^x_I + \epsilon^{xyz}\omega_{(1)}^yP^z_I = 2\iota_{k_I}K_{(2)}^x \,,
\end{equation}
where $\omega_{(1)}^x$ is the $Sp(1)\cong SU(2) \subset \SU(2)\times Sp(2) = \text{Hol}(G_{2(2)}/SO(4))$ part of the spin connection associated with the quaternionic K\"ahler metric $h_{uv}$, and $K_{(2)}^x$ is its curvature. The 2-forms $K^x_{(2)}$ are related to a triplet of complex structures $\mc{J}^x$ by $(\mc{J}^x)^{\ov u}_{\ph{\ov u}\ov v} =\delta^{\ov u\ov w}K^x_{\ov w\,\ov v}$ that satisfy the quaternionic algebra
\begin{equation}
\mc{J}^x\mc{J}^y = -\delta^{xy} + \epsilon^{xyz}\mc{J}^z \,.
\end{equation}
For the vector multiplets, we introduce homogeneous coordinates $X^I = (1,z^2)$ on $SU(1,1)/U(1)$, with $z$ the axion-dilaton defined in \eqref{axiondilaton4d}. The required pre-potential is given by 
\begin{equation}
\mc{F} = \sqrt{X^0(X^1)^3} \,.
\end{equation}
Accordingly, the K\"ahler potential is
\begin{equation}
K_V = -\log\left[i(z-\ov z)^3\right] + \log 2 \,,
\end{equation}
and the corresponding metric is 
\begin{equation}
G_{z\ov z} = 2\partial_z\partial_{\ov z}K_V = -\frac{6}{(z-\ov z)^2} \,.
\end{equation}
Next, we compute that 
\begin{equation}
\mc{N}_{IJ} = \frac{1}{2(z+\beta)}\begin{pmatrix}-z^3\ov z && 3\beta z \\ 3\beta z && 3\end{pmatrix} \,.
\end{equation}
Finally, defining $A^I_{(1)} = (A_{(1)}, -B_{(1)})$, we see that the portion of our Lagrangian given in \eqref{vectorus} is equivalent to the canonical form, \eqref{vectorcan}. Interestingly, our vector multiplet sector is identical to that of the $D=4$ theory obtained in \cite{Donos:2010ax}.

We now consider the hypermultiplets. The explicit parameterisation of the coset $G_{2(2)}/SO(4)$ is given in appendix \ref{G2parameterisation}. We identify the coordinates on $G_{2(2)}/SO(4)$ to be 
\begin{equation}
q^u = (\varphi_1,\varphi_2,\tau,\theta,\Theta,C,G,\Gamma) \,,
\end{equation}
and the quaternionic K\"{a}hler metric $h_{uv}$ is given in \eqref{G2metric}. In this form, it is apparent that \eqref{hyperus} is equivalent to \eqref{hypercan}. From the definitions of the covariant derivatives, we read off 
\begin{equation}\label{killing4d}
k_0 = -\partial_\tau + \theta\partial_\Theta - \frac{l}{4}\partial_C \,,\quad k_1 = \frac{1}{2}\partial_C \,,
\end{equation}
the two requisite Killing vectors of $G_{2(2)}/SO(4)$, along which we have an Abelian $\mb R^+\times\mb{R}^+$ gauging. All that remains is to match the scalar potential, which is done at the end of appendix \ref{G2parameterisation}. 

\subsection{Half-maximal truncation and $D=4$, $\mc{N}=1$ supersymmetry}

From \eqref{halfmax}, the half-maximal truncation of our $D=4$, $\mc N=2$ theory is obtained by setting
\begin{equation}
\Theta = \tau =  C =  G =  A_{(1)} =  B_{(1)} = 0 \,.
\end{equation} 
The theory describing the remaining degrees of freedom is a $D=4$, $\mc{N}=1$ supergravity coupled to three chiral multiplets, where the six scalar fields parameterise the K\"{a}hler manifold $\left[SU(1,1)/U(1)\right]^3$. To make the $\mc{N}=1$ supersymmetry explicit, we will recast the 4-dimensional theory in the canonical form. A review of the canonical Lagrangian of generic $D=4$, $\mc{N}=1$ supergravity coupled to matter is given in \cite{Freedman:2012zz}. The main step is to perform the following field redefinitions, 
\begin{equation}
\begin{split}
&{\beta}=\text{Re}\,z_1\,,\quad {\theta}=\text{Re}\,z_2\,,\quad \Gamma=\text{Re}\,z_3\,,\\
&{\varphi_0}=\log(\im z_1)\,,\quad \varphi_1=\frac{\sqrt{3}}{2}\log\left(\im z_2\im z_3\right) \,,\quad \varphi_2 = \frac{1}{2}\log\left(\frac{(\im z_2)^3}{\im z_3}\right) \,.
\end{split}
\end{equation} 
In terms of these complex variables, the Lagrangian in \eqref{4dus} is now recast into the canonical form
\begin{equation}
\mc{L}_{(4)} = R\vol_4-G_{\alpha\bar{\beta}}{{\ast}dz^{\alpha}}\wedge d\bar{z}^{\bar{\beta}}-P\vol_4\,,
\end{equation} 
where the three complex scalar fields $z_\alpha$ form the bosonic part of the three chiral multiplets in the reduced theory, and we removed the tildes from the 4-dimensional metric for notational convenience. The K\"{a}hler metric on the homogeneous space $\left[SU(1,1)/U(1)\right]^3$ is given by
\begin{equation}
G_{\alpha\bar{\beta}}=2\partial_{\alpha}\partial_{\bar{\beta}}\mc{K}\,,
\end{equation} 
where $\mc{K}$ is the K\"{a}hler potential
\begin{equation}
\mc{K}=-\log\left[i\left(z_1-\bar{z}_1\right)^3\left(z_2-\bar{z}_2\right)^3\left(z_3-\bar{z}_3\right)\right]\,.
\end{equation} 
The scalar potential $P$ is given by
\begin{equation}\label{eq:4dN=1potential}
P=4G^{\alpha\bar{\beta}}\partial_{\alpha}\mc{W}\partial_{\bar{\beta}}\mc{W}-\frac{3}{2}\mc{W}^2\,,\quad \mc{W}=-e^{\mc{K}/2}|W|\,,
\end{equation} 
where the holomorphic superpotential $W$ is
\begin{equation}
W=2\sqrt{2}{g}z_2 \left(3l-6z_1^2-2z_2^2\right)\,.
\end{equation} 
The fact that $W$ is independent of $z_3$ is crucial for the construction of a subclass of $AdS_3\times\mb{R}$ solutions, as we will show later. An equivalent writing of the scalar potential $P$ is 
\begin{equation}
P=e^{\mc{K}}\left(G^{\alpha\bar{\beta}}\nabla_{\alpha}W\nabla_{\bar{\beta}}\overline{W}-\frac{3}{2}W\overline{W}\right)\,,
\end{equation} 
where the K\"{a}hler covariant derivative is defined as $\nabla_{\alpha}(\cdot)=\partial_{\alpha}(\cdot)+(\cdot)\partial_{\alpha}\mc{K}$.

We have shown that our truncated theory is an $\mc{N}=1$ supergravity theory in $D=4$ coupled to three chiral multiplets with scalar manifold $\left[SU(1,1)/U(1)\right]^3$. We recognise that one of the $SU(1,1)/U(1)$ factors, the one parameterised by $z_1 = \beta+ie^{\varphi_0}$, is directly inherited from the $\mc N=2$ theory. In terms of the rest of the scalar manifold, we find that the truncation to the half-maximal sector decomposes 
\begin{equation}
\frac{G_{2(2)}}{SO(4)} \longrightarrow \frac{SU(1,1)}{U(1)}\times\frac{SU(1,1)}{U(1)} \,.
\end{equation} 
Equivalently, the non-linearly realised global symmetry of the scalar manifold under the half-maximal truncation, neglecting the common factor of $SU(1,1)$, is 
\begin{equation}
G_{2(2)}\longrightarrow SU(1,1)\times SU(1,1) \,.
\end{equation}

\subsection{Further consistent sub-truncations}\label{4dsubtruncations}

For the $D=4$, $\mc{N}=1$ theory, we notice that it is consistent to further set ${\beta}= {\theta}= 0$. There are two sub-truncations of particular interest (a) setting ${\lambda}=0$ with $l=0$, and (b) setting ${\Gamma}=0$. For truncation (a), the remaining fields consist of the metric, $\phi$, $\Phi$ and $\Gamma$. The scalar potential is independent of ${\Gamma}$, and the shift symmetry, $\Gamma \to \Gamma+c$ allows the construction of a class of supersymmetric $AdS_3\times \mb{R}$ solutions, which we will discuss in detail in the next section. For truncation (b), the remaining fields consist of the metric, $\phi$, $\lambda$ and $\Phi$. The $l=\pm1$ cases cover the solutions in \cite{Gauntlett:2001ur}, whereas the $l=0$ case covers the NS5-brane near-horizon, linear dilaton solution in $D=10$.

\subsection{Some solutions}

We consider a particular type of ansatz for the half-maximal truncation as in \cite{Bak:2003jk} describing warped $AdS_3\times \mb{R}$, which in canonical variables is given by
\begin{equation}
d{s}_4^2=e^{2A(r)}ds^2(AdS_3) + dr^2\,,\quad z^\alpha=z^\alpha(r) \,,
\end{equation}
where $ds^2(AdS_3)$ is the metric on $AdS_3$ of radius $L$. Applying the Killing spinor equations given in \cite{Arav:2018njv} to our ansatz, we obtain the following set of BPS equations
\begin{equation}
\begin{split}\label{4dBPSequationJanus}
\partial_rA+\frac{i\kappa}{L}e^{-A}&=\frac{1}{2}e^{i\omega}e^{\mc{K}/2}W\,,\\
\partial_r z^{\alpha}&=-e^{-i\omega}e^{\mc{K}/2}G^{\alpha\bar{\beta}}\nabla_{\bar{\beta}}\overline{W}\,,\\
\partial_r \omega+3\mc{C}_r&=-\frac{1}{2}e^{\mc{K}/2}\text{Im}(We^{i\omega})\,,
\end{split}
\end{equation} 
where $\kappa=\pm 1$ denotes the chirality condition, $\mc{C}_r=\frac{i}{6}\left(\partial_{\alpha}\mc{K}\partial_rz^\alpha-\partial_{\bar{\alpha}}\mc{K}\partial_r\bar{z}^{\bar{\alpha}}\right)$ is the K\"{a}hler connection, and $\omega=\omega(r)$ is a phase that appears in the expression for the Killing spinors. As discussed in Section \ref{4dsubtruncations}, we can further truncate the $D=4$ theory by setting $\text{Re}(z_1)=\text{Re}(z_2)=0$ and $\text{Im}(z_1)=\text{Im}(z_2)$ (or equivalently setting ${\beta}= {\theta}={\lambda}=0$) with $l=0$. Within this sub-truncation, we can construct analytic, supersymmetric solutions to \eqref{4dBPSequationJanus}, given by
\begin{equation}
\begin{split}
z_1&=z_2=ic_0\,,\quad z_3=  g \cos\omega (2\alpha_0 r -  gr^2\sin\omega ) + i(\alpha_0- g r \sin\omega)^2\,,\\
e^{A} &= \frac{|\alpha_0- g r \sin\omega|}{w_0 g L} \,,\quad \cos\omega = w_0\kappa\,,
\end{split}
\end{equation} 
where $w_0 \in (0,1]$, and $c_0$ and $\beta_0$ are real constants. When $w_0=1$, this simplifies to the following supersymmetric $AdS_3\times \mb{R}$, linear axion solution
\begin{equation}
z_1=z_2=ic_0\,,\quad z_3=2\kappa{g}\alpha_0 r+i\alpha_0^2\,,\quad e^{A}=\frac{\alpha_0}{{g}L}\,,\quad \cos\omega=\kappa\,,
\end{equation}
which is equivalent to ${\beta}={\theta}={\lambda}=0$, $e^{5{\phi}}=c_0\alpha_0$, ${\Phi}^{5/8}=c_0^{3/2}\alpha_0^{-1}$, ${\Gamma}=2\kappa{g}\alpha_0 r$. Using the uplift formulae in \cite{Cvetic:2000dm}, and taking $\Sigma_3=\mb{R}^3$ for convenience, we find that this linear axion solution in type IIA supergravity is given by 
\begin{equation}
\begin{split}
&d\hat s^2_{10} = \gamma^2\left(\frac{1}{L^2}ds^2(AdS_3) + ds^2(S^3)\right) + ds^2(\mb{R}^4) \,,\quad e^{2\hat\phi} = c_0^3\alpha_0^{-2} \,,\\
&\hat F_{(3)} = \frac{2\kappa}{ g^2L^3}\vol(AdS_3) - \frac{2}{ g^2}\vol(S^3) \,,
\end{split}
\end{equation} 
where $\gamma^2 = \alpha_0^{1/2}c_0^{-3/4} g^{-2}$, and $S^3$ is the unit 3-sphere. We can interpret this as the dyonic, Freund-Rubin $AdS_3\times S^3$ solution of a $D=6$ supergravity theory obtained by reducing the IIA theory on $\mb{R}^4$. In the 6-dimensional picture, this solution corresponds to the near horizon of the (anti)-self-dual string \cite{Duff:1993ye}. From this, it is clear that the solution preserves half of the supersymmetries.

For completeness, we also report the domain wall solution for $l=0$,
\begin{equation}
e^{5\phi} =  g r \,,\quad ds^2_4 = r^2 ds^2(\mb{R}^{1,2}) + dr^2\,,
\end{equation}
where $\beta = \theta=\Gamma =\lambda=0$ and $\log\Phi =-8\phi$. This uplifts to the NS5-brane near horizon, linear dilaton solution in $D=10$.


\section{Final comments}\label{conclusion}

We have constructed two new consistent Kaluza-Klein truncations of $D=10$ type \Romannum{2}A supergravity on (\romannum{1}) $\Sigma_2\times S^3$, where $\Sigma_2=S^2$, $\mb{R}^2$, $\mb{H}^2$ or a quotient thereof, and (\romannum{2}) $\Sigma_3\times S^3$, where $\Sigma_3=S^3$, $\mb{R}^3$, $\mb{H}^3$ or a quotient thereof, at the level of the bosonic fields. Instead of directly reducing the 10-dimensional theory on the corresponding cycles to obtain our new truncations, we show that they can be carried out starting from the $D=5$ and $D=4$ theories obtained from wrapped M5-brane truncations on the appropriate cycles using a singular limit procedure, known as the In\"{o}n\"{u}-Wigner contraction. The two new theories obtained in sections \ref{KK_truncations_Riemann} and \ref{KK_truncations_Slag} can be viewed as direct \say{cousins} of the 5 and 4-dimensional theories corresponding to the truncations associated with M5-branes wrapping $\Sigma_2$ \cite{Cheung:2019pge,Cassani:2019vcl} and $\Sigma_3$ \cite{Donos:2010ax}, in the sense that they have the same amount of supersymmetry and field content, but as we have shown, the precise details of the gauging and the vacuum structures of the theories are entirely different. 

There are more examples of wrapped NS5-brane truncations that can be obtained using our method. From the catalogue of wrapped M5-brane solutions given in \cite{Gauntlett:2003di}, we observe that it is possible to obtain wrapped NS5-brane truncations on: (1) $\Sigma_2\times\Sigma_2'$, a product of two Riemann surfaces inside two $CY_2$ spaces \footnote{The wrapped M5-brane truncation on a product of two Riemann surfaces was obtained in \cite{Karndumri:2015sia}.}; (2) $\Sigma_2\times\Sigma_3$ with $\Sigma_2$ and $\Sigma_3$ a Riemann surface and a Slag 3-cycle inside a $CY_2$ and $CY_3$ respectively; (3) a K\"{a}hler 4-cycle in a $CY_3$. IW contractions are of course not limited to just the $SO(5)$ and $ISO(4)$ gauged supergravities in $D=7$. For example, \cite{Hull:1984yy, Hull:1984vg,Hull:1984qz} obtained the variants of the $SO(8)$ gauged $\mc{N}=8$ supergravity theories in $D=4$ with gauge groups $ISO(7)$, interpreted as the IW contraction of the original $SO(8)$ gauge group, as well as $SO(p,q)$ $(p+q = 8)$ and its IW contraction about $SO(p)$. It is well-known that the $SO(8)$ gauged theory can be obtained by a consistent truncation of M-theory on $S^7$, as demonstrated in \cite{deWit:1986oxb}. By interpreting the $S^7$ as the internal 7-sphere of a stack of M2-branes, the $SO(8)$ gauged theory can be seen as the natural arena for wrapped M2-brane truncations. As such, the existence of the contracted $ISO(7)$ gauged theory suggests that the IW contractions can be used to relate wrapped D2-brane truncations from the corresponding wrapped M2-brane truncations in a similar fashion to the relation between wrapped NS5-brane and M5-brane truncations. These correspondences between the M2 and D2 truncations are purely within M-theory and its type \Romannum{2}A descendent, but can be seen to be related to the $S^6$ truncation of massive \Romannum{2}A, which yields the dyonic $ISO(7)$ gauged supergravity in $D=4$ \cite{Guarino:2015vca}. By setting the Romans mass to zero, the dyonic theory becomes the electric $ISO(7)$ theory in \cite{Hull:1984yy}.

The consistency of the IW contraction procedure also opens up the question of which lower-dimensional gauged supergravity theories are related via the IW contraction (or potentially other such procedures), and whether such relations are actually contingent on there being a higher-dimensional picture, as there is in our case. This can perhaps be answered more systematically using the language of generalised geometry along the lines discussed in \cite{Cassani:2019vcl,Cassani:2020cod,Hohm:2014qga,Malek:2020jsa}, which we would like to address in the future.

\subsection*{Acknowledgments}

We thank Jerome Gauntlett, Matthew Roberts, Christopher Rosen, Kellogg Stelle, David Tennyson and Daniel Waldram for discussions and useful comments on the draft. KCMC is supported by an Imperial College President's PhD Scholarship. 

\begin{appendix}

\section{Equations of motion for $D=7$ maximal $ISO(4)$ gauged supergravity}\label{7dequations}

In this section, we apply the IW contractions to obtain the equations of motion of maximal $ISO(4)$ gauged supergravity in $D=7$ from those of the maximal $SO(5)$ gauged theory. For clarity, we will write down the $SO(5)$ equations of motion. It is also more convenient to work with the scalar matrix $M_{AB} \equiv \tilde\Phi^{1/4}\tilde T_{AB}$ instead of $\tilde T_{AB}$. We note that $M_{AB}$ is not independent of $\tilde\Phi$, as $\det M = \tilde\Phi$. We begin with the $S^j_{(3)}$ equations,
\begin{equation}
D(T_{ij}{\ast S^j_{(3)})} = F^{ij}_{(2)}\wedge S^j_{(3)} \,.\label{eq1}
\end{equation}
Using the notation defined in \eqref{7dym}-\eqref{7dcombinations}, we find that \eqref{eq1} yields the following two equations of motion:
\begin{equation}
\tilde D\left(M_{AB}{\tilde{\ast} \tilde G^B_{(3)}}\right) = \tilde F^{AB}_{(2)} \wedge\tilde G^B_{(3)} - \tilde G^A_{(2)}\wedge \tilde S_{(3)} \,,
\end{equation} 
and 
\begin{equation}
d\left(\tilde\Phi^{-1}{\tilde{\ast} \tilde S_{(3)}}\right) = M_{AB}{\tilde{\ast}\tilde G^A_{(3)}}\wedge\tilde G^B_{(1)} + \tilde G^A_{(2)}\wedge\tilde G^A_{(3)} \,.
\end{equation} 
Next, we consider the non-Abelian Bianchi identities 
\begin{equation}
DS^i_{(3)} = gT_{ij}{\ast S^j_{(3)}} + \frac{1}{8}\epsilon_{ij_1\cdots j_4}F^{j_1j_2}_{(2)} \wedge F^{j_3j_4}_{(2)} \,.
\end{equation} 
These yield 
\begin{equation}\label{7dbianchi}
d\tilde S_{(3)} = \frac{1}{8}\epsilon_{ABCD}\tilde F^{AB}_{(2)} \wedge \tilde F^{CD}_{(2)}\,,
\end{equation}
and 
\begin{equation}
\tilde D\tilde G^A_{(3)} = \tilde g M_{AB}{\tilde{\ast}\tilde G^B_{(3)}} - \frac{1}{2}\epsilon_{ABCD}\tilde G^B_{(2)}\wedge \tilde F^{CD}_{(2)} - \tilde G^A_{(1)}\wedge\tilde S_{(3)} \,.
\end{equation} 
Following this, we consider the Yang-Mills equations 
\begin{equation}
\begin{split}
D\left(T^{-1}_{ik}T^{-1}_{jl}{\ast F^{ij}_{(2)}}\right) &= -2gT^{-1}_{i[k}{\ast DT_{l]i}} - \frac{1}{2}\epsilon_{i_1i_2i_3kl}T_{i_3j}F^{i_1i_2}_{(2)}\wedge{\ast S^j_{(3)}}- S^k_{(3)}\wedge S^l_{(3)} \,.
\end{split}
\end{equation}
The $(k,l)=(5,5)$ component gives a $0=0$ identity, the $(k,l)=(A,5)$ components give 
\begin{equation}
\tilde D\left(\tilde\Phi M^{-1}_{AB}{\tilde{\ast}\tilde G^B_{(2)}}\right) = \tilde g \tilde\Phi M_{AB}{\tilde{\ast} \tilde G^B_{(1)}}- \tilde S_{(3)}\wedge\tilde G^A_{(3)}  - \frac{1}{2}\epsilon_{AB_1B_2B_3}M_{B_3C}\tilde F^{B_1B_2}_{(2)}\wedge{\tilde{\ast} \tilde G^C_{(3)}} \,,
\end{equation} 
and the $(k,l)=(A,B)$ components give 
\begin{align*}
\tilde D\left(M^{-1}_{AC}M^{-1}_{BD}{\tilde{\ast}\tilde F^{CD}_{(2)}}\right) &= -2\tilde g M^{-1}_{C[A}{\tilde{\ast} DM_{B]C}} + \tilde\Phi M^{-1}_{AC}\tilde G^B_{(1)}\wedge{\tilde{\ast}\tilde G^C_{(2)}}-\tilde\Phi M^{-1}_{BC}\tilde G^A_{(1)}\wedge{\tilde{\ast}\tilde G^C_{(2)}} \\
&\ph{=} -\frac{1}{2}\epsilon_{ABCD}\left(\tilde\Phi^{-1}\tilde F^{CD}_{(2)}\wedge{\tilde{\ast}\tilde S_{(3)}} - 2M_{DE}\tilde G^C_{(2)}\wedge{\tilde{\ast}\tilde G^E_{(3)}}\right) \\
&\ph{=}- \tilde G^A_{(3)}\wedge \tilde G^B_{(3)}\,.\numberthis
\end{align*}
We now consider the scalar equations, which are given by 
\begin{equation}
\begin{split}
D\left(T^{-1}_{ik}{\ast DT_{kj}}\right) &= 2g^2(2T_{ik}T_{kj}-T_{kk}T_{ij})\vol_7 + T^{-1}_{im}T^{-1}_{kl}{\ast F^{ml}_{(2)}}\wedge F^{kj}_{(2)}  \\
&\ph{=}+T_{jk}{\ast S^k_{(3)}}\wedge S^i_{(3)}-\frac{1}{5}\delta_{ij}Q\,,
\end{split}
\end{equation}
where
\begin{equation}
Q = 2g^2\left(2T_{ij}T_{ij} - \left(T_{ii}\right)^2\right)\vol_7 + T^{-1}_{nm}T^{-1}_{kl}{\ast F^{ml}_{(2)}}\wedge F^{kn}_{(2)} + T_{kl}{\ast S^k_{(3)}}\wedge S^l_{(3)} \,.
\end{equation} 
Defining 
\begin{equation}
\begin{split}
\tilde Q &= 2\tilde g^2\left(2M_{AB}M_{AB}-(M_{AA})^2\right)\tilde{\vol}_7 - M^{-1}_{AB}M^{-1}_{CD}{\tilde{\ast} \tilde F^{AC}_{(2)}}\wedge \tilde F^{BD}_{(2)} \\
&\ph{=} -2\tilde\Phi M^{-1}_{AB}{\tilde{\ast}\tilde G^A_{(2)}}\wedge \tilde G^B_{(2)} + M_{AB}{\tilde{\ast}\tilde G^A_{(3)}}\wedge \tilde G^B_{(3)} + \tilde\Phi^{-1}{\tilde{\ast}\tilde S_{(3)}}\wedge\tilde S_{(3)} \,,
\end{split}
\end{equation}
which is the limit of $Q$ as $k\to0$, we find that the $(5,5)$, $(A,5)$ and $(A,B)$ components of the scalar equations respectively yield 
\begin{equation}
d(\tilde\Phi^{-1}{\tilde{\ast} d\tilde\Phi}) = \tilde\Phi M_{AB}{\tilde{\ast}\tilde G^A_{(1)}}\wedge\tilde G^B_{(1)} + \tilde\Phi M^{-1}_{AB}{\tilde{\ast} \tilde G^A_{(2)}}\wedge\tilde G^B_{(2)} - \tilde\Phi^{-1}{\tilde{\ast} \tilde S_{(3)}}\wedge \tilde S_{(3)} + \frac{1}{5}\tilde Q \,,
\end{equation}
\begin{equation}
\tilde D\left(\tilde\Phi M_{AB}\tilde{\ast}\tilde G^B_{(1)}\right) = \tilde\Phi M^{-1}_{BC}{\tilde{\ast}\tilde G^C_{(2)}}\wedge\tilde F^{AB}_{(2)} - M_{AB}{\tilde{\ast}\tilde G^B_{(3)}}\wedge\tilde S_{(3)} \,,
\end{equation} 
and 
\begin{align*}
\tilde D\left(M^{-1}_{AC}{\tilde{\ast}\tilde DM_{CB}}\right) &= 2\tilde g^2\left(2M_{AC}M_{CB}- M_{CC}M_{AB}\right)\tilde{\vol}_7 + M^{-1}_{AC}M^{-1}_{DE}{\tilde{\ast}\tilde F^{CE}_{(2)}}\wedge \tilde{F}^{DB}_{(2)} \\
&\ph{=} +\tilde\Phi M_{BC}{\tilde{\ast}\tilde G^C_{(1)}}\wedge \tilde G^A_{(1)} - \tilde\Phi M^{-1}_{AC}{\tilde{\ast} \tilde G^C_{(2)}}\wedge \tilde G^B_{(2)} + M_{BC}{\tilde{\ast}\tilde G^C_{(3)}}\wedge\tilde G^A_{(3)} \\
&\ph{=}- \frac{1}{5}\delta_{AB}\tilde Q\,.\numberthis
\end{align*}
Finally, we consider the Einstein equations 
\begin{equation}
R^{(7)}_{\mu\nu} = \frac{1}{4}T^{-1}_{ij}T^{-1}_{kl}D_\mu T_{jk}D_\nu T_{li} + \frac{1}{4}T^{-1}_{ik}T^{-1}_{jl}F^{ij}_{\mu\rho}F^{kl\rho}_\nu +\frac{1}{4}T_{ij}S^i_{\mu\rho_1\rho_2}S^{j\rho_1\rho_2}_{\nu} + \frac{1}{10}g_{\mu\nu}X\,,
\end{equation}
where 
\begin{equation}
X = -\frac{1}{4}T^{-1}_{ik}T^{-1}_{jl}F^{ij}_{\rho_1\rho_2}F^{kl\rho_1\rho_2}-\frac{1}{3}T_{ij}S^i_{\rho_1\rho_2\rho_3}S^{j\rho_1\rho_2\rho_3} + g^2\left(2T_{ij}T_{ij}-(T_{ii})^2\right) \,.
\end{equation} 
Defining 
\begin{equation}
\begin{split}
\tilde X &= -\frac{1}{4}M^{-1}_{AB}M^{-1}_{CD}\tilde F^{AC}_{\rho_1\rho_2}\tilde F^{BD\rho_1\rho_2}-\frac{1}{2}\tilde\Phi M^{-1}_{AB}\tilde G^A_{\rho_1\rho_2}\tilde G^{B\rho_1\rho_2}-\frac{1}{3}M_{AB}\tilde G^A_{\rho_1\rho_2\rho_3}\tilde G^{B\rho_1\rho_2\rho_3}\\
&\ph{=} -\frac{1}{3}\tilde\Phi\tilde S_{\rho_1\rho_2\rho_3}\tilde S^{\rho_1\rho_2\rho_3} + \tilde g^2\left(2M_{AB}M_{AB}-(M_{AA})^2\right)\,,
\end{split}
\end{equation}
which is the limit of $X$ as $k\to0$, we find, after some algebra, that 
\begin{equation}
\begin{split}
\tilde{R}^{(7)}_{\mu\nu} &= \frac{1}{4}M^{-1}_{AB}M^{-1}_{CD}\tilde D_\mu M_{BC}\tilde D_\nu M_{AD} + \frac{1}{4}\tilde\Phi^{-2}\tilde{\nabla}_\mu\tilde\Phi\tilde{\nabla}_\nu\tilde\Phi + \frac{1}{2}\tilde\Phi M_{AB}\tilde G^A_\mu\tilde G^B_\nu\\
&\ph{=}+ \frac{1}{4}M^{-1}_{AC}M^{-1}_{BD}\tilde F^{AB}_{\mu\rho}\tilde F^{CD\rho}_\nu +\frac{1}{2}\tilde\Phi M^{-1}_{AB}\tilde G^A_{\mu\rho}\tilde G^{B\rho}_\nu + \frac{1}{4}\tilde\Phi^{-1}\tilde S_{\mu\rho_1\rho_2}\tilde S_\nu^{\ph{\nu}\rho_1\rho_2} \\
&\ph{=}+ \frac{1}{4}M_{AB}\tilde G^A_{\mu\rho_1\rho_2}\tilde G^{B\rho_1\rho_2}_\nu + \frac{1}{10}\tilde{g}_{\mu\nu}\tilde X \,.
\end{split}
\end{equation}

\section{Equations of motion for wrapped NS5-brane truncations}

\subsection{$D=5$ Equations of motion}\label{5deomssection}

The equations of motion of the M5-brane $D=5$ theory in \cite{Cheung:2019pge} can be found in appendix B.1 of this paper. We are going to plug in our truncation ansatz \eqref{5d_identification} and set $k\rightarrow 0$ to obtain a new set of equations of motion. Using our definitions in \eqref{5ddefs}, equation (B.7) of \cite{Cheung:2019pge} yield the following three equations,
\begin{equation}
\begin{split}
&\tilde D\tilde K^a_{(2)} - \tilde g\tilde{R}^a\tilde H_{(3)} -\tilde g \tilde\psi^{a\alpha}(\tilde{\mc{T}}^{-1}\tilde G_{(3)})^\alpha \\
&=-\tilde g \tilde\Phi^{1/4}e^{-6\tilde\lambda-2\tilde\phi}\epsilon^{ab}\tilde\ast\tilde K^b_{(2)} + \epsilon^{ab}\tilde P^b_{(1)}\wedge\tilde{\mc{F}}_{(2)} + \epsilon^{ab}\epsilon^{\alpha\beta}\tilde D\tilde\psi^{b\alpha}\wedge\tilde J^\beta_{(2)} \,,
\end{split}
\end{equation}
\begin{equation}
\tilde D\left((\tilde{\mc{T}}^{-1}\tilde G_{(3)})^\alpha\right) = -\tilde Q^\alpha_{(1)}\wedge\tilde H_{(3)} +\epsilon_{\alpha\beta}\tilde J^\beta_{(2)}\wedge\tilde F_{(2)} + \tilde g \tilde\Phi^{1/4}e^{6\tilde\lambda-12\tilde\phi}\tilde\ast\tilde\sigma^\alpha_{(1)} \,,
\end{equation}
and
\begin{equation}\label{5dH3bianchi}
d\tilde H_{(3)} = \tilde{\mc{F}}_{(2)}\wedge \tilde F_{(2)} \,.
\end{equation} 
Next, the equations in (B.8) give 
\begin{equation}\label{tsigma1eqn}
\begin{split}
\tilde D\left((\tilde{\mc{T}}^{-1}\tilde\sigma_{(1)})^\alpha\right) &= -\tilde Q^\alpha_{(1)}\wedge\tilde X_{(1)}-2\tilde{g}\epsilon^{ab}\tilde{\psi}^{a\alpha}\tilde{K}^b_{(2)} +\tilde g\tilde\Phi^{1/4}e^{6\tilde\lambda+8\tilde\phi}\tilde\ast\tilde G^\alpha_{(3)}  \\
&\ph{=}+ 2\epsilon_{\alpha\beta}\left(\tilde D\tilde\psi^{a\beta}\wedge\tilde P^a_{(1)} + \frac{1}{2}\tilde g (l-\tilde\psi^2)\tilde J^\beta_{(2)} +g \epsilon_{ab}\tilde\psi^{a\beta}\tilde{R}^b\tilde F_{(2)}\right)\,,
\end{split}
\end{equation}
and 
\begin{equation}\label{5dXbianchi}
d\tilde X_{(1)} = \epsilon_{\alpha\beta}\left(\tilde D\tilde\psi^{a\alpha}\wedge \tilde D\tilde\psi^{a\beta} + \frac{1}{2}\tilde{g}\epsilon^{\alpha\beta}(l-\tilde\psi^2)\tilde{\mc{F}}_{(2)} + \tilde g \epsilon^{ab}\tilde\psi^{a\alpha}\tilde\psi^{b\beta}\tilde F_{(2)}\right) \,.
\end{equation} 
The equations in (B.9) give
\begin{equation}
\begin{split}\label{starKa2}
\tilde D\left(\tilde\Phi^{1/4}e^{-6\tilde\lambda-2\tilde\phi}\tilde\ast \tilde K^a_{(2)}\right) &= -\tilde F_{(2)}\wedge \tilde K^a_{(2)}  -\tilde g\tilde\Phi^{1/4} e^{6\tilde{\lambda}-12\tilde\phi}\epsilon^{ab}\tilde\psi^{b\alpha}\tilde{\ast}\tilde\sigma^\alpha_{(1)} \\
&{\ph{=}}-\epsilon^{ab}\left(D\tilde\psi^{b\alpha}\wedge(\tilde{\mc{T}}^{-1}\tilde G_{(3)})^\alpha+ \tilde P^b_{(1)}\wedge\tilde H_{(3)}\right)\,,
\end{split}
\end{equation}
\begin{equation}\label{starchialpha2}
\tilde D\left(\tilde\Phi^{1/4}e^{6\tilde\lambda-12\tilde\phi}\tilde{\ast}\tilde\sigma^\alpha_{(1)}\right)= \epsilon^{\alpha\beta}\tilde{\mc{F}}_{(2)}\wedge(\tilde{\mc{T}}^{-1}\tilde G_{(3)})^\beta + \tilde J^\alpha_{(2)}\wedge\tilde H_{(3)} \,,
\end{equation} 
and
\begin{equation}
\begin{split}
d\left(\tilde\Phi^{-1}e^{-12\tilde\phi}\tilde\ast\tilde X_{(1)}\right) &= \tilde\Phi^{1/4}e^{6\tilde\lambda-12\tilde\phi}\tilde Q^\alpha_{(1)}\wedge\tilde\ast\tilde \sigma^\alpha_{(1)} -\tilde{J}^{\alpha}_{(2)}\wedge \left(\tilde{\mc{T}}^{-1}\tilde{G}_{(3)}\right)^{\alpha} \,.
\end{split}
\end{equation}
The equations in (B.10) give
\begin{align*}\label{starh3alpha2}
\tilde D\left(\tilde{\Phi}^{1/4}e^{6\tilde{\lambda}+8\tilde\phi}\tilde\ast\tilde G^\alpha_{(3)} \right)&=\epsilon^{\alpha\beta}\tilde{\mc{F}}_{(2)}\wedge(\tilde{\mc{T}}^{-1}\tilde\sigma_{(1)})^\beta + \tilde J^\alpha_{(2)}\wedge\tilde X_{(1)} + 2\epsilon^{ab}D\tilde\psi^{a\alpha}\wedge \tilde K^b_{(2)} \numberthis \\
&{\ph{=}}+2\tilde g\epsilon^{ab}\tilde\psi^{a\alpha}\left(\tilde\psi^{b\beta}(\tilde{\mc{T}}^{-1}\tilde G_{(3)})^\beta+\tilde{R}^b\tilde H_{(3)}\right) + 2\tilde{g}\tilde\Phi^{1/4}e^{-6\tilde\lambda-2\tilde\phi}\tilde\psi^{a\alpha}\tilde\ast \tilde K^a_{(2)}\,,
\end{align*}
and
\begin{equation}\label{starH32}
\begin{split}
d\left(\tilde\Phi^{-1}e^{8\tilde\phi}\tilde\ast\tilde H_{(3)}\right) &= \tilde\Phi^{1/4}e^{6\tilde\lambda+8\tilde\phi}\tilde Q^\alpha_{(1)}\wedge\tilde\ast\tilde G^\alpha_{(3)} - \tilde J^\alpha_{(2)}\wedge(\tilde{\mc{T}}^{-1}\tilde\sigma_{(1)})^\alpha + 2\epsilon^{ab}\tilde P^a_{(1)}\wedge\tilde K^b_{(2)} \\
&\ph{=}+2\tilde g\tilde\Phi^{1/4}e^{-6\tilde\lambda-2\tilde\phi}\tilde{R}^a\tilde\ast\tilde K^a_{(2)} + 2\tilde g\epsilon^{ab}\tilde{R}^a\tilde\psi^{b\beta}(\tilde{\mc{T}}^{-1}\tilde G_{(3)})^\beta \,.
\end{split}
\end{equation}
Equation (B.11) gives
\begin{equation}
\begin{split}
&d\left(\tilde\Phi^{-1/2}e^{12\tilde\lambda+4\tilde\phi}\tilde\ast\tilde F_{(2)}\right)-2\tilde{g}\epsilon_{ab}\left(\tilde{\Phi}^{-1/2}e^{-6\tilde{\phi}}(\tilde{\mc{T}}^{-1}\tilde{\psi})^{a\alpha}{\tilde{\ast}\tilde{D}\tilde{\psi}^{b\alpha}}+\tilde{\Phi}^{1/4}e^{6\tilde{\lambda}-6\tilde{\phi}}\tilde{R}^a{\tilde{\ast}\tilde{P}^b_{(1)}}\right) \\
&+\tilde{\Phi}^{-1}e^{8\tilde{\phi}}\tilde{\mathcal{F}}_{(2)}\wedge {\tilde{\ast}\tilde{H}_{(3)}}-\tilde{\Phi}^{-1/4}e^{6\tilde{\lambda}+8\tilde{\phi}}\epsilon_{\alpha\beta}\tilde{J}^\alpha_{(2)}\wedge {\tilde{\ast}\tilde{G}^\beta_{(3)}}+\tilde{K}^a_{(2)}\wedge \tilde{K}^a_{(2)}\\
&+\tilde g\tilde{\Phi}^{-1}e^{-12\tilde{\phi}}\epsilon_{\alpha\beta}\epsilon^{ab}\tilde{\psi}^{a\alpha}\tilde{\psi}^{b\beta}{\tilde{\ast}\tilde{X}_{(1)}}+2\tilde g\tilde{\Phi}^{1/4}e^{6\tilde{\lambda}-12\tilde{\phi}}\epsilon_{\alpha\beta}\epsilon^{ab}\tilde{R}^{a}\tilde{\psi}^{b\alpha}{\tilde{\ast}\tilde{\sigma}^\beta_{(1)}}= 0 \,.
\end{split}
\end{equation}
Equation (B.12) gives
\begin{align*}
&\tilde D\left(\tilde\Phi^{3/4}e^{4\tilde\phi-6\tilde\lambda}\tilde\ast(\tilde{\mc{T}}^{-1}\tilde J_{(2)})^\alpha\right) - 2 \tilde g\tilde\Phi^{3/4}e^{6\tilde\lambda-6\tilde\phi}\tilde\psi^{a\alpha}\tilde\ast\tilde P^a_{(1)} + \tilde g\tilde\Phi^{5/4}e^{6\tilde\lambda}{\tilde{\ast}(\tilde{\mc{T}}\tilde Q_{(1)})^\alpha} \\
&-\epsilon_{\alpha\beta}\tilde\Phi^{1/4}\left[\tilde g(l-\tilde\psi^2)e^{6\tilde\lambda-12\tilde\phi}\tilde\ast\tilde\sigma^\beta_{(1)} + e^{6\tilde\lambda+8\tilde\phi}\tilde F_{(2)}\wedge\ast\tilde G^\beta_{(3)} - 2e^{-6\tilde\lambda-2\tilde\phi}\epsilon^{ab}\tilde D\tilde\psi^{a\beta}\wedge\tilde\ast \tilde{K}^b_{(2)}\right] \\
&+(\tilde{\mc{T}}^{-1}\tilde G_{(3)})^\alpha\wedge\tilde X_{(1)} - \tilde H_{(3)}\wedge(\tilde{\mc{T}}^{-1}\tilde\sigma_{(1)})^\alpha = 0 \,,\numberthis
\end{align*}
and
\begin{align*}
&d\left(\tilde\Phi^{-1/2}e^{4\tilde\phi-12\tilde\lambda}\tilde\ast\tilde{\mc{F}}_{(2)}\right) - \tilde\Phi^{3/4}e^{4\tilde\phi-6\tilde\lambda}\epsilon_{\alpha\beta}\tilde Q^\alpha_{(1)}\wedge{\tilde\ast (\tilde{\mc{T}}^{-1}\tilde J_{(2)})^\beta} - 2\tilde g\epsilon_{\alpha\beta}\tilde\Phi^{-1/2}e^{-6\tilde\phi}\tilde\psi^{a\alpha}\tilde{\mc{T}}^{-1}_{\beta\gamma}\tilde\ast\tilde D\tilde\psi^{a\gamma} \\
&+\tilde g\epsilon_{\alpha\beta}\tilde{\mc{T}}^{-1}_{\gamma\alpha}\tilde\ast\tilde D\tilde{\mc{T}}_{\beta\gamma} + \tilde g(l-\tilde\psi^2)\tilde\Phi^{-1}e^{-12\tilde\phi}\tilde\ast\tilde X_{(1)} + \tilde\Phi^{-1}e^{8\tilde\phi}\tilde F_{(2)}\wedge\tilde\ast\tilde H_{(3)} \\
&- 2\tilde\Phi^{1/4}e^{-6\tilde\lambda-2\tilde\phi}\epsilon^{ab}\tilde P^a_{(1)}\wedge\ast\tilde K^b_{(2)} + \epsilon_{\alpha\beta}(\tilde{\mc{T}}^{-1}\tilde G_{(3)})^\alpha\wedge(\tilde{\mc{T}}^{-1}\tilde\sigma_{(1)})^\beta = 0 \,.\numberthis
\end{align*}
Equation (B.13) gives
\begin{equation}
\begin{split}
&\tilde D\left(\tilde\Phi^{3/4}e^{6\tilde\lambda-6\tilde\phi}\tilde\ast\tilde P^a_{(1)}\right) - \tilde g^2\left[2\tilde\Phi^{3/4}e^{-6\tilde\lambda-16\tilde\phi}\epsilon^{ab}\epsilon^{cd}\tilde R^c(\tilde\psi^b\tilde{\mc{T}}^{-1}\tilde\psi^d) + \tilde\Phi^{5/4}e^{-6\tilde\lambda-10\tilde\phi}\tilde R^a\right]\tilde\vol_5 \\
&+\tilde\Phi^{1/4}e^{-6\tilde\lambda-2\tilde\phi}\tilde{\mc{F}}_{(2)}\wedge\epsilon^{ab}\tilde\ast\tilde K^b_{(2)} + \tilde\Phi^{1/4}e^{6\tilde\lambda-12\tilde\phi}\epsilon_{\beta\gamma}\tilde\ast\tilde\sigma^\beta_{(1)}\wedge\tilde D\tilde\psi^{a\gamma} + \tilde H_{(3)}\wedge\epsilon^{ab}\tilde K^b_{(2)} = 0 \,,
\end{split}
\end{equation}
and
\begin{align*}
&\tilde D\left(\tilde\Phi^{-1/2}e^{-6\tilde\phi}\tilde{\mc{T}}^{-1}_{\alpha\beta}\tilde\ast\tilde D\tilde\psi^{a\beta}\right) + \tilde{\Phi}^{3/4}e^{6\tilde{\lambda}-6\tilde{\phi}}\tilde Q_{(1)}^\alpha\wedge\tilde\ast\tilde P^a_{(1)} \\
&- \tilde g^2\left\{2\tilde\Phi^{-1/2}e^{-12\lambda-16\tilde\phi}\epsilon^{ab}\epsilon^{cd}(\tilde\psi^b\tilde{\mc{T}}^{-1}\tilde\psi^d)(\tilde{\mc{T}}^{-1}\tilde\psi)^{c\alpha}- \tilde\Phi^{-1/2}e^{12\tilde\lambda-16\tilde\phi}(l-\tilde\psi^2)\tilde\psi^{a\alpha}\right. \\
&\left. + e^{-10\tilde\phi}\left(e^{12\tilde\lambda}(\tilde{\mc{T}}\tilde\psi)^{a\alpha}-2\tilde\psi^{a\alpha} + e^{-12\tilde\lambda}(\tilde{\mc{T}}^{-1}\tilde\psi)^{a\alpha}\right)+ 2\tilde\Phi^{3/4}e^{-6\tilde\lambda-16\tilde\phi}\epsilon^{ab}\epsilon^{cd}\tilde R^b\tilde R^d(\tilde{\mc{T}}^{-1}\tilde\psi)^{c\alpha}\right\}\tilde\vol_5 \\
&- \epsilon_{\alpha\beta}\left[\tilde\Phi^{-1}e^{-12\tilde\phi}\tilde\ast X_{(1)}\wedge\tilde D\tilde\psi^{a\beta}-\tilde\Phi^{1/4}e^{6\tilde\lambda-12\tilde\phi}\tilde\ast\tilde\sigma^\beta_{(1)}\wedge\tilde P^a_{(1)} - \tilde\Phi^{1/4}e^{-6\tilde\lambda-2\tilde\phi}\tilde J^\beta_{(2)}\wedge\epsilon^{ab}\tilde\ast K^b_{(2)}\right] \\
&+ (\tilde{\mc{T}}^{-1}\tilde G_{(3)})^\alpha\wedge\epsilon^{ab}\tilde K^b_{(2)} = 0\,.\numberthis 
\end{align*}
Equations (B.14)-(B.15) give
\begin{align*}
&d\tilde{\ast}d\tilde{\lambda} -\frac{1}{24}\tilde\Phi^{5/4}e^{6\tilde\lambda}\tilde{\mc{T}}_{\alpha\beta}\tilde\ast\tilde Q_{(1)}^\alpha\wedge\tilde Q^\beta_{(1)} -\frac{1}{12}\tilde\Phi^{3/4}e^{6\tilde{\lambda}-6\tilde{\phi}}{\tilde{\ast}\tilde{P}^a_{(1)}\wedge\tilde{P}^a_{(1)}}-\frac{1}{24}\tilde\Phi^{1/4}e^{6\tilde{\lambda}-12\tilde{\phi}}\tilde{\mc{T}}^{-1}_{\alpha\beta}\tilde{\ast}\tilde{\sigma}^\alpha_{(1)}\wedge\tilde{\sigma}^\beta_{(1)}\\
&-\frac{1}{12}\tilde\Phi^{-1/2}e^{4\tilde{\phi}+12\tilde{\lambda}}\tilde{\ast}\tilde{F}_{(2)}\wedge \tilde{F}_{(2)}+\frac{1}{12}\tilde\Phi^{-1/2}e^{4\tilde{\phi}-12\tilde{\lambda}}\tilde{\ast}\tilde{\mc{F}}_{(2)}\wedge \tilde{\mc{F}}_{(2)} +\frac{1}{24}\tilde\Phi^{3/4}e^{4\tilde{\phi}-6\tilde{\lambda}}\tilde{\mc{T}}^{-1}_{\alpha\beta}\tilde{\ast}\tilde{J}^\alpha_{(2)}\wedge \tilde{J}^\beta_{(2)}\\
&+\frac{1}{12}\tilde{\Phi}^{1/4}e^{-6\tilde{\lambda}-2\tilde{\phi}}{\tilde{\ast}\tilde{K}^a_{(2)}}\wedge\tilde{K}^a_{(2)}-\frac{1}{24}\tilde\Phi^{1/4}e^{6\tilde{\lambda}+8\tilde{\phi}}\tilde{\mc{T}}^{-1}_{\alpha\beta}\tilde{\ast}\tilde{G}^\alpha_{(3)}\wedge\tilde{G}^\beta_{(3)}\\
&+\tilde{g}^2\left\{\frac{1}{6}e^{-10\tilde{\phi}}\left(e^{-12\tilde{\lambda}}(\tilde{\psi}\tilde{\mc{T}}^{-1}\tilde{\psi})-e^{12\tilde{\lambda}}(\tilde{\psi}\tilde{\mc{T}}\tilde{\psi})+\frac{1}{2}\tilde{\Phi}^{5/4}e^{-6\tilde{\lambda}}\tilde{R}^2\right)\right.\\
&\left.\phantom{+\tilde{g}^2\{}-\frac{1}{12}\tilde\Phi^{-1/2}e^{12\tilde{\lambda}-16\tilde{\phi}}(l-\tilde{\psi}^2)^2-\frac{1}{12}\tilde\Phi^{1/2}e^{12\tilde{\lambda}-4\tilde{\phi}}\left(2\text{Tr}(\tilde{\mc{T}}^2)-(\text{Tr}\tilde{\mc{T}})^2\right)\right.\\
&\left.\phantom{+\tilde{g}^2\{}+\frac{1}{6}\tilde\Phi^{-1/2}e^{-12\tilde{\lambda}-16\tilde{\phi}}\epsilon^{ab}\epsilon^{cd}(\tilde{\psi}^{a}\tilde{\mc{T}}^{-1}\tilde{\psi}^c)(\tilde{\psi}^{b}\tilde{\mc{T}}^{-1}\tilde{\psi}^d)\right.\\
&\left.\phantom{+\tilde{g}^2\{}+\frac{1}{6}\tilde\Phi^{3/4}e^{-6\tilde{\lambda}-16\tilde{\phi}}\epsilon^{ab}\epsilon^{cd}(\tilde{\psi}^{a}\tilde{\mc{T}}^{-1}\tilde{\psi}^c)\tilde{R}^b\tilde{R}^d\right\}\tilde{\text{vol}}_5=0 \,.\numberthis
\end{align*}
and
\begin{align*}
&d(\tilde\Phi^{-1}{\tilde{\ast}d\tilde{\Phi}})-\tilde\Phi^{5/4}e^{6\tilde\lambda}\tilde{\mc{T}}_{\alpha\beta}\tilde\ast\tilde Q_{(1)}^\alpha\wedge\tilde Q^\beta_{(1)} -\frac{6}{5}\tilde\Phi^{3/4}e^{6\tilde{\lambda}-6\tilde{\phi}}{\tilde{\ast}\tilde{P}^a_{(1)}\wedge\tilde{P}^a_{(1)}}-\frac{1}{5}\tilde\Phi^{1/4}e^{6\tilde{\lambda}-12\tilde{\phi}}\tilde{\mc{T}}^{-1}_{\alpha\beta}\tilde{\ast}\tilde{\sigma}^\alpha_{(1)}\wedge\tilde{\sigma}^\beta_{(1)}\\
&+\frac{4}{5}\tilde\Phi^{-1/2}e^{-6\tilde{\phi}}\tilde{\mc{T}}^{-1}_{\alpha\beta}\tilde{\ast}D\tilde{\psi}^{a\alpha}\wedge D\tilde{\psi}^{a\beta} + \frac{4}{5}\tilde\Phi^{-1}e^{-12\tilde\phi}\tilde\ast\tilde X_{(1)}\wedge\tilde X_{(1)} -\frac{3}{5}\tilde\Phi^{3/4}e^{-6\tilde\lambda+4\tilde\phi}\tilde{\mc{T}}^{-1}_{\alpha\beta}\tilde\ast \tilde J^\alpha_{(2)}\wedge\tilde J^\beta_{(2)}\\
&+\frac{2}{5}\tilde\Phi^{-1/2}e^{4\tilde{\phi}+12\tilde{\lambda}}\tilde{\ast}\tilde{F}_{(2)}\wedge \tilde{F}_{(2)}+\frac{2}{5}\tilde\Phi^{-1/2}e^{4\tilde{\phi}-12\tilde{\lambda}}\tilde{\ast}\tilde{\mc{F}}_{(2)}\wedge \tilde{\mc{F}}_{(2)} -\frac{2}{5}\tilde\Phi^{1/4}e^{-6\tilde\lambda-2\tilde\phi}\tilde\ast\tilde K^a_{(2)}\wedge\tilde K^a_{(2)} \\
&+\frac{4}{5}\tilde\Phi^{-1}e^{8\tilde\phi}\tilde\ast\tilde H_{(3)}\wedge \tilde H_{(3)} -\frac{1}{5}\tilde\Phi^{1/4}e^{6\tilde\lambda+8\tilde\phi}\tilde{\mc{T}}^{-1}_{\alpha\beta}\tilde\ast\tilde G^\alpha_{(3)}\wedge\tilde G^\beta_{(3)}\\
&+\tilde{g}^2\left\{\frac{2}{5}\tilde\Phi^{-1/2}e^{12\tilde{\lambda}-16\tilde{\phi}}(l-\tilde{\psi}^2)^2-\frac{2}{5}\tilde\Phi^{1/2}e^{-4\tilde{\phi}}\left(2e^{12\tilde{\lambda}}\text{Tr}(\tilde{\mc{T}}^2)-e^{12\tilde{\lambda}}(\text{Tr}\tilde{\mc{T}})^2-4\text{Tr}\tilde{\mc{T}}\right)\right.\\
&\left.\phantom{+\tilde{g}^2\{}-2\tilde\Phi^{5/4}e^{-6\tilde\lambda-10\tilde\phi}\tilde R^2+\frac{4}{5}\tilde\Phi^{-1/2}e^{-12\tilde{\lambda}-16\tilde{\phi}}\epsilon^{ab}\epsilon^{cd}(\tilde{\psi}^{a}\tilde{\mc{T}}^{-1}\tilde{\psi}^c)(\tilde{\psi}^{b}\tilde{\mc{T}}^{-1}\tilde{\psi}^d)\right.\\
&\left.\phantom{+\tilde{g}^2\{}-\frac{12}{5}\tilde\Phi^{3/4}e^{-6\tilde\lambda-16\tilde\phi}\epsilon^{ab}\epsilon^{cd}\tilde R^b\tilde R^d(\tilde\psi^a\tilde{\mc{T}}^{-1}\tilde \psi^c)\right\}\tilde{\text{vol}}_5=0\,,\numberthis
\end{align*}
and
\begin{align*}
&\tilde D\left(\tilde\Phi^{5/4}e^{6\tilde\lambda}\tilde\ast(\tilde{\mc{T}}\tilde Q_{(1)})^\alpha\right) - 2\tilde\Phi^{3/4}e^{6\tilde\lambda-6\tilde\phi}\tilde\ast \tilde P^a_{(1)}\wedge\tilde D\tilde\psi^{a\alpha} \\
&+\tilde\Phi^{3/4}e^{-6\tilde\lambda+4\tilde\phi}\epsilon^{\alpha\beta}\tilde\ast(\tilde{\mc{T}}^{-1}\tilde J_{(2)})^\beta\wedge\tilde{\mc{F}}_{(2)} + \tilde\Phi^{1/4}e^{6\tilde\lambda+8\tilde\phi}\tilde\ast\tilde G^\alpha_{(3)}\wedge\tilde H_{(3)} + \tilde\Phi^{1/4}e^{6\tilde\lambda-12\tilde\phi}\tilde\ast\tilde\sigma^\alpha_{(1)}\wedge\tilde X_{(1)} \\
&-2\tilde g^2e^{-6\tilde\lambda-10\tilde\phi}\left(\tilde\Phi^{5/4}\tilde\psi^{a\alpha}\tilde R^a -2\tilde\Phi^{3/4}e^{-6\tilde\phi}\epsilon^{ab}\epsilon^{cd}\tilde R^a(\tilde\psi^b\tilde{\mc{T}}^{-1}\tilde\psi^c)\tilde\psi^{d\alpha}\right)\tilde\vol_5 = 0 \,, \numberthis
\end{align*}
and
\begin{align*}
&\tilde{D}\left(\tilde{\mathcal{T}}^{-1}_{\alpha\gamma}{\tilde{\ast}\tilde{D}\tilde{\mathcal{T}}_{\gamma\beta}}\right)-\tilde\Phi^{5/4}e^{6\tilde{\lambda}}\Bigg(\tilde{\mc{T}}_{\beta\gamma}\delta_{\alpha\rho}-\frac{1}{2}\tilde{\mc{T}}_{\gamma\rho}\delta_{\alpha\beta}\Bigg){\tilde{\ast}\tilde{Q}_{(1)}^{\gamma}}\wedge \tilde{Q}_{(1)}^{\rho}\\
&+\tilde\Phi^{-1/2}e^{-6\tilde{\phi}}\Bigg(2\tilde{\mc{T}}^{-1}_{\alpha\gamma}\delta_{\beta\rho}-\tilde{\mc{T}}^{-1}_{\gamma\rho}\delta_{\alpha\beta}\Bigg){\tilde{\ast}\tilde{D}\tilde{\psi}^{a\gamma}}\wedge \tilde{D}\tilde{\psi}^{a\rho}-\tilde\Phi^{1/4}e^{6\tilde{\lambda}-12\tilde{\phi}}\Bigg(\tilde{\mc{T}}^{-1}_{\alpha\gamma}\delta_{\beta\rho}-\frac{1}{2}\tilde{\mc{T}}^{-1}_{\gamma\rho}\delta_{\alpha\beta}\Bigg){\tilde{\ast}\tilde{\sigma}_{(1)}^{\gamma}}\wedge \tilde{\sigma}_{(1)}^{\rho}\\
&+\tilde\Phi^{3/4}e^{-6\tilde{\lambda}+4\tilde{\phi}}\Bigg(\tilde{\mc{T}}^{-1}_{\alpha\gamma}\delta_{\beta\rho}-\frac{1}{2}\tilde{\mc{T}}^{-1}_{\gamma\rho}\delta_{\alpha\beta}\Bigg){\tilde{\ast}\tilde{J}_{(2)}^{\gamma}}\wedge \tilde{J}_{(2)}^{\rho}-\tilde\Phi^{1/4}e^{6\tilde{\lambda}+8\tilde{\phi}}\Bigg(\tilde{\mc{T}}^{-1}_{\alpha\gamma}\delta_{\beta\rho}-\frac{1}{2}\tilde{\mc{T}}^{-1}_{\gamma\rho}\delta_{\alpha\beta}\Bigg){\tilde{\ast}\tilde{G}_{(3)}^{\gamma}}\wedge \tilde{G}_{(3)}^{\rho}\\
&+\tilde{g}^2\Bigg\{e^{-10\tilde{\phi}}\Bigg[2e^{-12\tilde{\lambda}}(\tilde{\mc{T}}^{-1}\tilde{\psi})^{a\alpha}\tilde{\psi}^{a\beta}-2e^{12\tilde{\lambda}}\tilde{\psi}^{a\alpha}(\tilde{\mc{T}}\tilde{\psi})^{a\beta}-e^{-12\tilde{\lambda}}(\tilde{\psi}\tilde{\mc{T}}^{-1}\tilde{\psi})\delta_{\alpha\beta}+e^{12\tilde{\lambda}}(\tilde{\psi}\tilde{\mc{T}}\tilde{\psi})\delta_{\alpha\beta}\Bigg]\\
&\phantom{+\tilde{g}^2\Bigg\{}-2\tilde\Phi^{-1/2}e^{-12\tilde{\lambda}-16\tilde{\phi}}\Bigg[2\tilde{\mc{T}}^{-1}_{\alpha\gamma}\tilde{\mc{T}}^{-1}_{\rho\eta}\delta_{\beta\xi}-\tilde{\mc{T}}^{-1}_{\xi\gamma}\tilde{\mc{T}}^{-1}_{\rho\eta}\delta_{\alpha\beta}\Bigg](\epsilon^{ab}\tilde{\psi}^{a\gamma}\tilde{\psi}^{b\eta})(\epsilon^{cd}\tilde{\psi}^{c\rho}\tilde{\psi}^{d\xi})\\
&\phantom{+\tilde{g}^2\Bigg\{}+2\tilde\Phi^{3/4}e^{-6\tilde{\lambda}-16\tilde{\phi}}\Bigg[2\tilde{\mc{T}}^{-1}_{\alpha\gamma}\delta_{\beta\rho}-\tilde{\mc{T}}^{-1}_{\gamma\rho}\delta_{\alpha\beta}\Bigg](\epsilon^{ab}\tilde{\psi}^{a\gamma}\tilde{R}^{b})(\epsilon^{cd}\tilde{\psi}^{c\rho}\tilde{R}^{d})\\
&\phantom{+\tilde{g}^2\Bigg\{}+\tilde\Phi^{1/2}e^{-4\tilde{\phi}}\Bigg[2e^{12\tilde{\lambda}}\text{Tr}(\tilde{\mc{T}}^2)\delta_{\alpha\beta}-e^{12\tilde{\lambda}}(\text{Tr}\tilde{\mc{T}})^2\delta_{\alpha\beta}-2\text{Tr}\tilde{\mc{T}}\delta_{\alpha\beta}\\
&\phantom{+\tilde{g}^2\Bigg\{+\tilde\Phi^{1/2}e^{-4\tilde{\phi}}\Bigg[}-4e^{12\tilde{\lambda}}(\tilde{\mc{T}}^2)_{\alpha\beta}+2e^{12\tilde{\lambda}}\text{Tr}\tilde{\mc{T}}\tilde{\mc{T}}_{\alpha\beta}+4\tilde{\mc{T}}_{\alpha\beta}\Bigg]\Bigg\}\tilde{\text{vol}}_5=0\,.\numberthis
\end{align*}
Equations (B.16)-(B.17) give
\begin{align*}
&d\tilde{\ast}d\tilde{\phi}-\frac{1}{30}\tilde\Phi^{-1/2}e^{4\tilde{\phi}+12\tilde{\lambda}}\tilde{\ast}\tilde{F}_{(2)}\wedge \tilde{F}_{(2)}+\frac{1}{10}\tilde\Phi^{-1/2}e^{-6\tilde{\phi}}\tilde{\mc{T}}^{-1}_{\alpha\beta}\tilde{\ast}D\tilde{\psi}^{a\alpha}\wedge D\tilde{\psi}^{a\beta}\\
&+\frac{1}{10}\tilde\Phi^{3/4}e^{6\tilde{\lambda}-6\tilde{\phi}}{\tilde{\ast}\tilde{P}^a_{(1)}\wedge\tilde{P}^a_{(1)}}-\frac{1}{30}\tilde\Phi^{-1/2}e^{4\tilde{\phi}-12\tilde{\lambda}}\tilde{\ast}\tilde{\mc{F}}_{(2)}\wedge \tilde{\mc{F}}_{(2)}-\frac{1}{30}\tilde\Phi^{3/4}e^{4\tilde{\phi}-6\tilde{\lambda}}\tilde{\mc{T}}^{-1}_{\alpha\beta}\tilde{\ast}\tilde{J}^\alpha_{(2)}\wedge \tilde{J}^\beta_{(2)}\\
&+\frac{1}{30}\tilde{\Phi}^{1/4}e^{-6\tilde{\lambda}-2\tilde{\phi}}{\tilde{\ast}\tilde{K}^a_{(2)}}\wedge\tilde{K}^a_{(2)}+\frac{1}{10}\tilde\Phi^{-1}e^{-12\tilde{\phi}}\tilde{\ast}\tilde{X}_{(1)}\wedge\tilde{X}_{(1)}+\frac{1}{10}\tilde\Phi^{1/4}e^{6\tilde{\lambda}-12\tilde{\phi}}\tilde{\mc{T}}^{-1}_{\alpha\beta}\tilde{\ast}\tilde{\sigma}^\alpha_{(1)}\wedge\tilde{\sigma}^\beta_{(1)}\\
&-\frac{1}{15}\tilde\Phi^{-1}e^{8\tilde{\phi}}\tilde{\ast}\tilde{H}_{(3)}\wedge\tilde{H}_{(3)}-\frac{1}{15}\tilde\Phi^{1/4}e^{6\tilde{\lambda}+8\tilde{\phi}}\tilde{\mc{T}}^{-1}_{\alpha\beta}\tilde{\ast}\tilde{G}^\alpha_{(3)}\wedge\tilde{G}^\beta_{(3)}\\
&+\tilde{g}^2\left\{\frac{1}{6}e^{-10\tilde{\phi}}\left(e^{12\tilde{\lambda}}(\tilde{\psi}\tilde{\mc{T}}\tilde{\psi})-2(l+\tilde{\psi}^2)+e^{-12\tilde{\lambda}}(\tilde{\psi}\tilde{\mc{T}}^{-1}\tilde{\psi})+\tilde{\Phi}^{5/4}e^{-6\tilde{\lambda}}\tilde{R}^2\right)\right.\\
&\left.\phantom{+\tilde{g}^2\{}+\frac{2}{15}\tilde\Phi^{-1/2}e^{12\tilde{\lambda}-16\tilde{\phi}}(l-\tilde{\psi}^2)^2+\frac{1}{30}\tilde\Phi^{1/2}e^{-4\tilde{\phi}}\left(2e^{12\tilde{\lambda}}\text{Tr}(\tilde{\mc{T}}^2)-e^{12\tilde{\lambda}}(\text{Tr}\tilde{\mc{T}})^2-4\text{Tr}\tilde{\mc{T}}\right)\right.\\
&\left.\phantom{+\tilde{g}^2\{}+\frac{4}{15}\tilde\Phi^{-1/2}e^{-12\tilde{\lambda}-16\tilde{\phi}}\epsilon^{ab}\epsilon^{cd}(\tilde{\psi}^{a}\tilde{\mc{T}}^{-1}\tilde{\psi}^c)(\tilde{\psi}^{b}\tilde{\mc{T}}^{-1}\tilde{\psi}^d)\right.\\
&\left.\phantom{+\tilde{g}^2\{}+\frac{8}{15}\tilde\Phi^{3/4}e^{-6\tilde{\lambda}-16\tilde{\phi}}\epsilon^{ab}\epsilon^{cd}(\tilde{\psi}^{a}\tilde{\mc{T}}^{-1}\tilde{\psi}^c)\tilde{R}^b\tilde{R}^d\right\}\tilde{\text{vol}}_5=0 \,,\numberthis
\end{align*}
and
\begin{align*}
\tilde{R}^{(5)}_{mn}&=30\tilde\nabla_{m}\tilde{\phi}\tilde\nabla_{n}\tilde{\phi}+36\tilde\nabla_{m}\tilde{\lambda}\tilde\nabla_{n}\tilde{\lambda}+\frac{5}{16}\tilde\Phi^{-2}\tilde\nabla_m\tilde\Phi\tilde\nabla_n\tilde\Phi+\frac{1}{4}\tilde{\mc{T}}^{-1}_{\alpha\beta}\tilde{\mc{T}}^{-1}_{\gamma\rho}D_{m}\tilde{\mc{T}}_{\beta\gamma}D_{n}\tilde{\mc{T}}_{\rho\alpha}\\
&\phantom{=}+\tilde\Phi^{-1/2}e^{-6\tilde{\phi}}\tilde{\mc{T}}^{-1}_{\alpha\beta}D_m\tilde{\psi}^{a\alpha}D_n\tilde{\psi}^{a\beta}+\tilde\Phi^{3/4}e^{6\tilde{\lambda}-6\tilde{\phi}}\tilde{P}^a_{m}\tilde{P}^a_{n}+\frac{1}{2}\tilde\Phi^{5/4}e^{6\tilde\lambda}\tilde{\mc{T}}_{\alpha\beta}\tilde Q^\alpha_m\tilde Q^\beta_n\\
&\phantom{=}+\frac{1}{2}\Phi^{-1}e^{-12\tilde{\phi}}\tilde{X}_{m}\tilde{X}_{n}+\frac{1}{2}\tilde\Phi^{1/4}e^{6\tilde{\lambda}-12\tilde{\phi}}\tilde{\mc{T}}^{-1}_{\alpha\beta}\tilde{\sigma}^\alpha_{m}\tilde{\sigma}^\beta_{n}\\
&\phantom{=}+\frac{1}{2}\tilde\Phi^{-1/2}e^{4\tilde{\phi}+12\tilde{\lambda}}\Bigg((\tilde{F}_{(2)})_{ml}(\tilde{F}_{(2)})_{n}^{\phantom{n}l}-\frac{1}{6}\tilde{g}_{mn}(\tilde{F}_{(2)})_{ls}(\tilde{F}_{(2)})^{ls}\Bigg)\\
&\phantom{=}+\frac{1}{2}\tilde\Phi^{-1/2}e^{4\tilde{\phi}-12\tilde{\lambda}}\Bigg((\tilde{\mc{F}}_{(2)})_{ml}(\tilde{\mc{F}}_{(2)})_{n}^{\phantom{n}l}-\frac{1}{6}\tilde{g}_{mn}(\tilde{\mc{F}}_{(2)})_{ls}(\tilde{\mc{F}}_{(2)})^{ls}\Bigg)\\
&\phantom{=}+\frac{1}{2}\tilde\Phi^{3/4}e^{4\tilde{\phi}-6\tilde{\lambda}}\tilde{\mc{T}}^{-1}_{\alpha\beta}\Bigg((\tilde{J}^\alpha_{(2)})_{ml}(\tilde{J}^\beta_{(2)})_{n}^{\phantom{n}l}-\frac{1}{6}\tilde{g}_{mn}(\tilde{J}^\alpha_{(2)})_{ls}(\tilde{J}^\beta_{(2)})^{ls}\Bigg)\\
&\phantom{=}+\tilde\Phi^{1/4}e^{-2\tilde{\phi}-6\tilde{\lambda}}\Bigg((\tilde{K}^a_{(2)})_{ml}(\tilde{K}^a_{(2)})_{n}^{\phantom{n}l}-\frac{1}{6}\tilde{g}_{mn}(\tilde{K}^a_{(2)})_{ls}(\tilde{K}^a_{(2)})^{ls}\Bigg)\\
&\phantom{=}+\frac{1}{4}\tilde\Phi^{-1}e^{8\tilde{\phi}}\Bigg((\tilde{H}_{(3)})_{mls}(\tilde{H}_{(3)})_{n}^{\phantom{n}ls}-\frac{2}{9}\tilde{g}_{mn}(\tilde{H}_{(3)})_{lsr}(\tilde H_{(3)})^{lsr}\Bigg)\\
&\phantom{=}+\frac{1}{4}\tilde\Phi^{1/4}e^{8\tilde{\phi}+6\tilde\lambda}\tilde{\mc{T}}^{-1}_{\alpha\beta}\Bigg((\tilde{G}^\alpha_{(3)})_{mls}(\tilde{G}^\beta_{(3)})_{n}^{\phantom{n}ls}-\frac{2}{9}\tilde{g}_{mn}(\tilde{G}^\alpha_{(3)})_{lsr}(\tilde G^\beta_{(3)})^{lsr}\Bigg)\\
&\phantom{=}+\tilde{g}^2\tilde{g}_{mn}\Bigg\{\frac{1}{3}e^{-10\tilde{\phi}}\left(e^{12\tilde{\lambda}}(\tilde{\psi}\tilde{\mc{T}}\tilde{\psi})-2(l+\tilde{\psi}^2)+e^{-12\tilde{\lambda}}(\tilde{\psi}\tilde{\mc{T}}^{-1}\tilde{\psi})+\tilde{\Phi}^{5/4}e^{-6\tilde{\lambda}}\tilde{R}^2\right)\\
&\phantom{=}+\frac{1}{6}\tilde\Phi^{-1/2}e^{12\tilde{\lambda}-16\tilde{\phi}}(l-\tilde{\psi}^2)^2+\frac{1}{6}\tilde\Phi^{1/2}e^{-4\tilde{\phi}}\left(2e^{12\tilde{\lambda}}\text{Tr}(\tilde{\mc{T}}^2)-e^{12\tilde{\lambda}}(\text{Tr}\tilde{\mc{T}})^2-4\text{Tr}\tilde{\mc{T}}\right)\\
&\phantom{=}+\frac{1}{3}\tilde\Phi^{-1/2}e^{-12\tilde{\lambda}-16\tilde{\phi}}\epsilon^{ab}\epsilon^{cd}(\tilde{\psi}^{a}\tilde{\mc{T}}^{-1}\tilde{\psi}^c)(\tilde{\psi}^{b}\tilde{\mc{T}}^{-1}\tilde{\psi}^d)\\
&\phantom{=}+\frac{2}{3}\tilde\Phi^{3/4}e^{-6\tilde{\lambda}-16\tilde{\phi}}\epsilon^{ab}\epsilon^{cd}(\tilde{\psi}^{a}\tilde{\mc{T}}^{-1}\tilde{\psi}^c)\tilde{R}^b\tilde{R}^d\Bigg\}\,.\numberthis
\end{align*}

\subsection{$D=4$ Equations of motion}\label{4deomssection}

The equations of motion of the M5-brane $D=4$ theory in \cite{Donos:2010ax} can be found in appendix A.2 of this paper. We are going to plug in our truncation ansatz \eqref{4d_identification} and set $k\rightarrow 0$ to obtain a new set of equations of motion. Equation (A.11) gives 
\begin{equation}\label{B24d}
d\tilde B_{(2)} - \tilde g(\tilde\theta\tilde G_{(3)} + \tilde\rho\tilde H_{(3)}) + 2\tilde g \tilde\Phi^{1/4}e^{-4\tilde\lambda-8\tilde\phi}\tilde\ast\tilde C_{(1)} - d\tilde\beta \wedge\tilde F_{(2)} = 0 \,.
\end{equation}
Equation (A.12) gives 
\begin{equation}\label{C14d}
d\tilde C_{(1)} - \tilde g\tilde\beta\tilde B_{(2)} - \frac{\tilde g}{2} \tilde\Phi^{1/4}e^{-4\tilde\lambda+2\tilde\phi}\tilde\ast\tilde B_{(2)} - \frac{\tilde g}{4}(l-2\tilde\beta^2 - 2\tilde\theta^2)\tilde F_{(2)} - d\tilde\theta\wedge\tilde P_{(1)} = 0 \,.
\end{equation} 
Equation (A.13) gives 
\begin{equation}\label{H34d}
d\tilde H_{(3)} = 0 \,,
\end{equation}
and 
\begin{equation}\label{G34d}
d\tilde G_{(3)} + \tilde Q_{(1)}\wedge \tilde H_{(3)} - \tilde g^2\tilde\Phi^{1/4}e^{12\tilde\lambda-18\tilde\phi}\tilde\sigma\tilde\vol_4 = 0 \,.
\end{equation}
Equation (A.14) gives 
\begin{equation}\label{X4d}
d\tilde X + 6\tilde\theta\tilde\beta d\tilde\beta - \frac{3}{2}\left(l-2\tilde\beta^2-2\tilde\theta^2\right)d\tilde\theta = 0 \,,\
\end{equation}
and
\begin{equation}\label{sigma4d}
d\tilde\sigma + \tilde X \tilde Q_{(1)} - 6\tilde\beta\tilde\rho d\tilde\beta + 6\tilde\theta C_{(1)} +\frac{3}{2}\left(l-2\tilde\beta^2-2\tilde\theta^2\right)\tilde P_{(1)} +\tilde\Phi^{1/4}e^{12\tilde\lambda+12\tilde\phi}\tilde\ast\tilde G_{(3)} = 0 \,.
\end{equation}
Equation (A.15) gives 
\begin{equation}\label{starB24d}
d\left(\tilde\Phi^{1/4}e^{-4\tilde\lambda+2\tilde\phi}\tilde\ast\tilde B_{(2)}\right) + 2\tilde g\tilde\beta(\tilde\theta\tilde G_{(3)} + \tilde\rho\tilde H_{(3)}) - 4\tilde g \tilde\Phi^{1/4}e^{-4\tilde\lambda-8\tilde\phi}\tilde\beta\tilde\ast\tilde C_{(1)} + 2\tilde B_{(2)}\wedge d\tilde\beta = 0 \,.
\end{equation}
Equation (A.16) gives 
\begin{equation}\label{starC14d}
d\left(\tilde\Phi^{1/4}e^{-4\tilde\lambda-8\tilde\phi}\tilde\ast\tilde C_{(1)}\right) - \frac{1}{2}d\tilde\theta\wedge\tilde G_{(3)} - \frac{1}{2}\tilde P_{(1)}\wedge \tilde H_{(3)} - \frac{\tilde g^2}{2}\tilde\Phi^{1/4}e^{12\tilde\lambda-18\tilde\phi}\tilde\theta\tilde\sigma\tilde\vol_4 = 0 \,.
\end{equation}
Equation (A.17) gives 
\begin{equation}\label{starG34d}
d\left(\tilde\Phi^{1/4}e^{12\tilde\lambda+12\tilde\phi}\tilde\ast\tilde G_{(3)}\right) + 3\tilde g \tilde\Phi^{1/4}e^{-4\tilde\lambda+2\tilde\phi}\tilde\theta\tilde\ast\tilde B_{(2)} - 6\tilde C_{(1)}\wedge d\tilde\theta + 6\tilde g\tilde\beta\tilde\theta \tilde B_{(2)} + \tilde g\tilde X\tilde F_{(2)} = 0 \,,
\end{equation} 
and 
\begin{equation}\label{starH34d}
\begin{split}
&d\left(\tilde\Phi^{-1}e^{12\tilde\phi}\tilde\ast\tilde H_{(3)}\right) - \tilde\Phi^{1/4}e^{12\tilde\lambda+12\tilde\phi}\tilde Q_{(1)}\wedge\tilde\ast\tilde G_{(3)} + 3\tilde g \tilde\Phi^{1/4}e^{-4\tilde\lambda+2\tilde\phi}\tilde\rho\tilde\ast\tilde B_{(2)} \\
&- 6\tilde C_{(1)}\wedge\tilde P_{(1)} + 6\tilde g\tilde\beta\tilde\rho \tilde B_{(2)} -\tilde g\tilde \sigma\tilde F_{(2)} = 0\,.
\end{split}
\end{equation}
Equation (A.18) gives
\begin{equation}\label{starbeta4d}
\begin{split}
&d\left(\tilde\Phi^{-1/2}e^{-4\tilde\phi+8\tilde\lambda}\tilde\ast d\tilde\beta\right) + \tilde\Phi^{1/4}e^{-4\tilde\lambda+2\tilde\phi}\tilde F_{(2)}\wedge\tilde\ast \tilde B_{(2)} + \tilde B_{(2)}\wedge\tilde B_{(2)} \\
&+ \tilde g^2\tilde\beta\left\{4\tilde\Phi^{-1/2}e^{-14\tilde\phi-8\tilde\lambda}\tilde\theta^2 - 2\tilde\Phi^{-1}e^{-18\tilde\phi}\tilde\theta\tilde X - \tilde\Phi^{-1/2}e^{-14\tilde\phi+8\tilde\lambda}\big(l-2\tilde\beta^2-2\tilde\theta^2\big) \right.\\
&\ph{+ \tilde g^2\tilde\beta\quad}\left. + 4\tilde\Phi^{3/4}e^{-14\tilde\phi+4\tilde\lambda}\tilde\rho^2 + 2\tilde\Phi^{1/4}e^{-18\tilde\phi+12\tilde\lambda}\tilde\rho\tilde \sigma\right\}\tilde\vol_4 = 0 \,.
\end{split}
\end{equation}
Equation (A.19) gives 
\begin{equation}\label{starP14d}
\begin{split}
&d\left(\tilde\Phi^{3/4}e^{-4\tilde\phi+4\tilde\lambda}\tilde\ast\tilde P_{(1)}\right) - 4\tilde\Phi^{1/4}e^{-4\tilde\lambda-8\tilde\phi}\tilde\ast\tilde C_{(1)}\wedge d\tilde\theta + 2\tilde C_{(1)}\wedge\tilde H_{(3)} \\
&+\tilde g^2\Bigg\{4\tilde\Phi^{3/4}e^{-14\tilde\phi+4\tilde\lambda}\tilde\beta^2\tilde\rho +\tilde\Phi^{5/4}e^{-4\tilde\lambda-10\tilde\phi}\tilde\rho \\
&\ph{+\tilde g^2\{}- \frac{1}{2}\tilde\Phi^{1/4}e^{-18\tilde\phi+12\tilde\lambda}\big(l-2\tilde\beta^2-2\tilde\theta^2\big)\tilde\sigma\Bigg\}\tilde\vol_4 = 0 \,,
\end{split}
\end{equation}
and 
\begin{equation}\label{startheta4d}
\begin{split}
&d\left(\tilde\Phi^{-1/2}e^{-4\tilde\phi-8\tilde\lambda}\tilde\ast d\tilde\theta\right) +4\tilde\Phi^{1/4}e^{-8\tilde\phi-4\tilde\lambda}\tilde\ast\tilde C_{(1)}\wedge\tilde P_{(1)} - \tilde\Phi^{3/4}e^{-4\tilde\phi+4\tilde\lambda}\tilde\ast \tilde Q_{(1)}\wedge\tilde P_{(1)} \\\
&+ 2\tilde C_{(1)}\wedge\tilde G_{(3)} +\tilde g^2\left\{4e^{-14\tilde\phi-8\tilde\lambda}\tilde\beta^2\tilde\theta-\tilde\Phi^{-1/2}e^{-14\tilde\phi+8\tilde\lambda}\big(l-2\tilde\beta^2-2\tilde\theta^2\big)\tilde\theta \right. \\
&\left.+ e^{-10\tilde\phi}\big(e^{16\tilde\lambda}-2+e^{-16\tilde\lambda}\big)\tilde\theta +\frac{1}{2}\tilde\Phi^{-1}e^{-18\tilde\phi}\big(l-2\tilde\beta^2-2\tilde\theta^2\big)\tilde X \right\}\tilde\vol_4 = 0 \,.
\end{split}
\end{equation}
Equation (A.20) gives 
\begin{align*}\label{starF24d}
&d\left(e^{6\tilde\phi-12\tilde\lambda}\tilde\Phi^{3/4}\tilde\ast\tilde F_{(2)}\right) + 3\tilde g e^{-4\tilde\phi+4\tilde\lambda}\tilde\Phi^{3/4}\tilde\theta\tilde\ast \tilde P_{(1)} - \tilde g e^{12\tilde\lambda}\tilde\Phi^{5/4}\tilde\ast\tilde Q_{(1)} \numberthis \\
&\ph{-}-\tilde g\tilde\sigma\tilde H_{(3)} + \tilde g\tilde X\tilde G_{(3)} - 3\tilde g e^{-4\tilde\lambda-8\tilde\phi}\tilde\Phi^{1/4}\big(l-2\tilde\beta^2-2\tilde\theta^2\big)\tilde\ast\tilde C_{(1)} + 3e^{-4\tilde\lambda+2\tilde\phi}d\tilde\beta\wedge\tilde\ast\tilde B_{(2)} = 0 \,.
\end{align*}
Equation (A.21) and (A.22) gives
\begin{equation}\label{tau4d}
\begin{split}
&d\left(\tilde\Phi^{5/4}e^{12\tilde\lambda}\tilde\ast \tilde Q_{(1)}\right) - 3\tilde\Phi^{3/4}e^{4\tilde\lambda-4\tilde\phi}d\tilde\theta\wedge\tilde\ast\tilde P_{(1)} + \tilde\Phi^{1/4}e^{12\tilde\phi+12\tilde\lambda}\tilde H_{(3)}\wedge\tilde\ast\tilde G_{(3)} \\
&+\tilde g^2\left\{3\tilde\Phi^{5/4}e^{-4\tilde\lambda-10\tilde\phi}\tilde\theta\tilde\rho + 12\tilde\Phi^{3/4}e^{4\tilde\lambda-14\tilde\phi}\tilde\beta^2\tilde\theta\tilde\rho - \tilde\Phi^{1/4}e^{12\tilde\lambda-18\tilde\phi}\tilde X\tilde\sigma\right\}\tilde\vol_4 = 0\,,
\end{split}
\end{equation}
and
\begin{align*}\label{lambda4d}
&d\tilde\ast d\tilde\lambda + \frac{1}{8}\tilde\Phi^{-1/2}e^{8\tilde\lambda-4\tilde\phi}\tilde\ast d\tilde\beta\wedge d\tilde\beta - \frac{1}{8}\tilde\Phi^{-1/2}e^{-8\tilde\lambda-4\tilde\phi}\tilde\ast d\tilde\theta\wedge d\tilde\theta + \frac{1}{16}\tilde\Phi^{-3/4}
e^{4\tilde\lambda-4\tilde\phi}\tilde\ast\tilde P_{(1)}\wedge\tilde P_{(1)} \\
&+ \frac{1}{16}\tilde\Phi^{1/4}e^{12\tilde\lambda+12\tilde\phi}\tilde\ast\tilde G_{(3)}\wedge\tilde G_{(3)} - \frac{1}{16}e^{-12\tilde\lambda+6\tilde\phi}\tilde\ast\tilde F_{(2)}\wedge\tilde F_{(2)} - \frac{1}{16}\tilde\Phi^{1/4}e^{-4\tilde\lambda+2\tilde\phi}\tilde\ast\tilde B_{(2)}\wedge\tilde B_{(2)} \\
&-\frac{1}{4}\tilde\Phi^{1/4}e^{-4\tilde\lambda-8\tilde\phi}\tilde\ast\tilde C_{(1)}\wedge\tilde C_{(1)} + \frac{1}{16}\tilde\Phi^{5/4}e^{12\tilde\lambda}\tilde\ast\tilde Q_{(1)}\wedge\tilde Q_{(1)} \\
&+ \tilde g^2\left\{\frac{1}{32}\tilde\Phi^{-1/2}e^{8\tilde\lambda-14\tilde\phi}(l-2\tilde\beta^2-2\tilde\theta^2)^2+\frac{1}{8}\tilde\Phi^{1/2}e^{-6\tilde\phi}(e^{-8\tilde\lambda}-2e^{8\tilde\lambda}+e^{24\tilde\lambda})\right.\\
&\left. + \frac{1}{4}e^{-10\tilde\phi}(e^{16\tilde\lambda}-e^{-16\tilde\lambda})\tilde\theta^2 - \frac{1}{16}\tilde\Phi^{5/4}e^{-10\tilde\phi-4\tilde\lambda}\tilde\rho^2- \frac{1}{2}\tilde\Phi^{-1/2}e^{-8\tilde\lambda-14\tilde\phi}\tilde\beta^2\tilde\theta^2\right.\\
&\left.  + \frac{1}{4}\tilde\Phi^{3/4}e^{4\tilde\lambda-14\tilde\phi}\tilde\beta^2\tilde\rho^2 + \frac{1}{16}\tilde\Phi^{1/4}e^{12\tilde\lambda-18\tilde\phi}\tilde\sigma^2\right\}\tilde\vol_4 = 0 \,,\numberthis
\end{align*}
and
\begin{equation}\label{Phi4d}
\begin{split}
&d(\tilde\Phi^{-1}\tilde\ast d\tilde\Phi) - \frac{6}{5}\tilde\Phi^{-1/2}e^{8\tilde\lambda-4\tilde\phi}\tilde\ast d\tilde\beta\wedge d\tilde\beta - \frac{6}{5}\tilde\Phi^{-1/2}e^{-8\tilde\lambda-4\tilde\phi}\tilde\ast d\tilde\theta\wedge d\tilde\theta \\
&+ \frac{9}{5}\tilde\Phi^{-3/4}e^{4\tilde\lambda-4\tilde\phi}\tilde\ast\tilde P_{(1)}\wedge\tilde P_{(1)} + \frac{1}{5}\tilde\Phi^{1/4}e^{12\tilde\lambda+12\tilde\phi}\tilde\ast\tilde G_{(3)}\wedge\tilde G_{(3)}  \\
& + \frac{3}{5}e^{-12\tilde\lambda+6\tilde\phi}\tilde\ast\tilde F_{(2)}\wedge\tilde F_{(2)}+ \frac{3}{5}\tilde\Phi^{1/4}e^{-4\tilde\lambda+2\tilde\phi}\tilde\ast\tilde B_{(2)}\wedge\tilde B_{(2)}\\
&+\frac{12}{5}\tilde\Phi^{1/4}e^{-4\tilde\lambda-8\tilde\phi}\tilde\ast\tilde C_{(1)}\wedge\tilde C_{(1)} + \tilde\Phi^{5/4}e^{12\tilde\lambda}\tilde\ast\tilde Q_{(1)}\wedge\tilde Q_{(1)}\\
& - \frac{4}{5}\tilde\Phi^{-1}e^{12\tilde\phi}\tilde\ast H_{(3)}\wedge\tilde H_{(3)} + \tilde g^2\left\{-\frac{3}{10}\tilde\Phi^{-1/2}e^{8\tilde\lambda-14\tilde\phi}(l-2\tilde\beta^2-2\tilde\theta^2)^2\right.\\
&\left.-\frac{2}{5}\tilde\Phi^{1/2}e^{-6\tilde\phi}(3e^{-8\tilde\lambda}+6e^{8\tilde\lambda}-e^{24\tilde\lambda}) +3\tilde\Phi^{5/4}e^{-10\tilde\phi-4\tilde\lambda}\tilde\rho^2 \right.\\
&\left. - \frac{24}{5}\tilde\Phi^{-1/2}e^{-8\tilde\lambda-14\tilde\phi}\tilde\beta^2\tilde\theta^2 + \frac{36}{5}\tilde\Phi^{3/4}e^{4\tilde\lambda-14\tilde\phi}\tilde\beta^2\tilde\rho^2\right.\\
&\left.-\frac{4}{5}\tilde\Phi^{-1}e^{-18\tilde\phi}\tilde X^2 + \frac{1}{5}\tilde\Phi^{1/4}e^{12\tilde\lambda-18\tilde\phi}\tilde\sigma^2\right\}\tilde\vol_4 = 0 \,.
\end{split}
\end{equation}
Equation (A.23) gives
\begin{align*}\label{phi4d}
&d\tilde\ast d\tilde\phi-\frac{1}{10}\tilde\Phi^{-1/2}e^{8\tilde\lambda-4\tilde\phi}\tilde\ast d\tilde\beta\wedge d\beta - \frac{1}{10}\tilde\Phi^{-1/2}e^{-8\tilde\lambda-4\tilde\phi}\tilde\ast d\tilde\theta\wedge d\tilde\theta  \\
&- \frac{1}{10}\tilde\Phi^{3/4}e^{4\tilde\lambda-4\tilde\phi}\tilde\ast\tilde P_{(1)}\wedge\tilde P_{(1)}+\frac{1}{10}\tilde\Phi^{-1}e^{12\tilde\phi}\tilde\ast\tilde H_{(3)}\wedge\tilde H_{(3)} \\
&+ \frac{1}{10}\tilde\Phi^{1/4}e^{12\tilde\lambda+12\tilde\phi}\tilde\ast\tilde G_{(3)}\wedge\tilde G_{(3)} + \frac{1}{20}\tilde\Phi^{3/4}e^{-12\tilde\lambda+6\tilde\phi}\tilde\ast\tilde F_{(2)}\wedge \tilde F_{(2)} \\
&+\frac{1}{20}\tilde\Phi^{1/4}e^{-4\tilde\lambda+2\tilde\phi}\tilde\ast\tilde B_{(2)}\wedge \tilde B_{(2)} - \frac{4}{5}\tilde\Phi^{1/4}e^{-4\tilde\lambda-8\tilde\phi}\tilde\ast\tilde C_{(1)}\wedge C_{(1)} \\
&+ \tilde g^2\left\{\frac{l}{2}e^{-10\tilde\phi}-\frac{7}{80}\tilde\Phi^{-1/2}e^{8\tilde\lambda-14\tilde\phi}(l-2\tilde\beta^2-2\tilde\theta^2)\right.\\
&\left.+\frac{1}{20}\tilde\Phi^{1/2}e^{-6\tilde\phi}(3e^{-8\tilde\lambda}-e^{24\tilde\lambda}+6e^{8\tilde\lambda}) - \frac{1}{4}\tilde\Phi^{5/4}e^{-10\tilde\phi-4\tilde\lambda}\tilde\rho^2\right.\\
&\left.-\frac{1}{4}e^{-10\tilde\phi}(e^{16\tilde\lambda}-2+e^{-16\tilde\lambda})\tilde\theta^2  -\frac{7}{5}\tilde\Phi^{-1/2}e^{-8\tilde\lambda-14\tilde\phi}\tilde\beta^2\tilde\theta^2\right.\\
&\left.-\frac{7}{5}\tilde\Phi^{3/4}e^{4\tilde\lambda-14\tilde\phi}\tilde\beta^2\tilde\rho^2 -\frac{3}{20}\tilde\Phi^{-1}e^{-18\tilde\phi}\tilde X^2 - \frac{3}{20}\tilde\Phi^{1/4}e^{12\tilde\lambda-18\tilde\phi}\tilde\sigma^2\right\}\tilde\vol_4 = 0 \,,\numberthis
\end{align*}
and
\begin{align*}
\tilde R^{(4)}_{mn} &= 30\tilde\nabla_{m}\tilde\phi \tilde\nabla_{n}\tilde\phi + 48\tilde\nabla_{m}\tilde\lambda\tilde\nabla_{n}\tilde\lambda + \frac{5}{16}\tilde\Phi^{-2}\tilde\nabla_{m}\tilde\Phi \tilde\nabla_{n}\tilde\Phi  \\
&\quad + \frac{3}{2}\tilde\Phi^{-1/2}e^{8\tilde\lambda-4\tilde\phi}\tilde\nabla_{m}\tilde\beta\tilde\nabla_{n}\tilde\beta+ \frac{3}{2}\tilde\Phi^{-1/2}e^{-8\tilde\lambda-4\tilde\phi}\tilde\nabla_{m}\tilde\theta\tilde\nabla_{n}\tilde\theta \\
&\quad+ \frac{3}{2}\tilde\Phi^{3/4}e^{4\tilde\lambda-4\tilde\phi}\tilde P_{m}\tilde P_{n} + \frac{1}{2}\tilde\Phi^{5/4}e^{12\tilde\lambda}\tilde Q_{m}\tilde Q_{n}+6\tilde\Phi^{1/4}e^{-4\tilde\lambda-8\tilde\phi}\tilde C_{m}\tilde C_{n} \\
&\quad + \frac{1}{2}\tilde\Phi^{3/4}e^{-12\tilde\lambda+6\tilde\phi}\left((\tilde F_{(2)})_{ml}(\tilde F_{(2)})_{n}^{\ph{n}l} - \frac{1}{4}\tilde g_{mn}(\tilde F_{(2)})_{ls}(\tilde F_{(2)})^{ls}\right)\\
&\quad +\frac{3}{2}\tilde\Phi^{1/4}e^{-4\tilde\lambda+2\tilde\phi}\left((\tilde B_{(2)})_{ml}(\tilde B_{(2)})_{n}^{\ph{n}l} - \frac{1}{4}\tilde g_{mn}(\tilde B_{(2)})_{ls}(\tilde B_{(2)})^{ls}\right) \\
&\quad+\frac{1}{4}\tilde\Phi^{-1}e^{12\tilde\phi}\left((\tilde H_{(3)})_{mls}(\tilde H_{(3)})_{n}^{\ph{n}ls} - \frac{1}{3}\tilde g_{mn}(\tilde H_{(3)})_{lsr}(\tilde H_{(3)})^{lsr}\right) \\
&\quad+ \frac{1}{4}\tilde\Phi^{1/4}e^{12\tilde\lambda+12\tilde\phi}\left((\tilde G_{(3)})_{mls}(\tilde H_{(3)})_{n}^{\ph{n}ls} - \frac{1}{3}\tilde g_{mn}(\tilde G_{(3)})_{lsr}(\tilde G_{(3)})^{lsr}\right) \\
&\quad -\frac{\tilde g^2}{2}\left\{3le^{-10\tilde\phi}-\frac{3}{8}\tilde\Phi^{-1/2}e^{8\tilde\lambda-14\tilde\phi}(l-2\tilde\beta^2-2\tilde\theta^2)^2 -\frac{3}{2}e^{-10\tilde\phi}(e^{16\tilde\lambda}-2+e^{-16\tilde\lambda})\tilde\theta^2\right.\\
&\quad \left. + \frac{1}{2}\tilde\Phi^{1/2}e^{-6\tilde\phi}(3e^{-8\tilde\lambda}-e^{24\tilde\lambda}+6e^{8\tilde\lambda})- \frac{3}{2}\tilde\Phi^{5/4}e^{-10\tilde\phi-4\tilde\lambda}\tilde\rho^2 - 6\tilde\Phi^{3/4}e^{4\tilde\lambda-14\tilde\phi}\tilde\beta^2\tilde\rho^2 \right.\\
&\quad \left.-6\tilde\Phi^{-1/2}e^{-8\tilde\lambda-14\tilde\phi}\tilde\beta^2\tilde\theta^2- \frac{1}{2}\tilde\Phi^{-1}e^{-18\tilde\phi}\tilde X^2 - \frac{1}{2}\tilde\Phi^{1/4}e^{12\tilde\lambda-18\tilde\phi}\tilde\sigma^2 \right\}\tilde g_{mn} \,.\numberthis
\end{align*}

\section{Minimal representations of $A^{100}_{5,17}$ and $A^{0}_{5,18}$}\label{appendix:min_rep}

We provide here an explicit representation of the generators in \eqref{eq:weird_generators},
\begin{equation}
\begin{split}
&\mathfrak{g}_1= \begin{pmatrix} 0 & 0 & 0& 0\\ 0 & 0 & 0& 1  \\0 & 0 & 0& 0 \\0 & 0 & 0& 0\\\end{pmatrix} \,,\quad\mathfrak{g}_2= \begin{pmatrix} 0 & 0 & 0& 1\\ 0 & 0 & 0& 0  \\0 & 0 & 0& 0 \\0 & 0 & 0& 0\\\end{pmatrix} \,,\quad \mathfrak{g}_3= \begin{pmatrix} 0 & 0 & 0& 0\\ 0 & 0 & 1& 0  \\0 & 0 & 0& 0 \\0 & 0 & 0& 0\\\end{pmatrix} \,,\\
&\mathfrak{g}_4= \begin{pmatrix} 0 & 0 & 1& 0\\ 0 & 0 & 0& 0  \\0 & 0 & 0& 0 \\0 & 0 & 0& 0\\\end{pmatrix} \,,\quad\mathfrak{g}_5= \begin{pmatrix} 0 & 1 & 0& 0\\ -1 & 0 & 0& 0  \\0 & 0 & 0& -l \\0 & 0 & 0& 0\\\end{pmatrix} \,.
\end{split}
\end{equation}

\section{Matching with $\mc{N} = 4$ supergravity}\label{app:matching_N=4}
In this section, we present some of the details on how to match the truncated $D=5$ theory of section \ref{KK_truncations_Riemann} with the canonical language of $\mc{N}=4$ theory in \cite{Schon:2006kz}.

The parametrisation of the coset $SO(5,3)/(SO(5)\times SO(3))$ is given in \eqref{eq:mcV_parametrisation}, and we find that the Lie algebra valued Maurer-Cartan one form is given by
\begin{equation}
\begin{split}
d\mc{V}\cdot \mc{ V}^{-1} &= \frac{1}{\sqrt{2}}d\varphi_1\mc{H}+\frac{1}{\sqrt{2}}d\varphi_2\hat{\mc{H}}+\frac{1}{\sqrt{2}}d{\varphi}_3 H^3+e^{2\varphi_1}d\rho\,\mc{E}\\
&\quad +e^{{\varphi}_3+{\varphi}_1+{\varphi}_2}\left( Q^3_{(1)} +{\rho}\tilde Q^4_{(1)}\right)T^2+e^{{\varphi}_3-{\varphi}_1+{\varphi}_2} Q^4_{(1)} T^3+e^{-2\varphi_2} X_{(1)}T^4\\
&\quad +e^{{\varphi}_3-{\varphi}_1-{\varphi}_2}({\mc{T}}^{-1}\tilde{\sigma}_{(1)})^4T^5-e^{{\varphi}_3+{\varphi}_1-{\varphi}_2}\left[({\mc{T}}^{-1}{\sigma}_{(1)})^3+{\rho}({\mc{T}}^{-1}{\sigma}_{(1)})^4\right]T^6\\
&\quad+\sqrt{2}e^{-\varphi_1-\varphi_2}d{\psi}^{13}T^7+\sqrt{2}e^{\varphi_1-\varphi_2}(d{\psi}^{14}-\rho d{\psi}^{13})T^8-\sqrt{2}e^{{\varphi}_3}{P}^1_{(1)}T^{9}\\
&\quad +\sqrt{2}e^{-\varphi_1-\varphi_2}d{\psi}^{23}T^{10}+\sqrt{2}e^{\varphi_1-\varphi_2}(d{\psi}^{24}-\rho d{\psi}^{23})T^{11}-\sqrt{2}e^{{\varphi}_3}{P}^2_{(1)}T^{12} \,,
\end{split}
\end{equation} 
where $\varphi_2,{\varphi}_3$ are defined in \eqref{eq:definitions_of_varphi2_and_Delta} and
\begin{equation}
\begin{split}
&{R}^a={\Psi}^a+{\tau}^\alpha{\psi}^{a\alpha}\,,\quad  X_{(1)}=d\Xi+\epsilon_{\alpha\beta}{\psi}^{a\alpha}d{\psi}^{a\beta}\,,\quad  P^a_{(1)} = d{\Psi}^a + \tau^\alpha d\psi^{a\alpha} \,,\\
& Q^\alpha_{(1)} = d\tau^\alpha \,,\quad ({\mc{T}}^{-1}\sigma_{(1)})^\alpha = d\xi^\alpha -{\tau}^\alpha d{\Xi} - \epsilon_{\alpha\beta}\left(\psi^{a\beta}\psi^{a\gamma} Q^\gamma_{(1)}+2 R^a d\psi^{a\beta} \right) \,,\\
\end{split}
\end{equation} 
are the ungauged versions of ${X}_{(1)}$, $  P^a_{(1)}$, $ Q^\alpha_{(1)}$ and $({\mc{T}}^{-1}\sigma_{(1)})^\alpha $. The Maurer-Cartan one form can be decomposed as $d\mc{V}\cdot \mc{ V}^{-1}=\mathcal{ Q}_{(1)}^{(0)}+\mathcal{P}_{(1)}^{(0)}$, where $\mathcal{ Q}_{(1)}^{(0)}$ lies in the Lie algebra of $SO(5)\times SO(3)$, and $\mathcal{P}_{(1)}^{(0)}$ lies in its complement. The ungauged kinetic term of the scalar fields in the coset $SO(5,3)/(SO(5)\times SO(3))$ is equal to
\begin{equation}
\begin{split}
\frac{1}{8}{\ast d\mc{M}_{MN}}\wedge d\mc{M}^{MN}=&-\frac{1}{2}\text{Tr}\left({{\ast}\mc{ P}}_{(1)}^{(0)}\wedge \mc{P}_{(1)}^{(0)}\right)\\
=&-\frac{1}{4}\text{Tr}\Bigg({{\ast} \Bigg[d\mc{V}\cdot\mc{V}^{-1}}\Bigg]\wedge\Bigg[d\mc{V}\cdot\mc{ V}^{-1}+\Big(d\mc{V}\cdot\mc{ V}^{-1}\Big)^T\Bigg]\Bigg)\,.
\end{split}
\end{equation}
The scalar manifold of the reduced theory is $SO(1,1)\times  SO(5,3)/(SO(5)\times SO(3))$, and the ungauged kinetic term of all of the scalar fields can be recast into
\begin{equation}\label{eq:ungauged_ke_scalar}
\begin{split}
\mc{L}^{S}_{\mc{N}=4}=&-3\Sigma^{-2}{\ast}d\Sigma\wedge d\Sigma+\frac{1}{8}{\ast}d\mc{M}_{MN}\wedge d\mc{M}^{MN}\,,
\end{split}
\end{equation}
where the $SO(1,1)$ part of the scalar manifold is described by the real scalar field $\Sigma$, via
\begin{align}
\Sigma={\Phi}^{1/8}e^{-{\phi}-3{\lambda}}\,.
\end{align}
To incorporate the gauging, we need to use the covariant derivative given in \eqref{eq:5dN=4_cov_derv} which we denote as $D=d+{g}\mathfrak{A}$ with
\begin{align}\label{eq:5d_precise_gauging}
\mathfrak{A}={A}_{(1)}\mathfrak{g}_0+{\mathscr{A}}^3_{(1)}\mathfrak{g}_1+{\mathscr{A}}^4_{(1)}\mathfrak{g}_2+{{V}}^3_{(1)}\mathfrak{g}_3+{{V}}^4_{(1)}\mathfrak{g}_4+{\mc{A}}_{(1)}\mathfrak{g}_5\,.
\end{align}
Now we can decompose the gauged version of the Maurer-Cartan one form $D\mathcal{V}\cdot\mathcal{V}^{-1}=\mathcal{P}+\mathcal{Q}$.
In particular we have 
$\mathcal{P}=\mathcal{P}^{(0)}+\frac{{g}}{2}\left[\mathcal{V}\cdot \mathfrak{A}\cdot \mathcal{V}^{-1}+\left(\mathcal{V}\cdot \mathfrak{A}\cdot \mathcal{V}^{-1}\right)^T\right]$, which lies in the complement of the Lie algebra of $SO(5)\times SO(3)$. Finally, we find that the gauged scalar kinetic terms are recovered precisely after evaluating $-\frac{1}{2}\text{Tr}(\ast \mathcal{P}\wedge\mathcal{P})$.

We provide here the explicit expression of the matrix $\mc{M}_{MN}$ which is defined in \eqref{emmmat}, 
\begin{align}\label{eq:Mmn_matrix}
\mc{M}=
\left(
\begin{array}{ccc}
\mc{T}_3^{-1}&\mc{T}_3^{-1}\cdot \mc{S}^T&\mc{T}_3^{-1}\cdot \mc{Y}\\
\mc{S}\cdot \mc{T}_3^{-1}&\mc{S}\cdot\mc{T}_3^{-1}\cdot\mc{S}^T+\mathds{1}_2&\mc{S}\cdot\mc{T}_3^{-1}\cdot\mc{Y}+\mc{S}\\
\mc{Y}^T\cdot \mc{T}_3^{-1}&\mc{Y}^T\cdot\mc{T}_3^{-1}\cdot\mc{S}^T+\mc{S}^T&\mc{Y}^T\cdot\mc{T}_3^{-1}\cdot \mc{Y}+\mc{S}^T\cdot\mc{S}+\mc{T}_3
\end{array}
\right)\,,
\end{align}
where
\begin{align}
\begin{split}
&V_3=
\left(
\begin{array}{ccc}
e^{{\varphi}_1+{\varphi}_2}&e^{{\varphi}_1+{\varphi}_2}{\rho}&e^{{\varphi}_1+{\varphi}_2}({\tau}^3+{\rho}{\tau}^4)\\
0&e^{-{\varphi}_1+{\varphi}_2}&e^{-{\varphi}_1+{\varphi}_2}{\tau}^4\\
0&0&e^{-{\varphi}_3}
\end{array}
\right)\,,\quad
\mc{S}=\left(
\begin{array}{ccc}
 \sqrt{2} {\psi}^{13} & \sqrt{2} {\psi}^{14}& -\sqrt{2}{\Psi}^1\\
 \sqrt{2} {\psi}^{23}& \sqrt{2} {\psi}^{24} & -\sqrt{2}{\Psi}^2 \\
\end{array}
\right)\,,
\end{split}
\end{align}
and
\begin{align}
\mc{Y}_{ij}=\frac{1}{2}\mc{S}^a_i\mc{S}^a_j+\left(
\begin{array}{ccc}
0&{\Xi}&{\xi}^4+{R}^a{\psi}^{a3}\\
-{\Xi}&0&-{\xi}^3+{R}^a{\psi}^{a4}\\
-{\xi}^4-{R}^a{\psi}^{a3}&{\xi}^3-{R}^a{\psi}^{a4}&0
\end{array}
\right)_{ij}\,.
\end{align}
and we also define $\mc{T}_3\equiv V_3^T\cdot V_3$. To calculate the $\mc{N} = 4$ scalar potential \eqref{eq:N=4_potential_SW}, we have to make use of the embedding tensor specified in \eqref{eq:embedding_tensor_comp} and we find the following non-vanishing contributions
\begin{align*}
&-\frac{1}{2}f_{MNP}f_{QRS}\Sigma^{-2}\Bigg(\frac{1}{12}\mathcal{M}^{MQ}\mathcal{M}^{NR}\mathcal{M}^{PS}-\frac{1}{4}\mathcal{M}^{MQ}\eta^{NR}\eta^{PS}+\frac{1}{6}\eta^{MQ}\eta^{NR}\eta^{PS}\Bigg)\\
&=-\frac{1}{2}\Phi^{1/2}e^{-4{\phi}}\left(2e^{12{\lambda}}\text{Tr}({\mc{T}}^2)-e^{12{\lambda}}(\text{Tr}{\mc{T}})^2\right)-\frac{1}{2}\Phi^{-1/2}e^{12{\lambda}-16{\phi}}(l-{\psi}^2)^2-e^{-10{\phi}}e^{12{\lambda}}({\psi}{\mc{T}}{\psi})\,,\\
&-\frac{1}{8}\xi_{MN}\xi_{PQ}\Sigma^4\Bigg(\mathcal{M}^{MP}\mathcal{M^{NQ}}-\eta^{MP}\eta^{NQ}\Bigg)\\
&=-\tilde\Phi^{-1/2}e^{-12{\lambda}-16{\phi}}\epsilon^{ab}\epsilon^{cd}({\psi}^{a}{\mc{T}}^{-1}{\psi}^c)({\psi}^{b}{\mc{T}}^{-1}{\psi}^d)-2\Phi^{3/4}e^{-6{\lambda}-16{\phi}}\epsilon^{ab}\epsilon^{cd}({\psi}^{a}{\mc{T}}^{-1}{\psi}^c){R}^b{R}^d\\
&\quad-e^{-10{\phi}}\left(e^{-12{\lambda}}({\psi}\tilde{\mc{T}}^{-1}{\psi})+{\Phi}^{5/4}e^{-6{\lambda}}{R}^2\right)\,,\\
&-\frac{1}{3\sqrt{2}}f_{MNP}\xi_{QR}\Sigma\mathcal{M}^{MNPQR}=\,2e^{-10{\phi}}(l+{\psi}^2)+2{\Phi}^{1/2}e^{-4{\phi}}\text{Tr}{\mc{T}}\,.\numberthis
\end{align*}
Combining these contributions, we are able to recover the scalar potential in our truncated theory \eqref{5doriginalpotential}.

We now turn to the vector sector. Using the identifications in \eqref{eq:canonical_field_strengths} and $\mc{M}_{MN}$ in \eqref{eq:Mmn_matrix}, we find that the $\mc{N}=4$ kinetic terms of the vectors in \eqref{eq:ke_vectors_SW} matches with the kinetic terms of the vectors of the truncated theory \eqref{eq_ke_vectors_after_redef} (after applying the field redefinitions in \eqref{dualG3} and \eqref{dualH3}). For the topological part of the Lagrangian, we find that the non-zero contributions to $L^T_{\mc{N}=4}$ \eqref{eq:n4lag5_topological} are (up to a total derivative),
\begin{align*}
&-\frac{1}{\sqrt{2}}gZ^{\mathcal{M}\mathcal{N}}\mathcal{B}_{\mathcal{M}}\wedge D\mathcal{B}_{\mathcal{N}}= \frac{1}{2 g}\epsilon_{ab} L^a_{(2)}\wedge D L^b_{(2)} \,, \\
&\frac{1}{2\sqrt{2}}{g}d_{\mathcal{M}\mathcal{N}\mathcal{P}}X_{\mathcal{Q}\mathcal{R}}^{\phantom{\mathcal{Q}\mathcal{R}}\mathcal{M}}\mathcal{A}^{\mathcal{N}}\wedge \mathcal{A}^{\mathcal{Q}}\wedge \mathcal{A}^{\mathcal{R}}\wedge d\mathcal{A}^{\mathcal{P}}\\
&=-\frac{ g l}{2}\epsilon_{\alpha\beta} V^\alpha_{(1)}\wedge V^\beta_{(1)}\wedge{\mc{A}}_{(1)}\wedge F_{(2)}-{g}{\mc{A}}_{(1)}\wedge {\mathscr{A}}^\alpha_{(1)}\wedge V^\alpha_{(1)}\wedge F_{(2)} \,,\numberthis\\
&\frac{\sqrt{2}}{3}d_{\mathcal{M}\mathcal{N}\mathcal{P}}\mathcal{A}^{\mathcal{M}}\wedge d\mathcal{A}^{\mathcal{N}}\wedge d\mathcal{A}^{\mathcal{P}}\\
&=- \epsilon_{\alpha\beta} \left({d\left[{\mathscr{A}}^\alpha_{(1)}-l\epsilon_{\alpha\beta}{V}^\beta_{(1)}\right]}\right)\wedge V^\beta_{(1)}\wedge\tilde F_{(2)}\\
&\quad -{F}_{(2)}\wedge{\mc{F}}_{(2)}\wedge\left(\mathscr{{B}}_{(1)}- \tau^\alpha\left[{\mathscr{A}}^\alpha_{(1)}-l\epsilon_{\alpha\beta}{V}^\beta_{(1)}- \frac{l}{2 g}\epsilon_{\alpha\beta}d\tau^\beta\right] + \frac{1}{2 g}\tau^2 d\Xi -\Xi\tau^\alpha V^\alpha_{(1)}+ \frac{1}{ g}\xi^\alpha Q^\alpha_{(1)} \right)\,.
\end{align*}
Combining these contributions, we recover the topological part of the Lagrangian in our truncated theory \eqref{eq:5dtopological_useful}.

\section{Parametrisation of $SO(4,2)/(SO(4)\times SO(2))$ and the quaternionic K\"{a}hler structure}\label{coset_parametrisation}

In order to parametrise the coset $SO(4,2)/(SO(4)\times SO(2))$, we again exponentiate a suitable solvable subalgebra of the Lie algebra. Following, for example \cite{Lu:1998xt}, the two Cartan generators $H^i$ and the six positive root generators are given by 
\begin{equation}
\begin{split}
H^{1}=\sqrt{2}(E_{11}-E_{55}), \quad &T^{1}=E_{56}-E_{21},\quad T^{2}=E_{16}-E_{25},\quad T^{3}=E_{13}+E_{35}\,,\\
H^{2}=\sqrt{2}(E_{22}-E_{66}),\quad &T^{4}=E_{14}+E_{45},\quad T^{5}=E_{23}+E_{36},\quad T^{6}=E_{24}+E_{46}\,,
\end{split}
\end{equation} 
where $E_{ij}$ denotes the $6\times6$ matrix with 1 in the $(i,j)$ position and 0 elsewhere. We note that $\text{Tr}(T^i(T^j)^T)=2\delta^{ij}$, and $\text{Tr}(H^mH^n)=4\delta^{mn}$ with $(H^m)^T=H^m$. To make contact with the scalar fields in the truncated theory, we first find an explicit embedding of the coset $SL(2,\mb{R})/SO(2)$ inside $SO(4,2)/(SO(4)\times SO(2))$ which can be achieved by defining
\begin{equation}
\mc{H}=H^2-H^1,\quad \mc{E}=T^1\,.
\end{equation} 
In addition, we define $\hat{\mc{H}}=-H^1-H^2$ which commutes with the above two generators. We introduce three scalar fields $\{\tilde\varphi_1,\tilde\varphi_2,\tilde\rho\}$ to form the following coset
\begin{equation} 
\mc{V}_{(s)}=e^{\frac{1}{\sqrt{2}}\varphi_1\mc{H}+\frac{1}{\sqrt{2}}\varphi_2\hat{\mc{H}}}e^{\rho\mc{E}}=\left(
\begin{array}{ccc}
 e^{-\varphi _2} V^{-T}&0 &0 \\
 0 & \mathds{1}_2 & 0 \\
 0 & 0 & e^{\varphi _2} V 
\end{array}
\right)\,,
\end{equation} 
where the $2\times 2$ matrix $\tilde{V}$ parametrises the coset $SL(2,\mb{R})/SO(2)$ in the standard upper triangular gauge
\begin{equation} 
 V=\left(
\begin{array}{cc}
 e^{\varphi_1} & e^{\varphi_1} \rho \\
 0 & e^{-\varphi_1}
\end{array} 
\right) \,.
\end{equation} 
We can identify the scalar fields in the $2\times 2$ matrix ${\mc{T}}_{\alpha\beta}$ in the truncated theory as
\begin{equation}
{\mc{T}}_{\alpha\beta}=( V^T  V)_{\alpha\beta}
=\left(
\begin{array}{cc}
 e^{2\varphi_1} & e^{2\varphi_1} \rho \\
  e^{2\varphi_1}\rho  & e^{-2\varphi_1}+e^{2\varphi_1}\rho^2
\end{array}
\right) \,,
\end{equation} 
and the scalar field $\varphi_2$ is identified as
\begin{equation}\label{eq:definition_of_varphi2}
\varphi_2=3{\phi}+\frac{1}{4}\log\Phi \,.
\end{equation} 
Collecting our results, the parametrisation of the coset $SO(4,2)/(SO(4)\times SO(2))$ is
\begin{equation}
\mc{V}
=\mc{V}_{(s)}\cdot\text{exp}\Bigg[\Xi T^2+\sqrt{2}{\psi}^{11}T^{3}+\sqrt{2}{\psi}^{21}T^{4}+\sqrt{2}{\psi}^{12}T^{5}+\sqrt{2}{\psi}^{22}T^{6}\Bigg] \,.
\end{equation} 
To calculate the kinetic terms of the scalar fields, we make use of the Lie algebra valued Maurer-Cartan one form. It is given by
\begin{equation}
\begin{split}
d\mc{V}\cdot \mc{V}^{-1} &= \frac{1}{\sqrt{2}}d\varphi_1\mc{H}+\frac{1}{\sqrt{2}}d\varphi_2\hat{\mc{H}}+e^{2\varphi_1}d\rho\,\mc{E}+e^{-2\varphi_2} X_{(1)}T^2\\
&\quad+\sqrt{2}e^{-\varphi_1-\varphi_2}d{\psi}^{11}T^3+\sqrt{2}e^{-\varphi_1-\varphi_2}d{\psi}^{21}T^4+\sqrt{2}e^{\varphi_1-\varphi_2}(d{\psi}^{12}-\rho d{\psi}^{11})T^5\\
&\quad +\sqrt{2}e^{\varphi_1-\varphi_2}(d{\psi}^{22}-\rho d{\psi}^{21})T^{6} \,,
\end{split}
\end{equation} 
where $\varphi_2$ is defined in \eqref{eq:definition_of_varphi2} and
\begin{equation}\label{eq:definition_of_X_(1)} 
X_{(1)}=d\Xi+\epsilon_{\alpha\beta}{\psi}^{a\alpha}d{\psi}^{a\beta}\,,
\end{equation} 
is equal to the ungauged version of the 1-form $X_{(1)}$. Again, the Maurer-Cartan one form can be decomposed as $d\mc{ V}\cdot \mc{ V}^{-1}=\mathcal{ Q}^{(0)}_{(1)}+\mathcal{P}_{(1)}^{(0)}$, where $\mathcal{ Q}_{(1)}^{(0)}$ lies in the Lie algebra of $SO(4)\times SO(2)$, and $\mathcal{ P}_{(1)}^{(0)}$ lies in its complement. The ungauged kinetic term of the scalar fields in the hypermultiplet sector is equal to
\begin{equation}
-\frac{1}{2}\text{Tr}\left({{\ast}\mc{ P}}_{(1)}^{(0)}\wedge \mc{ P}_{(1)}^{(0)}\right)=-\frac{1}{4}\text{Tr}\Bigg({{\ast} \Bigg[d\mc{ V}\cdot\mc{ V}^{-1}}\Bigg]\wedge\Bigg[d\mc{ V}\cdot\mc{ V}^{-1}+\Big(d\mc{V}\cdot\mc{ V}^{-1}\Big)^T\Bigg]\Bigg)\,.
\end{equation}
The scalar manifold of the reduced theory is $SO(1,1)\times SO(1,1)\times  SO(4,2)/(SO(4)\times SO(2))$, and the ungauged kinetic term of all of the scalar fields can be recast into
\begin{equation}\label{eq:ungauged_ke_scalar}
\begin{split}
\mc{L}^{kin}_{scalar}=&-\frac{3}{16}{{\ast}(8d{\phi}-\Phi^{-1}d\tilde\Phi)}\wedge (8d{\phi}-\Phi^{-1}d\Phi)-36{{\ast} d{\lambda}}\wedge d{\lambda}\\
&-\frac{1}{4}\text{Tr}\Bigg({{\ast} \Bigg[d\mc{ V}\cdot\mc{ V}^{-1}}\Bigg]\wedge\Bigg[d\mc{ V}\cdot\mc{ V}^{-1}+\Big(d\mc{V}\cdot\mc{ V}^{-1}\Big)^T\Bigg]\Bigg)\,.
\end{split}
\end{equation}
In terms of the $\mc{N}=2$ canonical language in \eqref{eq:5d_canonical_lagrangian}, we can make the following identifications
\begin{equation}
\begin{split}\label{eq:scalarcoset_metric}
g_{xy}{{\ast} d\phi^{x}}\wedge d\phi^y&=\frac{3}{8}{{\ast}(8d{\phi}-\Phi^{-1}d\Phi)}\wedge (8d{\phi}-\Phi^{-1}d\Phi)+72{{\ast} d{\lambda}}\wedge d{\lambda}\,,\\
g_{XY}{{\ast} dq^X}\wedge dq^Y&=\frac{1}{2}\text{Tr}\Bigg({{\ast} \Bigg[d\mc{ V}\cdot\mc{ V}^{-1}}\Bigg]\wedge\Bigg[d\mc{ V}\cdot\mc{ V}^{-1}+\Big(d\mc{V}\cdot\mc{ V}^{-1}\Big)^T\Bigg]\Bigg)\\
&=4{{\ast} d\varphi_1}\wedge d\varphi_1+4{{\ast} d\varphi_2}\wedge d\varphi_2+e^{4\varphi_1}{{\ast} d\rho}\wedge d\rho+e^{-4\varphi_2}{{\ast}  X_{(1)}}\wedge  X_{(1)}\\
&\phantom{=}+2e^{-2\varphi_1-2\varphi_2}{{\ast} d{\psi}^{11}}\wedge d{\psi}^{11}+2e^{-2\varphi_1-2\varphi_2}{{\ast} d{\psi}^{21}}\wedge d{\psi}^{21}\\
&\phantom{=}+2e^{2\varphi_1-2\varphi_2}{{\ast}{(d{\psi}^{12}-\rho d{\psi}^{11})}}\wedge (d{\psi}^{12}-\rho d{\psi}^{11})\\
&\phantom{=}+2e^{2\varphi_1-2\varphi_2}{{\ast}{(d{\psi}^{22}-\rho d{\psi}^{21})}}\wedge {(d{\psi}^{22}-\rho d{\psi}^{21})}\,,
\end{split}
\end{equation}
which agrees with the ungauged kinetic terms of our truncated theory. In order to display the quaternionic K\"{a}hler structure of $SO(4,2)/(SO(4)\times SO(2))$, we introduce the vielbeins
\begin{equation}
\begin{split}
&f^1=2d\varphi_1\,,\quad f^2=2d\varphi_2\,,\quad f^3=e^{2\varphi_1}d\rho\,,\quad f^4=e^{-2\varphi_2} X_{(1)}\,,\\
&f^5=\sqrt{2}e^{-\varphi_1-\varphi_2}{d{\psi}^{11}}\,,\quad f^6=\sqrt{2}e^{-\varphi_1-\varphi_2}{d{\psi}^{21}}\,,\\
&f^7=\sqrt{2}e^{\varphi_1-\varphi_2}{{(d{\psi}^{12}-\rho d{\psi}^{11})}}\,,\quad f^8=\sqrt{2}e^{\varphi_1-\varphi_2}{{(d{\psi}^{22}-\rho d{\psi}^{21})}} \,.
\end{split}
\end{equation}
Let $M_{AB} = \hat E_{AB}-\hat E_{BA}$ be the generators of $\mf{so}(8)$, where $\hat E_{AB}$ is the $8\times 8$ matrix with 1 in the $(A, B)$ position and 0 elsewhere. Then, the associated spin connection 1-form is given by
\begin{equation}
\omega_{(1)} = f^3 Z_1 + f^4 Z_2 + f^5 Z_3 + f^6 Z_4 + f^7 Z_5 + f^8 Z_6 \,,
\end{equation}
where we have defined the combinations of the $\mf{so}(8)$ generators as 
\begin{equation}
\begin{split}
Z_1 &= -\frac{1}{2}(2M_{13} + M_{57} + M_{68})\,,\quad Z_2 = \frac{1}{2}(2M_{24}-M_{57}-M_{68})\,,\\
Z_3 &= \frac{1}{2}(M_{15}+M_{25}+M_{37}-M_{47}) \,,\quad Z_4 = \frac{1}{2}(M_{16}+M_{26}+M_{38}-M_{48})\,,\\
Z_5 &= \frac{1}{2}(M_{35}+M_{45}+M_{27}-M_{17}) \,,\quad Z_6 = \frac{1}{2}(M_{36}+M_{46}+M_{28}-M_{18}) \,.
\end{split}
\end{equation}
By defining an additional generator $Z_7 = (M_{56}+M_{78})/2$, the set of generators $\{Z_a\}$, with $a\in\{1,\dots,7\}$ close into the subalgebra $\mf{so}(4)\oplus\mf{u}(1)\cong\mf{so}(3)\oplus\mf{so}(3)\oplus\mf{u}(1)$. The $\mf{u}(1)$ generator is $Y =Z_1 + Z_2$. The first set of $\mf{so}(3)$ generators are
\begin{equation}
L_1=\frac{1}{2}(Z_1-Z_2)+Z_7,\quad L_2=\frac{1}{\sqrt{2}}\left(Z_3+Z_6\right),\quad L_3=\frac{1}{\sqrt{2}}\left(Z_5-Z_4\right)\,,
\end{equation}
and the second set of $\mf{so}(3)$ generators are
\begin{equation}
K_1=\frac{1}{2}(Z_1-Z_2)-Z_7,\quad K_2=\frac{1}{\sqrt{2}}\left(Z_3-Z_6\right),\quad K_3=\frac{1}{\sqrt{2}}\left(Z_4+Z_5\right) \,.
\end{equation}
These generators satisfy the canonical commutation relations
\begin{equation}
[L_i,L_j] = \epsilon_{ijk}L_k \,,\quad [L_i,K_j] = 0 \,,\quad [K_i,K_j] = \epsilon_{ijk}K_k \,,
\end{equation} 
for $i\in\{1,2,3\}$. In terms of our new generators, the spin connection is
\begin{equation}
\begin{split}
\omega_{(1)} &= \frac{f^3+f^4}{2}Y+\frac{f^3-f^4}{2}\left(L_1+K_1\right)\\
&\quad+\frac{f^5+f^8}{\sqrt{2}}L_2+\frac{f^5-f^8}{\sqrt{2}}K_2+\frac{f^7-f^6}{\sqrt{2}}L_3+\frac{f^6+f^7}{\sqrt{2}}K_3\,.
\end{split}
\end{equation}
From this, it is straightforward to calculate the curvature 2-form $R_{(2)}$ and the Ricci tensor, and we find that the metric is indeed Einstein with $R_{XY}=-2g_{XY}$. The holonomy group of an 8-dimensional quaternionic K\"{a}hler manifold is $SU(2)\times Sp(2)$, and the $SU(2)$ factor is an important feature of the quaternionic structure. We can use either $L_i$ or $K_i$ as the $SU(2)$ symmetry generators, and we choose $K_i$ here. If instead we use $L_i$, the $I=1$ component of the embedding coordinates in \eqref{embeddingcoordinate} must change from $h^{I=1}$ to $-h^{I=1}$. The $SU(2)$ factor of the curvature 2-form is defined as the projection $\boldsymbol{R} = -\tr(R_{(2)}\boldsymbol K)/2$, and is related to the triplet of complex structures $\boldsymbol{J}$ by $\boldsymbol{R}=-\boldsymbol{J}/2$. Explicitly, we find that 
\begin{equation}
\begin{split}
{J}^1&= -(f^{13}+f^{24}+f^{56}+f^{78})\,,\\
{J}^2&= \frac{1}{\sqrt{2}}(f^{15} + f^{18} + f^{25}- f^{28}-f^{36}+f^{37}-f^{46}-f^{47})\,,\\
{J}^3&= \frac{1}{\sqrt{2}}(f^{16}-f^{17}+f^{26}+f^{27}+f^{35}+f^{38}+f^{45}-f^{48})\,,
\end{split}
\end{equation}
where $f^{ij} = f^i\wedge f^j$. A little algebra shows that the complex structures obey the quaternionic algebra $J^iJ^j = -\delta^{ij}+\epsilon^{ijk}J^k$.

We are now ready to show that the scalar potential of our truncated theory is consistent with $\mc{N} = 2$ supersymmetry. The scalar potential for a general $\mc{N} = 2$, $D = 5$ gauged supergravity with no tensor multiplets and no Fayet-Iliopoulos terms is given by \cite{Bergshoeff:2004kh}
\begin{equation}\label{eq:5d_pot}
\mc{L}_{\mc{N}=2}^{pot}=4{g}^2\left(4\vec{P}\cdot\vec{P}-2\vec{P}^{x}\cdot \vec{P}_{x}-2W_{x}W^x-2\mc{N}_{A}\mc{N}^{A}\right)\,.
\end{equation}
The first two terms involve the moment maps for the Killing vectors $k^X_I$ defined via
\begin{equation}
\vec{P}_I=\frac{1}{2}\vec{J}_{X}^{\phantom{X}Y}\nabla_{Y}k^X_I\,,
\end{equation}
where $\vec{J}$ is the triplet of complex structures, with $\vec{P}$ and $\vec{P}_{x}$ are defined to be
\begin{equation}
\vec{P}\equiv  \frac{1}{2}h^I \vec{P}_{I}\,,\quad  \vec{P}_{x}\equiv - \frac{\sqrt{3}}{2}\partial_x h^I\vec{P}_{I}\,,
\end{equation}
and indices are raised and lowered using the metrics $g_{xy}$ and $a_{IJ}$ as discussed in section \ref{5dN=2SUSY}. Here, $\vec{J}$ is represented in terms of the Pauli matrices $\vec{\sigma}$, while $\boldsymbol{J}$ is represented by $\boldsymbol{K}$. This means that 
\begin{equation}
\vec{J}_{X}^{\phantom{X}Y}=\frac{1}{2}\boldsymbol{J}_{X}^{\phantom{X}Y}\,.
\end{equation}
We now turn to the term $W_{x}$ in the scalar potential, which is defined to be
\begin{equation}
W^x\equiv -\frac{3}{4}\bar{f}_{JI}^{\phantom{JI}P}h^Jh^{I}h^x_{P}\,,
\end{equation}
where $\bar{f}_{JI}^{\phantom{JI}P}$ are the structure constants of the gauge algebra. For our truncated theory, the algebra is Abelian, so $\bar{f}_{JI}^{\phantom{JI}P}=0$, and hence $W^x=0$. The final term in the scalar potential is given by
\begin{equation}
\mc{N}_{A}\mc{N}^{A}\equiv \frac{3}{16}h^Ik_I^Xg_{XY}h^Jk^Y_J\,.
\end{equation}
After explicitly evaluating the terms in \eqref{eq:5d_pot} using the ingredients in this appendix as well as those in section \ref{5dN=2SUSY}, we recover the scalar potential in our truncated theory.

\section{Parameterisation of $G_{2(2)}/SO(4)$ and the quaternionic K\"{a}hler structure}\label{G2parameterisation}

To construct a coset parameterisation of $G_{2(2)}/SO(4)$, we exponentiate the Borel subalgebra of $\mf{g}_2$. From \cite{Donos:2010ax}, an explicit realisation of the $\mf{g}_2$ generators in the fundamental representation is given by
\begin{align*}
&H_1 = \tfrac{1}{\sqrt{2}}(E_{11}-E_{22}+2E_{33}-2E_{55}+E_{66}-E_{77}) \,, & &H_2 = E_{11}+E_{22}-E_{66}-E_{77} \,,\\
&E_1 = -2E_{16}-2E_{27} \,,& &F_1 = -\tfrac{1}{2}(E_{61}+E_{72}) \,,\\
&E_2 = \tfrac{1}{2\sqrt{3}}(2E_{41}-E_{52}-E_{63}+2E_{74})\,, & &F_2 = \tfrac{2}{\sqrt{3}}(E_{14}-E_{25}-E_{36}+E_{47}) \,,\\
&E_3 = \tfrac{1}{\sqrt{3}}(E_{13}-2E_{24}+2E_{46}-E_{57})\,, & &F_3 = \tfrac{1}{\sqrt{3}}(E_{31}-E_{42}+E_{64}-E_{75}) \,,\\
&E_4 = -\tfrac{1}{\sqrt{3}}(E_{21}+E_{43}+E_{54}+E_{76})\,, & &F_4 = -\tfrac{1}{\sqrt{3}}(E_{12}+2E_{34}+2E_{45}+E_{67})\,,\\
&E_5=\tfrac{1}{2}(-E_{51}+E_{73}) \,, & &F_5 = -2E_{15}+2E_{37} \,,\\
&E_6 =-E_{23}-E_{56} \,, & &F_6 = -E_{32}-E_{65} \,,\numberthis
\end{align*}
where $E_{MN}$ is the $7\times7$ matrix with 1 in the $M^{\text{th}}$ row and $N^{\text{th}}$ column. $H_a$, $a\in\{1,2\}$, are the Cartan generators, and $E_i$ and $F_i$, $i\in\{1,\dots,6\}$, are the positive-root and negative-root generators respectively. The positive roots are given by 
\begin{equation}
\begin{split}
&\alpha_1 = (0,2) \,, \\
&\alpha_3 = (-\tfrac{1}{\sqrt{3}},1) \,,\\
&\alpha_5 = (-\sqrt{3},-1) \,,
\end{split}
\quad
\begin{split}
&\alpha_2 = (-\tfrac{1}{\sqrt{3}},-1) \,,\\
&\alpha_4 = (-\tfrac{2}{\sqrt{3}},0) \,,\\
&\alpha_6 = (-\sqrt{3},1) \,.
\end{split}
\end{equation}
With these ingredients, we construct the following coset representative of $G_{2(2)}/SO(4)$,
\begin{equation}
\mc{V} = e^{\frac{1}{2}\varphi\cdot H}e^{-\tau E_1}e^{\sqrt{3}(-\theta E_2+\Theta E_3)}e^{\left(\Gamma-\tau[G+2\theta^2\left(\Theta+\tau\theta/2\right)]-\theta\Theta^2\right)E_6}e^{2\sqrt{3}CE_4-GE_5} \,,
\end{equation}
where $\varphi\cdot H =\varphi_aH_a$, and
\begin{equation}
q^u = (\varphi_1,\varphi_2,\tau,\theta,\Theta,C,G,\Gamma)
\end{equation}
are the coordinates on $G_{2(2)}/SO(4)$. The metric on $G_{2(2)}/SO(4)$ is given by
\begin{equation}\label{metricdef}
h_{uv}dq^udq^v = \frac{1}{4}\tr(P_{(1)}^2) \,,
\end{equation}
where $P_{(1)}$ is the $\mf{g}_2$-valued 1-form
\begin{equation}
P_{(1)} = \frac{1}{2}\left(d\mc{V}\cdot\mc{V}^{-1} + \sharp(d\mc{V}\cdot\mc{V}^{-1})\right) \,.
\end{equation} 
Here, the operator $\sharp$ is the generalised transpose, and is defined by
\begin{equation}
\sharp (H_a) = H_a \,,\quad \sharp(E_i) = F_i \,,\quad \sharp(F_i) = E_i \,,
\end{equation}
and our generators satisfy $\tr(H_a\sharp(H_b)) = 4\delta_{ab}$ and $\tr(E_i\sharp(E_i)) = 2\delta_{ij}$. By defining the 1-forms
\begin{equation}
\begin{split}
\mc{F}^1_{(1)} &= d\tau \,,\\
\mc{F}^2_{(1)} &= \sqrt{3}d\theta \,,\\
\mc{F}^3_{(1)} &= \sqrt{3}\left(d\Theta + \tau d\theta\right) \,,\\
\mc{F}^4_{(1)} &= 2\sqrt{3}\left(dC + \tfrac{1}{2}\theta d\Theta - \tfrac{1}{2}\Theta d\theta\right) \,,\\
\mc{F}^5_{(1)} &= dG -6\theta dC + \theta(\Theta d\theta - \theta d\Theta) \,,\\
\mc{F}^6_{(1)} &= d\Gamma - 3\rho(2dC + \theta(d\Theta + \tau d\theta))- (G+2\theta^2\rho)d\tau \,,
\end{split}
\end{equation}
where we recall the definition $\rho = \Theta +\tau\theta$, we find that \eqref{metricdef} gives the following expression for the metric
\begin{equation}\label{G2metric}
h_{uv}dq^udq^v = \frac{1}{4}(d\varphi_1)^2 + \frac{1}{4}(d\varphi_2)^2+ \frac{1}{4}\sum_{i=1}^6e^{\alpha_i\cdot\varphi}(\mc{F}^i_{(1)})^2 \,,
\end{equation}
which matches the ungauged kinetic terms of our theory. 

To match the potential terms, we will require the moment maps defined in \eqref{momentum}, which in turn require the spin connection and curvature associated to the metric $h_{uv}$. First, we define the vielbeins
\begin{equation}
f^1 = \frac{1}{2}d\varphi_1 \,,\quad f^2 = \frac{1}{2}d\varphi_2 \,,\quad f^{i+2} = \frac{1}{2}e^{\frac{1}{2}\alpha_i\cdot\varphi}\mc{F}^i_{(1)} \,.
\end{equation}
Let $M_{\ov u\,\ov v} = E_{\ov u\,\ov v} - E_{\ov v\,\ov u}$, with $E_{\ov u\,\ov v}$ the $8\times 8$ matrix with 1 in the $\ov u^{\text{th}}$ row and $\ov v^{\text{th}}$ column be the generators of $\mf{so}(8)$. The spin connection is then given by
\begin{equation}
\omega_{(1)} = f^3Z_1 + f^4Z_2 + f^5Z_3 + f^6Z_4 + f^7Z_5 + f^8Z_6 \,,
\end{equation}
where we defined the combinations of $\mf{so}(8)$ generators as
\begin{equation}
\begin{split}
Z_1 &= -2M_{23}+M_{45}+M_{78} \,, \\
Z_2 &= \tfrac{1}{\sqrt{3}}M_{14}+\tfrac{2}{\sqrt{3}}M_{56} + M_{24}-M_{35}-M_{67} \,,\\
Z_3 &= \tfrac{1}{\sqrt{3}}M_{15}-\tfrac{2}{\sqrt{3}}M_{46} - M_{25}- M_{34}-M_{68} \,, \\
Z_4 &= \tfrac{2}{\sqrt{3}}M_{16}- \tfrac{2}{\sqrt{3}}M_{45} + M_{47} + M_{58} \,,\\
Z_5 &= \sqrt{3}M_{17} + M_{27} - M_{38} + M_{46} \,, \\
Z_6 &= \sqrt{3}M_{18} - M_{28} - M_{37} + M_{56} \,.
\end{split}
\end{equation}
These close into an $\mf{so}(4) \cong \mf{so}(3)\oplus\mf{so}(3)$ subalgebra of $\mf{so}(8)$. More specifically, the first set of $\mf{so}(3)$ generators are
\begin{equation}
J_1 = \frac{\sqrt{3}}{4}\left(Z_4 + \frac{1}{\sqrt{3}}Z_1\right) \,,\quad J_2 = \frac{1}{4}\left(Z_5+\sqrt{3}Z_3\right)\,,\quad J_3 = \frac{\sqrt{3}}{4}\left(Z_2 - \frac{1}{\sqrt{3}}Z_6\right)  \,,
\end{equation} 
and the second set of $\mf{so}(3)$ generators are
\begin{equation}
L_1 = \frac{3}{4}\left(Z_1 - \frac{1}{\sqrt{3}}Z_4\right) \,,\quad L_2 = \frac{\sqrt{3}}{4}\left(Z_3 - \sqrt{3}Z_5\right) \,,\quad L_3 = \frac{3}{4}\left(Z_6 + \frac{1}{\sqrt{3}}Z_2\right) \,.
\end{equation}
These generators satisfy the canonical commutation relations
\begin{equation}
[J_x,J_y]=\epsilon_{xyz}J_z\,,\quad [J_x,L_y] = 0\,,\quad [L_x,L_y]=\epsilon_{xyz}L_z \,.
\end{equation}
In terms of these generators, the spin connection is 
\begin{equation}
\omega^{}_{(1)} = \omega^x_{(1)}J_x + \Delta^x_{(1)}L_x \,,
\end{equation}
where we defined 
\begin{equation}
\begin{split}
\omega^1_{(1)} &= f^3 + \sqrt{3}f^6 \,, \\
\omega^2_{(1)} &= \sqrt{3}f^5 + f^7 \,, \\
\omega^3_{(1)} &= \sqrt{3}f^4 - f^8 \,.
\end{split}
\quad
\begin{split}
\Delta^1_{(1)} &= f^3 - \tfrac{1}{\sqrt{3}}f^6 \,, \\
\Delta^2_{(1)} &= \tfrac{1}{\sqrt{3}}f^5 - f^7 \,, \\
\Delta^3_{(1)} &= \tfrac{1}{\sqrt{3}}f^4 + f^8 \,, 
\end{split}
\end{equation}
The curvatures associated to $\omega^x_{(1)}$ and $\Delta^x_{(1)}$ are defined by 
\begin{equation}
\begin{split}
&-2K^x_{(2)} \equiv d\omega^x_{(1)} + \frac{1}{2}\epsilon^{xyz}\omega^y_{(1)}\wedge\omega^z_{(1)} \,,\\
&-2D^x_{(2)} \equiv d\Delta^x_{(1)} + \frac{1}{2}\epsilon^{xyz}\Delta^y_{(1)}\wedge\Delta^z_{(1)}  \,.
\end{split}
\end{equation}
Explicitly, they are 
\begin{equation}
\begin{split}
K^1_{(2)} &= -\frac{1}{2}\left(2f^{23}-2f^{16}+f^{45}-\sqrt{3}f^{47}-\sqrt{3}f^{58}-f^{78}\right) \,,\\
K^2_{(2)} &= -\frac{1}{2}\left(f^{46}-f^{15}+\sqrt{3}f^{25}+\sqrt{3}f^{34}-\sqrt{3}f^{17}-f^{27}+\sqrt{3}f^{68}+f^{38}\right) \,,\\
K^3_{(2)} &= -\frac{1}{2}\left(f^{37}-f^{14}-f^{28}-\sqrt{3}f^{24}+\sqrt{3}f^{18}-f^{56}+\sqrt{3}f^{67}-2f^{37}+\sqrt{3}f^{35}\right) \,,
\end{split}
\end{equation}
and
\begin{equation}
\begin{split}
D^1_{(2)} &= -\frac{1}{2}\left(2f^{23}+\frac{2}{3}f^{16}-\frac{5}{3}f^{45} + \frac{1}{\sqrt{3}}f^{47} + \frac{1}{\sqrt{3}}f^{58} - f^{78}\right ) \,,\\
D^2_{(2)} &= -\frac{1}{2}\left(f^{27}- \frac{1}{3}f^{15} + \frac{1}{\sqrt{3}}f^{25} + \frac{1}{\sqrt{3}}f^{34} +\sqrt{3}f^{17}+\frac{5}{3}f^{46}-f^{38}+\frac{1}{\sqrt{3}}f^{68} \right)\,,\\
D^3_{(2)} &= -\frac{1}{2}\left(f^{28} - \frac{1}{3}f^{14}-\frac{1}{\sqrt{3}}f^{24}-\frac{5}{3}f^{56}+f^{37}-\sqrt{3}f^{18} + \frac{1}{\sqrt{3}}f^{35}+\frac{1}{\sqrt{3}}f^{67} \right) \,,
\end{split}
\end{equation}
where $f^{ij}\equiv f^i\wedge f^j$. From this, we find that $(\mc{J}^x)^{\ov u}_{\ph{\ov u}\ov v} =\delta^{\ov u\ov w}K^x_{\ov w\,\ov v}$ is a triplet of complex structures satisfying the quaternionic algebra 
\begin{equation}
\mc{J}^x\mc{J}^y = -\delta^{xy} + \epsilon^{xyz}\mc{J}^z \,,
\end{equation}
meaning that $K^x_{(2)}$ are indeed the curvatures corresponding to the $SU(2)$ factor of the $SU(2)\times Sp(2)$ holonomy of $G_{2(2)}/SO(4)$. The curvatures $D^x_{(2)}$ corresponding to the connection $\Delta^x_{(1)}$ do not lead to a quaternionic structure. 

With the quaternionic structure made explicit, we are ready to compute the moment maps defined in \eqref{momentum}. For the Killing vectors $k_0$ and $k_1$ written in \eqref{killing4d}, we have 
\begin{equation}
\begin{split}
P^1_0 &= -\frac{3}{4}e^{\frac{1}{2}\alpha_4\cdot\varphi}(l-2\theta^2) - \frac{1}{2}e^{\frac{1}{2}\alpha_1\cdot\varphi} \,,\\
P^2_0 &= \frac{1}{2}e^{\frac{1}{2}\alpha_5\cdot\varphi}\left(\frac{3}{2}l-\theta^2\right)\theta + \frac{3}{2}e^{\frac{1}{2}\alpha_3\cdot\varphi}\theta \,,\\
P^3_0 &= -\frac{1}{2}e^{\frac{1}{2}\alpha_6\cdot\varphi}\left(G + \left(\frac{3}{2}l-\theta^2\right)\rho\right) \,,
\end{split}
\end{equation}
and 
\begin{equation}
\begin{split}
P^1_1 &= \frac{3}{2}e^{\frac{1}{2}\alpha_4\cdot\varphi}\,,\\
P^2_1 &= -\frac{3}{2}e^{\frac{1}{2}\alpha_5\cdot\varphi}\theta\,,\\
P^3_1 &= \frac{3}{2}e^{\frac{1}{2}\alpha_6\cdot\varphi}\rho \,,
\end{split}
\end{equation}
respectively. A bit of algebra shows that the potential given in \eqref{potentialcan} does indeed match the potential given in \eqref{4dpotentialterm}.

\end{appendix}


\addcontentsline{toc}{section}{References}

\begin{thebibliography}{99}


\bibitem{Gauntlett:2001ps}
J.~P.~Gauntlett, N.~Kim, D.~Martelli and D.~Waldram,
``Wrapped five-branes and N=2 superYang-Mills theory,''
Phys. Rev. D \textbf{64} (2001), 106008
doi:10.1103/PhysRevD.64.106008
[arXiv:hep-th/0106117 [hep-th]].
    
\bibitem{Bigazzi:2001aj}
F.~Bigazzi, A.~L.~Cotrone and A.~Zaffaroni,
``N=2 gauge theories from wrapped five-branes,''
Phys. Lett. B \textbf{519} (2001), 269-276
doi:10.1016/S0370-2693(01)01100-5
[arXiv:hep-th/0106160 [hep-th]].

\bibitem{Gauntlett:2001ur}
J.~P.~Gauntlett, N.~Kim, D.~Martelli and D.~Waldram,
``Five-branes wrapped on SLAG three cycles and related geometry,''
JHEP \textbf{11} (2001), 018
doi:10.1088/1126-6708/2001/11/018
[arXiv:hep-th/0110034 [hep-th]].

\bibitem{Witten:1988ze}
E.~Witten,
``Topological Quantum Field Theory,''
Commun. Math. Phys. \textbf{117} (1988), 353
doi:10.1007/BF01223371

\bibitem{Bershadsky:1995qy}
M.~Bershadsky, C.~Vafa and V.~Sadov,
``D-branes and topological field theories,''
Nucl. Phys. B \textbf{463} (1996), 420-434
doi:10.1016/0550-3213(96)00026-0
[arXiv:hep-th/9511222 [hep-th]].


\bibitem{Maldacena:2000mw}
J.~M.~Maldacena and C.~Nunez,
``Supergravity description of field theories on curved manifolds and a no go theorem,''
Int. J. Mod. Phys. A \textbf{16} (2001), 822-855
doi:10.1142/S0217751X01003937
[arXiv:hep-th/0007018 [hep-th]].

\bibitem{Maldacena:2000yy}
J.~M.~Maldacena and C.~Nunez,
``Towards the large N limit of pure N=1 superYang-Mills,''
Phys. Rev. Lett. \textbf{86} (2001), 588-591
doi:10.1103/PhysRevLett.86.588
[arXiv:hep-th/0008001 [hep-th]].

\bibitem{Brinne:2000fh}
B.~Brinne, A.~Fayyazuddin, S.~Mukhopadhyay and D.~J.~Smith,
``Supergravity M5-branes wrapped on Riemann surfaces and their QFT duals,''
JHEP \textbf{12} (2000), 013
doi:10.1088/1126-6708/2000/12/013
[arXiv:hep-th/0009047 [hep-th]].

\bibitem{Acharya:2000mu}
B.~S.~Acharya, J.~P.~Gauntlett and N.~Kim,
``Five-branes wrapped on associative three cycles,''
Phys. Rev. D \textbf{63} (2001), 106003
doi:10.1103/PhysRevD.63.106003
[arXiv:hep-th/0011190 [hep-th]].

\bibitem{Gauntlett:2000ng}
J.~P.~Gauntlett, N.~Kim and D.~Waldram,
``M Five-branes wrapped on supersymmetric cycles,''
Phys. Rev. D \textbf{63} (2001), 126001
doi:10.1103/PhysRevD.63.126001
[arXiv:hep-th/0012195 [hep-th]].

\bibitem{Nieder:2000kc}
H.~Nieder and Y.~Oz,
``Supergravity and D-branes wrapping special Lagrangian cycles,''
JHEP \textbf{03} (2001), 008
doi:10.1088/1126-6708/2001/03/008
[arXiv:hep-th/0011288 [hep-th]].

\bibitem{Gauntlett:2001jj}
J.~P.~Gauntlett and N.~Kim,
``M five-branes wrapped on supersymmetric cycles. 2.,''
Phys. Rev. D \textbf{65} (2002), 086003
doi:10.1103/PhysRevD.65.086003
[arXiv:hep-th/0109039 [hep-th]].

\bibitem{Nunez:2001pt}
C.~Nunez, I.~Y.~Park, M.~Schvellinger and T.~A.~Tran,
``Supergravity duals of gauge theories from F(4) gauged supergravity in six-dimensions,''
JHEP \textbf{04} (2001), 025
doi:10.1088/1126-6708/2001/04/025
[arXiv:hep-th/0103080 [hep-th]].

\bibitem{Edelstein:2001pu}
J.~D.~Edelstein and C.~Nunez,
``D6-branes and M theory geometrical transitions from gauged supergravity,''
JHEP \textbf{04} (2001), 028
doi:10.1088/1126-6708/2001/04/028
[arXiv:hep-th/0103167 [hep-th]].

\bibitem{Gomis:2001vg}
J.~Gomis and T.~Mateos,
``D6 branes wrapping Kahler four cycles,''
Phys. Lett. B \textbf{524} (2002), 170-176
doi:10.1016/S0370-2693(01)01389-2
[arXiv:hep-th/0108080 [hep-th]].

\bibitem{Hernandez:2001bh}
R.~Hernandez,
``Branes wrapped on coassociative cycles,''
Phys. Lett. B \textbf{521} (2001), 371-375
doi:10.1016/S0370-2693(01)01219-9
[arXiv:hep-th/0106055 [hep-th]].

\bibitem{Schvellinger:2001ib}
M.~Schvellinger and T.~A.~Tran,
``Supergravity duals of gauge field theories from SU(2) x U(1) gauged supergravity in five-dimensions,''
JHEP \textbf{06} (2001), 025
doi:10.1088/1126-6708/2001/06/025
[arXiv:hep-th/0105019 [hep-th]].

\bibitem{Naka:2002jz}
M.~Naka,
``Various wrapped branes from gauged supergravities,''
[arXiv:hep-th/0206141 [hep-th]].

\bibitem{Gauntlett:2001qs}
J.~P.~Gauntlett, N.~Kim, S.~Pakis and D.~Waldram,
``Membranes wrapped on holomorphic curves,''
Phys. Rev. D \textbf{65} (2002), 026003
doi:10.1103/PhysRevD.65.026003
[arXiv:hep-th/0105250 [hep-th]].

\bibitem{Benini:2013cda}
F.~Benini and N.~Bobev,
``Two-dimensional SCFTs from wrapped branes and c-extremization,''
JHEP \textbf{06} (2013), 005
doi:10.1007/JHEP06(2013)005
[arXiv:1302.4451 [hep-th]].


\bibitem{Suh:2018tul}
M.~Suh,
``Supersymmetric AdS$_{6}$ black holes from F(4) gauged supergravity,''
JHEP \textbf{01} (2019), 035
doi:10.1007/JHEP01(2019)035
[arXiv:1809.03517 [hep-th]].

\bibitem{Kim:2019fsg}
N.~Kim and M.~Shim,
``Wrapped Brane Solutions in Romans $F(4)$ Gauged Supergravity,''
Nucl. Phys. B \textbf{951} (2020), 114882
doi:10.1016/j.nuclphysb.2019.114882
[arXiv:1909.01534 [hep-th]].

\bibitem{Cvetic:2003xr}
M.~Cvetic, G.~W.~Gibbons and C.~N.~Pope,
``A String and M theory origin for the Salam-Sezgin model,''
Nucl. Phys. B \textbf{677} (2004), 164-180
doi:10.1016/j.nuclphysb.2003.10.016
[arXiv:hep-th/0308026 [hep-th]].

\bibitem{Cheung:2019pge}
K.~C.~M. Cheung, J.~P.~Gauntlett and C.~Rosen,
``Consistent KK truncations for M5-branes wrapped on Riemann surfaces,''
Class. Quant. Grav. \textbf{36} (2019) no.22, 225003
doi:10.1088/1361-6382/ab41b3
[arXiv:1906.08900 [hep-th]].

\bibitem{Cassani:2019vcl}
D.~Cassani, G.~Josse, M.~Petrini and D.~Waldram,
``Systematics of consistent truncations from generalised geometry,''
JHEP \textbf{11} (2019), 017
doi:10.1007/JHEP11(2019)017
[arXiv:1907.06730 [hep-th]].

\bibitem{Donos:2010ax}
A.~Donos, J.~P.~Gauntlett, N.~Kim and O.~Varela,
``Wrapped M5-branes, consistent truncations and AdS/CMT,''
JHEP \textbf{12} (2010), 003
doi:10.1007/JHEP12(2010)003
[arXiv:1009.3805 [hep-th]].

\bibitem{Freedman:1978ra}
D.~Z.~Freedman and J.~H.~Schwarz,
``N=4 Supergravity Theory with Local SU(2) x SU(2) Invariance,''
Nucl. Phys. B \textbf{137} (1978), 333-339
doi:10.1016/0550-3213(78)90526-6



\bibitem{Chamseddine:1997mc}
A.~H.~Chamseddine and M.~S.~Volkov,
``NonAbelian solitons in N=4 gauged supergravity and leading order string theory,''
Phys. Rev. D \textbf{57} (1998), 6242-6254
doi:10.1103/PhysRevD.57.6242
[arXiv:hep-th/9711181 [hep-th]].

\bibitem{Cowdall:1998bu}
P.~M.~Cowdall and P.~K.~Townsend,
``Gauged supergravity vacua from intersecting branes,''
Phys. Lett. B \textbf{429} (1998), 281-288
[erratum: Phys. Lett. B \textbf{434} (1998), 458-458]
doi:10.1016/S0370-2693(98)00445-6
[arXiv:hep-th/9801165 [hep-th]].


\bibitem{Patera:1976ud}
J.~Patera, R.~T.~Sharp, P.~Winternitz and H.~Zassenhaus,
``Invariants of Real Low Dimension Lie Algebras,''
J. Math. Phys. \textbf{17} (1976), 986-994
doi:10.1063/1.522992


\bibitem{Nastase:1999cb}
H.~Nastase, D.~Vaman and P.~van Nieuwenhuizen,
``Consistent nonlinear K K reduction of 11-d supergravity on AdS(7) x S(4) and selfduality in odd dimensions,''
Phys. Lett. B \textbf{469} (1999), 96-102
doi:10.1016/S0370-2693(99)01266-6
[arXiv:hep-th/9905075 [hep-th]].

\bibitem{Nastase:1999kf}
H.~Nastase, D.~Vaman and P.~van Nieuwenhuizen,
``Consistency of the AdS(7) x S(4) reduction and the origin of selfduality in odd dimensions,''
Nucl. Phys. B \textbf{581} (2000), 179-239
doi:10.1016/S0550-3213(00)00193-0
[arXiv:hep-th/9911238 [hep-th]].


\bibitem{Boonstra:1998mp}
H.~J.~Boonstra, K.~Skenderis and P.~K.~Townsend,
``The domain wall / QFT correspondence,''
JHEP \textbf{01} (1999), 003
doi:10.1088/1126-6708/1999/01/003
[arXiv:hep-th/9807137 [hep-th]].


\bibitem{Cvetic:2000dm}
M.~Cvetic, H.~Lu and C.~N.~Pope,
``Consistent Kaluza-Klein sphere reductions,''
Phys. Rev. D \textbf{62} (2000), 064028
doi:10.1103/PhysRevD.62.064028
[arXiv:hep-th/0003286 [hep-th]].

\bibitem{Cvetic:2000ah}
M.~Cvetic, H.~Lu, C.~N.~Pope, A.~Sadrzadeh and T.~A.~Tran,
``S**3 and S**4 reductions of type IIA supergravity,''
Nucl. Phys. B \textbf{590} (2000), 233-251
doi:10.1016/S0550-3213(00)00466-1
[arXiv:hep-th/0005137 [hep-th]].

\bibitem{Schon:2006kz}
J.~Schon and M.~Weidner,
``Gauged N=4 supergravities,''
JHEP \textbf{05} (2006), 034
doi:10.1088/1126-6708/2006/05/034
[arXiv:hep-th/0602024 [hep-th]].

\bibitem{DallAgata:2001wgl}
G.~Dall'Agata, C.~Herrmann and M.~Zagermann,
``General matter coupled N=4 gauged supergravity in five-dimensions,''
Nucl. Phys. B \textbf{612} (2001), 123-150
doi:10.1016/S0550-3213(01)00367-4
[arXiv:hep-th/0103106 [hep-th]].


\bibitem{Awada:1985ep}
M.~Awada and P.~K.~Townsend,
``$N=4$ Maxwell-einstein Supergravity in Five-dimensions and Its SU(2) Gauging,''
Nucl. Phys. B \textbf{255} (1985), 617-632
doi:10.1016/0550-3213(85)90156-7

\bibitem{Lu:1998xt}
H.~Lu, C.~N.~Pope and K.~S.~Stelle,
``M theory / heterotic duality: A Kaluza-Klein perspective,''
Nucl. Phys. B \textbf{548} (1999), 87-138
doi:10.1016/S0550-3213(99)00086-3
[arXiv:hep-th/9810159 [hep-th]].




\bibitem{Ghanam:2014}
R.~Ghanam and G.~Thompson,
``Minimal matrix representations of five-dimensional Lie algebras,''
[arXiv:1402.4879 [math.DG]].


\bibitem{Ceresole:2000jd}
A.~Ceresole and G.~Dall'Agata,
``General matter coupled N=2, D = 5 gauged supergravity,''
Nucl. Phys. B \textbf{585} (2000), 143-170
doi:10.1016/S0550-3213(00)00339-4
[arXiv:hep-th/0004111 [hep-th]].

\bibitem{Gunaydin:1999zx}
M.~Gunaydin and M.~Zagermann,
``The Gauging of five-dimensional, N=2 Maxwell-Einstein supergravity theories coupled to tensor multiplets,''
Nucl. Phys. B \textbf{572} (2000), 131-150
doi:10.1016/S0550-3213(99)00801-9
[arXiv:hep-th/9912027 [hep-th]].


\bibitem{Bergshoeff:2004kh}
E.~Bergshoeff, S.~Cucu, T.~de Wit, J.~Gheerardyn, S.~Vandoren and A.~Van Proeyen,
``N = 2 supergravity in five-dimensions revisited,''
Class. Quant. Grav. \textbf{21} (2004), 3015-3042
doi:10.1088/0264-9381/23/23/C01
[arXiv:hep-th/0403045 [hep-th]].

%

\bibitem{Andrianopoli:1996vr}
L.~Andrianopoli, M.~Bertolini, A.~Ceresole, R.~D'Auria, S.~Ferrara and P.~Fre',
``General matter coupled N=2 supergravity,''
Nucl. Phys. B \textbf{476} (1996), 397-417
doi:10.1016/0550-3213(96)00344-6
[arXiv:hep-th/9603004 [hep-th]].


\bibitem{Andrianopoli:1996cm}
L.~Andrianopoli, M.~Bertolini, A.~Ceresole, R.~D'Auria, S.~Ferrara, P.~Fre and T.~Magri,
``N=2 supergravity and N=2 superYang-Mills theory on general scalar manifolds: Symplectic covariance, gaugings and the momentum map,''
J. Geom. Phys. \textbf{23} (1997), 111-189
doi:10.1016/S0393-0440(97)00002-8
[arXiv:hep-th/9605032 [hep-th]].

\bibitem{Freedman:2012zz}
D.~Z.~Freedman and A.~Van Proeyen, (2012)
``Supergravity,'' Cambridge University Press.

\bibitem{Bak:2003jk}
D.~Bak, M.~Gutperle and S.~Hirano,
``A Dilatonic deformation of AdS(5) and its field theory dual,''
JHEP \textbf{05} (2003), 072
doi:10.1088/1126-6708/2003/05/072
[arXiv:hep-th/0304129 [hep-th]].

\bibitem{Arav:2018njv}
I.~Arav, J.~P.~Gauntlett, M.~Roberts and C.~Rosen,
``Spatially modulated and supersymmetric deformations of ABJM theory,''
JHEP \textbf{04} (2019), 099
doi:10.1007/JHEP04(2019)099
[arXiv:1812.11159 [hep-th]].

\bibitem{Duff:1993ye}
M.~J.~Duff and J.~X.~Lu,
``Black and super p-branes in diverse dimensions,''
Nucl. Phys. B \textbf{416} (1994), 301-334
doi:10.1016/0550-3213(94)90586-X
[arXiv:hep-th/9306052 [hep-th]].

\bibitem{Gauntlett:2003di}
J.~P.~Gauntlett,
``Branes, calibrations and supergravity,''
Clay Math. Proc. \textbf{3} (2004), 79-126
[arXiv:hep-th/0305074 [hep-th]].

\bibitem{Karndumri:2015sia}
P.~Karndumri and E.~\'O.~Colg\'ain,
``3D supergravity from wrapped M5-branes,''
JHEP \textbf{03} (2016), 188
doi:10.1007/JHEP03(2016)188
[arXiv:1508.00963 [hep-th]].

\bibitem{Hull:1984yy}
C.~M.~Hull,
``A New Gauging of $N=8$ Supergravity,''
Phys. Rev. D \textbf{30} (1984), 760
doi:10.1103/PhysRevD.30.760

\bibitem{Hull:1984vg}
C.~M.~Hull,
``Noncompact Gaugings of $N=8$ Supergravity,''
Phys. Lett. B \textbf{142} (1984), 39
doi:10.1016/0370-2693(84)91131-6

\bibitem{Hull:1984qz}
C.~M.~Hull,
``More Gaugings of $N=8$ Supergravity,''
Phys. Lett. B \textbf{148} (1984), 297-300
doi:10.1016/0370-2693(84)90091-1

\bibitem{deWit:1986oxb}
B.~de Wit and H.~Nicolai,
``The Consistency of the S**7 Truncation in D=11 Supergravity,''
Nucl. Phys. B \textbf{281} (1987), 211-240
doi:10.1016/0550-3213(87)90253-7

\bibitem{Guarino:2015vca}
A.~Guarino and O.~Varela,
``Consistent $ \mathcal{N}=8 $ truncation of massive IIA on S$^{6}$,''
JHEP \textbf{12} (2015), 020
doi:10.1007/JHEP12(2015)020
[arXiv:1509.02526 [hep-th]].


\bibitem{Cassani:2020cod}
D.~Cassani, G.~Josse, M.~Petrini and D.~Waldram,
``$\mathcal{N} $ = 2 consistent truncations from wrapped M5-branes,''
JHEP \textbf{02} (2021), 232
doi:10.1007/JHEP02(2021)232
[arXiv:2011.04775 [hep-th]].

\bibitem{Hohm:2014qga}
O.~Hohm and H.~Samtleben,
``Consistent Kaluza-Klein Truncations via Exceptional Field Theory,''
JHEP \textbf{01} (2015), 131
doi:10.1007/JHEP01(2015)131
[arXiv:1410.8145 [hep-th]].

\bibitem{Malek:2020jsa}
E.~Malek and V.~Vall Camell,
``Consistent truncations around half-maximal AdS$_5$ vacua of 11-dimensional supergravity,''
[arXiv:2012.15601 [hep-th]].



\end{thebibliography}
\end{document}